 \documentclass[final,5p,times,twocolumn,authoryear]{elsarticle}

\usepackage{amssymb}
\usepackage{tabularx}
\usepackage{color,graphicx}
\usepackage{verbatim}

\usepackage[switch]{lineno}
\bibpunct{(}{)}{;}{a}{,}{,}

\usepackage{ulem}
\usepackage{pstricks}
\usepackage{color}

\usepackage{hyperref}
\hypersetup{
    bookmarks=true,         
    unicode=true,           
    pdftoolbar=true,        
    pdfmenubar=true,        
    pdffitwindow=true,      
    pdfauthor={Devogele et al.},
    pdfsubject={Planetary Science},
    pdfkeywords={},         
    pdfnewwindow=true,      
    colorlinks=true,        
    linkcolor=gray,         
    citecolor=blue,         
    filecolor=gray,         
    urlcolor=gray           
}

\journal{Special Issue: Asteroids}

\begin{document}

\begin{frontmatter}



\title{New polarimetric and spectroscopic evidence of anomalous enrichment in spinel-bearing Calcium-Aluminium-rich Inclusions among L-type asteroids}


\author{M. Devog\`{e}le$^{1,2}$, P. Tanga$^{2}$, A. Cellino$^{3}$, Ph. Bendjoya$^{2}$, J.-P. Rivet$^{2}$, J. Surdej$^{1}$, D. Vernet$^{4}$, J. M. Sunshine$^{5}$, S. J. Bus$^{6}$, L. Abe$^{2}$, S. Bagnulo$^{7}$, G. Borisov$^{7,8}$, H. Campins$^{9}$, B. Carry$^{2,10}$, J. Licandro$^{11,12}$,W. McLean$^{7,13}$, N. Pinilla-Alonso$^{14}$  \\
}

\address{$^{1}$Universit\'{e} de Li\`{e}ge, Space sciences, Technologies and Astrophysics Research (STAR) Institute, All\'{e}e du 6 Ao\^{u}t 19c, Sart Tilman, 4000 Li\`{e}ge, Belgium \\
$^{2}$Universit\'e C{\^o}te d'Azur, Observatoire de la C{\^o}te d'Azur, CNRS, Laboratoire Lagrange UMR7293, Nice, France\\
$^{3}$INAF - Osservatorio Astrofisico di Torino, Pino Torinese, Italy\\
$^{4}$Universit\'e C{\^o}te d'Azur, Obs. de la C\^ote d'Azur, UMR7293 CNRS Laboratoire Optique, Bv de l'Observatoire, CS 34229, 06304 Nice, France\\
$^{5}$Department for Astronomy, University of Maryland, College Park, MD 20742-2421, USA \\
$^{6}$Institute for Astronomy, 640 N. Aohoku Place, Hilo, HI 96720, United States \\
$^{7}$Armagh Observatory and Planetarium, College Hill, Armagh BT61 9DG, UK\\
$^{8}$Institute of Astronomy and NAO, 72 Tsarigradsko Chaussee Blvd., BG-1784 Sofia, Bulgaria \\
$^{9}$Physics Department, University of Central Florida, P.O. Box 162385, Orlando, FL 32816-2385, USA \\
$^{10}$IMCCE, Observatoire de Paris, PSL Research University, CNRS, Sorbonne Universit\'es, UPMC Univ Paris 06, Univ. Lille, France\\
$^{11}$Instituto de Astrof\'{\i}sica de Canarias (IAC), C/V\'{\i}a L\'{a}ctea s\/n, 38205 La Laguna, Spain \\
$^{12}$Departamento de Astrof\'{\i}sica, Universidad de La Laguna, 38206 La Laguna, Tenerife, Spain \\
$^{13}$Department of Physics, University of Warwick, Gibbet Hill Road, Coventry CV4 7AL, UK).\\
$^{14}$Florida Space Institute, University of Central Florida, Orlando, FL 32816, USA \\}

\begin{abstract}
Asteroids can be classified into several groups based on their spectral
reflectance. Among these groups, the one belonging to the  L-class
in the taxonomic classification based on visible and near-infrared
spectra exhibit several peculiar properties. First, their near-infrared
spectrum is characterized by a strong absorption band interpreted as
the diagnostic of a high content of the FeO bearing spinel
mineral. This mineral is one of the main constituents of
Calcium-Aluminum-rich Inclusions (CAI) the oldest mineral
compounds found in the solar system. In polarimetry, they possess an
uncommonly large value of the inversion angle incompatible with all
known asteroid belonging to other taxonomical classes. Asteroids found
to possess such a high inversion angle are commonly called Barbarians
based on the first asteroid on which this property was first
identified, (234)~Barbara.
In this paper we present the results of an extensive campaign of
polarimetric and spectroscopic observations of L-class objects. We
have derived phase-polarization curves for a sample of 7 Barbarians,
finding a variety of inversion angles ranging between 25 and
30$^{\circ}$. Spectral reflectance data exhibit variations in terms of
spectral slope and absorption features in the near-infrared. We
analyzed these data using a Hapke model to obtain some
inferences about the relative abundance of CAI and other mineral
compounds. By combining spectroscopic and polarimetric results, we
find evidence that the polarimetric inversion angle is directly
correlated with the presence of CAI, and the peculiar polarimetric
properties of Barbarians are primarily a consequence of their anomalous composition.
\end{abstract}


\end{frontmatter}


\section{Introduction}
\label{sec:Intro}

The first attempts to classify asteroids, mostly based on multi-color photometry and spectroscopy at visible wavelengths, led to the identification of the so-called S and C ``complexes''. Later on, taxonomy based on low-resolution spectra in the visible domain by \citet{Bus_1999}, \citet{Xu_1995}, and \citet{Bus_Bin} (we will refer to it as SMASS taxonomy hereafter) resulted in the partition of the S-complex into several sub-classes, based on differences in spectral slope and drop of reflectance at wavelengths above 0.72-0.76~${\rm \mu}$m. Among them, the L-class includes asteroids having the smallest drop of reflectance and a relatively steep slope. The first goal of taxonomy is to differentiate asteroids based on their composition. However, the differences between the L and other classes (S, K, and A) identified using visible wavelengths data, are sometimes not very sharp and can lead to compositional misclassification. Another SMASS class, similar to the L, but exhibiting a slightly steeper spectral slope, was also introduced and named Ld \citep{Bus_Bin}.

More recently, \citet{Dem_2009} extended the SMASS taxonomy to the near-infrared region (from 0.82 to 2.45~${\rm \mu}$m), including the whole 1~${\rm \mu}$m silicate absorption band and extending up to the other major absorption band of silicates around 2~${\rm \mu}$m. Based on the observed behaviour in this larger wavelength range, most of the previously identified L-class asteroids retained an L-classification. These asteroids were found to exhibit a strong absorption feature at wavelengths around 2~${\rm \mu}$m and an almost absence of the 1~${\rm \mu}$m band. However, several differences were also found with respect to the SMASS classification. Some objects previously classified as K- and A- were found to belong to the new L-class, whereas some SMASS L-class were moved to S-, D- or X-class in the new classification. The SMASS Ld-class was found to be almost fully contained in the new L- and D-class \citep{Dem_2009}. For sake of clarity, in the rest of this paper any references to the SMASS taxonomy will be referred to as (SMASS), while the more recent DeMeo taxonomy will be referred to as (DM).



\begin{figure}
\centering
\includegraphics[width=8cm]{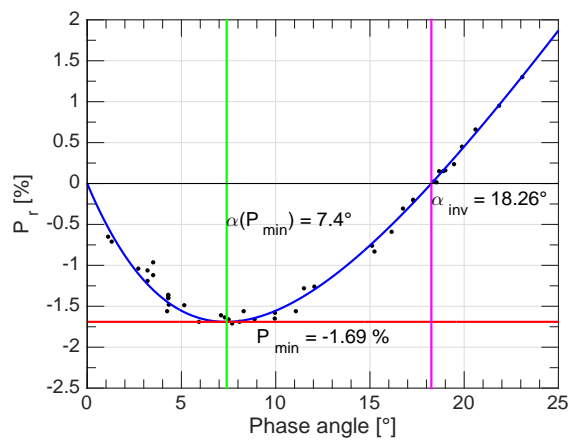}
\caption{Example of the typical phase-polarization curve of (1)~Ceres}
\label{fig:PP_Curve}
\end{figure}

The degree of linear polarization of sunlight scattered by asteroid surfaces is a function of the phase angle, namely the angle between the directions asteroid-Sun and asteroid-observer. The resulting phase-polarization curves share common general morphology, with variations that are mostly albedo-dependent \citep[for a recent analysis of the subject, see][and references therein]{Cel_2015a}. The plane of linear polarization is almost always found to be either coincident with or perpendicular to the scattering plane  \citep{Dol_1989,Muinonenetal2002}. The state of linear polarization of asteroids is usually described using the $P_{\rm r}$ parameter, whose module is equal to the measured degree of linear polarization, but adding a sign, to indicate if the plane of polarization is found to be perpendicular (positive sign) or parallel (negative sign) to the scattering plane. Observations show that the range of phase angle for which $P_{\rm r}$ is negative extends from zero up to an \textit{inversion angle} ($\alpha_{\rm inv}$) generally found around $20^\circ$. This region of the phase-polarization curve is commonly called the ``negative polarization branch''. At larger phase angles, the sign of $P_{\rm r}$ becomes positive (``positive polarization''). The details of the morphology of phase-polarization curves (the maximum value of negative polarization $P_{\rm min}$, at which phase angle this is occurring $\alpha(P_{\rm min})$, value of the inversion angle $\alpha_{\rm inv}$ and slope of the curve around it) are not only diagnostic of the albedo \citep{Cel_2015a}, but have also been found to be useful to discriminate among different taxonomic classes based on reflectance spectra \citep{Penttila05}. Fig. \ref{fig:PP_Curve} shows the typical asteroid phase-polarization curve of the asteroid (1)~Ceres. The locations of $P_{\rm min}$, $\alpha(P_{\rm min})$, and $\alpha_{\rm inv}$ are displayed and labelled.

\citet{b5} reported the discovery of the anomalous polarimetric properties exhibited by the asteroid (234)~Barbara. The phase-polarization curve of this object possesses an unusually wide negative polarization branch, extending up to an inversion angle around $30^{\circ}$. Such behaviour had not been previously observed and was not predicted by theoretical models \citep{Shk_1994}. Because (234) Barbara had been classified as Ld in the SMASS taxonomy, other asteroids of this or similar taxonomic classes were subsequently observed and 18 objects sharing the same polarimetric behaviour of Barbara were found by \citet{b6}, \citet{b31}, \citet{Gil_2011}, \citet{Gil_2014}, \citet{Cel_2014}, \citet{Bag_2015}, and \citet{Dev_2017a}. They were collectively named \textit{Barbarians} after the asteroid (234)~Barbara. Barbarians are a class of rare objects which do not exhibit any preferred location within the asteroid main belt, apart from the presence of the Watsonia dynamical family \citep{Cel_2014}, and possibly a few more, which members are Barbarians. In this paper, we will call ``Barbarian'' any asteroid exhibiting an inversion angle above $25^{\circ}$, independently on any other physical or dynamical property. 

Tentative explanations have been proposed to explain the unusual polarimetric properties of the Barbarians. They include peculiar surface composition and/or texture properties, and/or the presence of large concavities that might introduce an unusual distribution of the directions of the incidence and emergence angles of scattered sunlight, with respect to the case of purely convex bodies. Barbara itself, in particular, has been extensively studied and was actually found to have a fairly irregular shape including large-scale concavities. The rotation period was also found to be unusually long ($P = 26.4744 \pm 0.0001$ h) \citep{Tan_2015} compared to other asteroids of the same size. Other known Barbarian asteroids also exhibit slow rotation rates. In \citet{Dev_2017b} we studied in more detail the role of possible concavities and also the significance of an anomalous abundance of slow rotators among Barbarians. However, it seems that we do not have enough evidence yet to draw any definitive conclusion. 


Available reflectance spectra suggest that Barbarians have spinel-enriched surfaces. The first indication of this came from an analysis of the reflectance spectra of (387) Aquitania and (980) Anacostia performed by \citet{Bur_1992}, well before the discovery of the Barbarian polarimetric properties of these two objects, by \citet{b6} and \citet{b31}. \citet{Bur_1992} noted that these two asteroids, previously classified as S, show reflectance spectra clearly different from that of typical S-class asteroid. Both present a strong 2~${\rm \mu}$m absorption feature, and a nearly absent absorption feature around 1~${\rm \mu}$m while typical S-class spectrum is characterized by a 1~${\rm \mu}$m band stronger than the 2~${\rm \mu}$m one. These authors interpreted such behaviour as being due to the presence on the surface of unusually high amounts of the spinel mineral. Spinels ($\left[{\rm Fe,Mg}\right]{\rm Al_2O_4}$) are important components of the conglomerate of different element called Calcium Aluminium-rich Inclusion (CAI) found in meteorites. Even a small fraction, typically from $10\%$-$30\%$, of FeO enriched aluminous spinel (${\rm Mg Al_2 O_4}$) in CAIs can produce a strong absorption feature around 2~${\rm \mu}$m, similar to what we see in Barbarian spectra.

The CO chondrite meteorites exhibit the highest known abundance of CAI, but never exceeding $13\%$ in volume. CV3 chondrites possess the greatest diversity of CAIs, but abundances are lower than $10\%$. \citet{Bur_1992} had originally suggested an abundance between $5$ and $10\%$ of CAIs on the surfaces of (387)~Aquitania and (980)~Anacostia considering an immature regolith. More recently, considering mature regolith, an analysis of the spectra of asteroids (234), (387) and (980) compared with laboratory spectra led \citet{b15,Sun_2008a} to conclude that a fraction of spinel-bearing CAIs of the order of $\sim30\%$ in volume is needed to fit the observed near-infrared spectra of these asteroids. No known example of such high CAI abundances can be found in current meteorite collections. A high concentration of a particular type of CAI, called fluffy-type A, suggests a formation in an environment characterized by a high concentration of CAIs, and an absence of strong thermal alteration after formation.


FeO enriched aluminous spinel has a relatively high and variable real part of the refractive index in the visible part of the spectrum \citep{Hos_2008}. It is found to be $n=1.83$ in blue light, and decreases at longer wavelengths, down to $n=1.78$ in the red and near-infrared (NIR), where it is almost constant. \citet{Sun_2008b} suggested that a high refractive index might possibly be responsible for an uncommonly large inversion angle of polarization. In the case that a high abundance of spinel-bearing CAIs can be proven to be the correct explanation of the wide negative polarization branch observed for Barbarians, we would have good reasons to believe that these asteroids accreted in a nebula rich of refractory materials, and contain the most ancient mineral assemblage currently found in the inner solar system.

Based on the facts mentioned above, we started an observational campaign of L-class asteroids (both SMASS and DM) and known Barbarian asteroids. We present in Section \ref{sec:Obs} our new polarimetric observations done in the framework of Calern Asteroid Polarimetric Survey (CAPS) \citep{Dev_2017a} and at the Rozhen observatory as well as new spectroscopic observations carried out at the NASA Infra Red Telescope Facility (IRTF) using the near-infrared spectrograph SpeX \citep{Ray_2003}. The section \ref{sec:Data_Anal} is devoted to the description of the models used to analyse the phase-polarization curve and the spectra of L-type asteroids. A Hapke model involving space weathering process is also described. In Section \ref{sec:Res_Disc}, the results relative to the spectral classification, the phase-polarization curve, the geometric albedo, the spectral fitting, asteroid families, and the identification of new Barbarians are presented. The Section \ref{Sec:Inter_Disc} is devoted to the interpretation and discussion of the relation between (DM) L-type and the Barbarians, the composition of L-class asteroids, some interpretation for the high polarimetric inversion angle of the Barbarians. Finally, Section \ref{sec:Conc_Persp} presents our conclusions and perspectives for future works.

\section{Observations}
\label{sec:Obs}
In this work, 36 targets were observed in polarimetry, spectroscopy or both. They were selected on the basis of satisfying one or more of the following criteria: 
\begin{itemize}
\item Being a known Barbarian;
\item Belonging to (SMASS) L- or Ld-class;
\item Belonging to (DM) L-class;
\item Being a member of one of the following dynamical families
 known or suspected to include Barbarians and/or L-class (SMASS or DM) members: Watsonia,
  Henan \citep{Nes_2015} and, Tirela \citep{Mot_2008}, renamed Klumpkea by
  \citet{Mil_2014}.
\end{itemize} 

\subsection{Polarimetric observations}
A high-priority goal of this work was to understand the reason of the abnormally large inversion angle of Barbarian. However, we included also some targets which were already known to be non-Barbarians. These targets belong to (SMASS) taxonomic classes which have been found in the past to include Barbarians. Their characterization allows us to better understand the relationship between spectroscopy and polarimetry. 

The polarimetric data were acquired in two distinct observatories. Most of them were taken at the C2PU (Centre p\'edagogique Plan\`ete et Univers) facility of the Calern station of the Observatoire de la C\^ote d'Azur (Nice, France). The ToPol (Torino Polarimeter) was used to carry out the observations, that were part of the CAPS program started in early 2015.
The ToPol is mounted on the Cassegrain focus (F/12.5) of the $1.04$~m West telescope of the C2PU facility. It involves a Wedged-Double Wollaston prism, a configuration yielding the polarimetric reduced Stokes parameters $q$ and $u$ in one single exposure. All data were processed using classical aperture photometry, but using a curve of growth procedure \citep{Bag_2011, Bag_2015}. This procedure consist of measuring the fluxes of the four replicas of the target using gradually increasing aperture size. The optimal aperture size is then selected by visual inspection of the values of $q$ and $u$ as a function of the aperture size. Full description of the instrument and the reduction techniques used for this instrument are described in \citet{SPIE2012} and \citet{Dev_2017a}.

Some fainter targets were observed using the 2-Channel-Focal-Reducer Rozhen (FoReRo2) with a polarimetric mode retarder half-wave plate mounted at the 2 m telescope of the Bulgarian National Astronomical Observatory (Rozhen, Bulgaria). See \citet{Joc_2000} for a full description of the instrument. A single Wollaston prism is used to measure on each CCD acquisition either the $q$ or $u$ reduced Stokes parameter. A retarder wave-plate was recently added to more easily rotate the observed polarization angles. The retarder wave plate is not  described in the original paper.

Table \ref{Pola_meas} lists all the polarimetric observations
presented in this paper. All the observations were done between
December 20, 2014 and  January 14, 2017. All measurements were done in
the standard $V$ Johnson-Cousins band.

\subsection{Near-infrared spectroscopic observations}
\label{sec:NIR_Spec}
The new spectroscopic data presented in this work were obtained during two nights (September 21, 2014 and January 22, 2015). The asteroids were observed from 0.8 to 2.5~${\rm\mu}$m, using the SpeX instrument \citep{Ray_2003} in the low resolution (R $\sim$ 200) PRISM mode mounted on the 3-meter NASA InfraRed Telescope Facility (IRTF) telescope on Mauna Kea. All the targets were observed near the meridian and solar analogue stars were observed near the target just after or before the target to calibrate out telluric absorptions and to correct for differences from the solar spectrum. A nodding procedure was used for each set of exposures. This procedure consists in acquiring a pair of spectra at two distinct locations on the CCD field (referred to A and B positions). A $0.8 \times 15$ arcseconds slit aligned north-south was used for all the observations. Flat field images were obtained by illuminating an integrating sphere. Spectra of an argon lamp were also taken immediately before or after the observation of the targets for wavelength calibration.

The extraction and first reduction of the spectra were carried out using the IRTF pipeline SpexTool \citep{Cus_2004}. This pipeline performs sky subtraction using the A-B pair, corrects for the flat field, calibrates the wavelength using spectra taken at the beginning and end of each target observation, and finally extracts the reduced spectra. The removal of telluric absorptions was performed using the ATmospheric TRAnsmission (ATRAN) model \citep{Lord_1992} on each individual spectrum. This correction constitutes a very important step since the water vapour has strong absorption bands around 1.4 and 2~${\rm\mu}$m.

The final spectrum of an asteroid is constructed by averaging all the individual observed spectra. A sigma clipping procedure is used to reject outliers that may occur due to cosmic rays contamination. 

The new obtained NIR spectra were merged with SMASS visible spectra whenever available (see Sec.~\ref{sssec:V-IR Merging} for details on the merging procedure). 

Table \ref{Sum} summarizes all the spectroscopic observations
presented in this work. For each target, the magnitude, number of AB
pairs, total exposure time and airmass at mid observation are listed. The solar analogue star used for the reduction of each individual target
is also listed so as its airmass at mid observation. All the reduced
spectra, merged with the visible part when available, are displayed in ~\ref{app:spec}. 

\begin{table*}
\centering
\begin{tabular}{| lccccc | cc |}
\hline
								\multicolumn{6}{|c|}{Asteroids}					&					\multicolumn{2}{c|}{Solar Analogues}	\\ 
Name								&	$m_v$	&	\#AB	pairs	&	$t_{\rm exp}$ [s]	&	Run	& Airmass & Name & Airmass  	\\
\hline
(12)~Victoria						&	11.9	&	5					&	200			&	2			&	1.30	& SAO42382 				&	1.86 \\
(122)~Gerda						&	13.9	&	5					&	900			&	2			&	1.31	& BS4486 					&	1.18 \\ 
(172)~Baucis					&	12.3	&	6					&	1080			&	1			&	1.48	& SA 115-271 	&	1.15	\\
(458)~Hercynia					&	13.0	&	8					&	960			&	1			&	1.22	& SA 93-101		& 	1.07\\
(611)~Valeria					&	11.4 	&	5					&	600			&	1			&	1.12	& SA 93-101		&	1.42	\\
(753)~Tiflis						&	14.8	&	5					&	1000		&	2			&	1.05	& Hyades 64 				&	1.07 \\
(1372)~Haremari				&	14.6	&	7					&	1400		&	2 			&	1.08	& SA 98-978 		&	1.10\\
(2354)~Lavrov					&	15.8	&	6					&	1440		&	1			&	1.30	& SA 115-271 	&	1.15\\ 
(4917)~Yurilvovia				&	16.2	&	8					&	1880		&	1			&	1.04	& SA 115-271 	&	1.06\\
(8250)~Cornell					&  17.7	&	10					&	2400		&  2     		&	1.14	& Hyades 64					&	1.07 \\
(15552)~Sandashounkan	&	17.1	&	10					&	2400		&	1			&	1.05	& SA 93-101 		&	1.07\\    
(19369)~1997 YO				&	15.9	&	10					&	2400		&	2			&	1.06	& SA 102-1081 	&	1.07\\
(26219)~1997 WO21			&	16.9	&	14					&	3360		&	2			&	1.26	& SA 102-1081 	&	1.07\\
(67255)~2000 ET109		&	16.5	&	10					&	2400		&	1			&	1.05	& SA 115-271 	&	1.06\\ 
\hline
\end{tabular}
\caption{Observing conditions at the IRTF telescope. The first column corresponds to the number and the name of the observed asteroid, the second column gives the V magnitude of the target at the time of observation, \#AB pairs stands for the number of AB pairs taken, $t_{\rm exp}$ is the total exposure time, the first run corresponds to the night of September 21, 2014 and the second run to the night of January 22, 2015, the Airmass columns give the airmass at mid-observation. The ``Solar analogues'' frame gives the solar analogue star used for calibration, and its airmass at mid-observation.}
\label{Sum}
\end{table*}

\section{Data analysis} 
\label{sec:Data_Anal}
In this section, we present the data analysis tools that were used to interpret the polarimetric and the spectroscopic data.

\subsection{Phase-polarization curve}
\label{ssec:Phase_Pol}
As already discussed in Section \ref{sec:Intro}, the phase-polarization curves exhibited by asteroids share a general morphology characterized by the presence of a so-called negative polarization branch, in which $P_{\rm r}$ has negative values. Negative polarization reaches an extreme value (conventionally called $P_{\rm min}$ in the literature) at a phase angle $\alpha(P_{\rm min})$. For increasing phase angles $P_{\rm r}$ decreases in absolute value up to the inversion angle $\alpha_{\rm min}$ around $20^{\circ}$ for regular (non-Barbarian) asteroids, where it becomes null again. Beyond the inversion angle, $P_{\rm r}$ takes positive values and shows generally a linear increase up to the extreme values of phase angle that can be possible for asteroid orbits (see Fig. \ref{fig:PP_Curve}).

A frequently used model of the phase-polarization curve is the so-called ``Exponential-Linear'' model \citep{Mui_2009}: 
\begin{equation}
\label{eq:Pr2}
P_{\rm r}(\alpha) = A \cdot [e^{(-\alpha/B)} -1] + C \cdot \alpha
\end{equation}
where $\alpha$ is the phase angle, and $A$, $B$, and $C$ are parameters to be derived by least-squares techniques. In a couple of recent papers \citep{Cel_2015a, paper2} the computation of the best-fit parameters for a large number of asteroids  was carried out using a genetic algorithm. Having determined the values of $A$, $B$, and $C$, for any given object, the inversion angle was derived by some simple numerical method, while $P_{\rm min}$ and $\alpha(P_{\rm min})$ were derived through a computation of the first derivative of Eq.~\ref{eq:Pr2} using the values of the parameters determined by means of the genetic algorithm.

In this paper, the strategy described by \citet{Cel_2015a, paper2} is applied to a larger dataset which includes recent observations obtained at Calern and Rozhen. In addition, to increase the robustness of our results, we derived again the best-fit values of the unknown parameters in Eq.\ref{eq:Pr2} using another, independent approach. In particular, the inversion angle of the phase-polarization curve was analytically determined by evaluating 
\begin{equation}
\label{eq:Alpha_0}
\alpha_{0} =\frac{ BC {W}\left(-\frac{A\exp{\frac{-A}{BC}}}{BC}\right) +A}{C}
\end{equation}
for values of phase angle different from 0, where W($x$) is the positive branch of the Lambert function. An estimation of the $1\sigma$ uncertainty of the inversion angle computed in this way was obtained by using a Markov Chain Monte Carlo fitting procedure (MCMC). The estimation of the $1\sigma$ uncertainty was then done by fitting a normal distribution to the histogram of the derived $\alpha_{0}$ values. The same procedure was used for the other parameters such as the degree of minimum polarization $P_{\rm min}$ and the phase angle at which it is occurring. 

The results of the computations of polarimetric parameters based on the two independent approaches described above were found to be in excellent mutual agreement, the differences being found to be within the formal error bars computed by the two algorithms.

\subsection{Spectral fitting}
\label{ssec:Fit_Spectra}

\subsubsection{Visible-NIR spectral merging}
\label{sssec:V-IR Merging}
\begin{figure*}
\includegraphics[width=18cm]{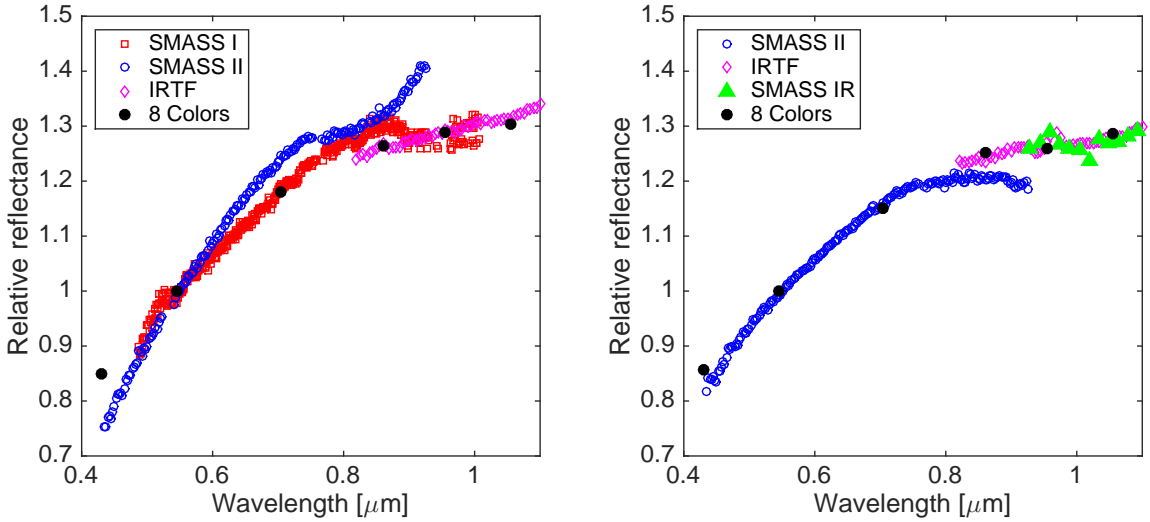}
\caption{Example of Visible-Near-infrared merging issues for
  (234)~Barbara (left part) and (236)~Honoria (right part). SMASS I
  data are represented by red squares, SMASS II by blue circles, SMASS
  IR by green triangles, IRTF by purple diamonds, and ECAS data by black dots.}
\label{fig:Barb_reco}
\end{figure*}

In our analysis of the spectral reflectance properties, the first step was to merge together available spectra covering the visible and NIR regions. Note that we did not limit our analysis to the objects observed by us at IRTF, but also analysed L-type spectra taken from the literature. The major sources of visible spectra are the SMASS \citep{Bus_1999} and SMASS II \citep{Bus_Bin} data-bases, while for the NIR, most available spectra have been taken at IRTF in the framework of the MIT-UH-IRTF Joint Campaign for NEO Spectral Reconnaissance and also the new data presented in this paper. Merging visible and NIR spectra turned out to be a non-straightforward task. In particular, we found almost systematically noticeable differences in spectral slope between the red end of available visible spectra and the blue end of NIR data. This is not an unusual problem in asteroid spectroscopy.
Moreover, the same object can sometimes exhibit bands that are visible in one spectrum, but not in others taken at different epochs. As an example of a difficult case, in the left part of Fig.~\ref{fig:Barb_reco} we show the case of the asteroid (234)~Barbara which was observed both in SMASS and SMASS II as well as, previously, in the 8-colors Asteroid Surveys (ECAS, \citet{8C}). We can see that the red ends of the visible spectra are quite different. We can also notice that none of these spectra merge well with NIR data. However, the older ECAS data, which cover a wider range of NIR wavelengths, links very well to IRTF data from the MIT-UH-IRTF Joint Campaign and are compatible with SMASS I data in the blue part of the spectrum.
In the case of asteroid (236)~Honoria (see the right part of Fig.~\ref{fig:Barb_reco}), SMASS II and ECAS data are available as well as two spectra in the near-infrared from SMASS IR \citep{Bur_2002} and IRTF (MIT-UH-IRTF Joint Campaign). One can see that in this case, the two NIR spectra show a more reasonable mutual agreement with each other and are also compatible with available ECAS data. 

In general we find that the NIR spectra tend to show less variability than the visible spectra. In our procedure, we always multiply the NIR spectrum so as its blue-end merges with the red-end of the visible spectrum at the same wavelength (0.82~${\rm \mu}$m) simply removing visible data covering longer wavelengths. This is justified by the fact that the red end of visible spectra corresponds to a drop in sensitivity of the CCD detectors. However, different merging methods can lead to spectra showing different behaviour near the merging point. As a general rule, therefore, we base our assessment of the quality of the fit by looking at the morphology of the absorption features in the NIR region beyond merging wavelength. Especially, in some cases, an apparent absorption feature around 0.7-0.8~${\rm \mu}$m cannot be modelled by our procedure and is not taken into account.

\subsubsection{Fitting techniques}

The general approach we adopted to model the obtained spectra was to use a combination of a small number of candidate mineral components using a simplified Hapke spectral mixing model \citep{Hap_1981,Hap_1984, Hap_1986}. 
The idea is to linearly combine the spectrum of several candidate end-members to find satisfactory best-fits of the observed asteroid spectrum. 

A technical problem with the adopted approach is that, working in terms of spectral reflectance ($r$), the spectra of intimately mixed materials do not combine linearly \citep{Nas_1974}. On the other hand, working in terms of single scattering albedo ($w$), the combination of different spectra is linear even when the materials are intimately mixed. The fitting procedure was then carried out in the following steps: 
\begin{enumerate}
\item Convert spectral reflectance of the end-member spectra into single scattering albedo.
\item Combine linearly the single scattering albedo of the different end-members.
\item Convert back single scattering albedo into spectral reflectance.
\item Compare the combined reflectance spectrum with the spectrum of the asteroid to be fit. 
\item Repeat the above steps using an optimization procedure until an acceptable fit is obtained. 
\end{enumerate}

As the result of the fitting procedure, the relative abundance of each end-member in the obtained mixture is simply the coefficient used to linearly combine the single scattering albedo of the considered end-members. 

The spectral reflectance of asteroid spectra is usually normalized to the value measured at the wavelength of 0.55~${\rm \mu}$m. The absolute reflectance could be therefore computed, in principle, by equalling the reflectance at 0.55~${\rm \mu}$m to the geometric albedo ($p_{\rm V}$) of the object, when this is known. According to its definition, $p_{\rm V}$ is equal to the ratio between the object brightness measured at zero phase angle and that of an ideal, flat and perfectly Lambertian disk,  having the same projected surface of the object, where the brightness is measured at the wavelength of 0.55~${\rm \mu}$m, and both the asteroid and the Lambertian disk are assumed to be located at unit distance from the Sun and from the observer. 

However, the albedo of asteroid is usually known with a noticeable
uncertainty. Most asteroid albedo come from the WISE \citep{Mas_2011}, NEOWISE \citep{Mai_2016}, IRAS \citep{Ted_2002,Rya_2010}, and AKARI \citep{Usui_2012} surveys. Comparing the derived values of these surveys leads to differences as high as $20\%$ and even more in many cases. Based on the
difficulty in assigning a well determined value of geometric albedo to
our objects, we allow it to vary in our optimization procedure.  

Step (1) of the fitting procedure described above makes use of the equation linking the bidirectional reflectance $r_c$ and the single scattering albedo $w$, according to \citep{Hap_1981}. This equation can be written as: 
\begin{equation}
\label{eq:hapke}
r_c = \frac{w}{4}\frac{1}{\mu_0+\mu}\left[\left(1+B(g)\right)P(g)+H(\gamma,\mu_0)H(\gamma,\mu)-1\right],
\end{equation}  
where $\mu_0$ and $\mu$ are respectively the cosine of the incidence and emergence angles, $g$ is the phase angle (\textit{i.e.} the angle between the incident and reflected beams),
and
\begin{equation}
\gamma =  \sqrt{1-w}.
\label{eq:gamma}
\end{equation}

Eq.~\ref{eq:hapke} involves the three functions $B(g)$, $P(g)$ and $H(\gamma,x)$ which deserve some explanations. $B(g)$ is the backscatter function which defines the increase in brightness of a rough surface with decreasing phase. This effect is known as the opposition effect. According to \citet{Mus_1989} this function can be set to zero for phase angles greater than $15^{\circ}$. Since all the laboratory spectra used in this work were taken at a phase angle around $30^{\circ}$, $B(g)$ can be safely set to zero. $P(g)$ is the single particle phase function. \citet{Mus_1989} found that if one assumes an isotropic scattering (\textit{i.e} putting $P(g)$ equal to one for all $g$), the resulting errors are of the order of a few percent. Since we do not know this function for all the end-members, we set $P(g)$ to one. $H(\gamma,x)$ is the so-called Chandrasekhar isotropic $H$ function which can be approximated by the analytical expression: 
\begin{equation}
H(\gamma,x) = \frac{1+2x}{1+2\gamma x}
\label{eq:H}
\end{equation}
where $x$ represents $\mu$ or $\mu_0$.

Laboratory spectra are normalized to a reference spectrum, taken at the same incidence and emergence angles, for which the single scattering albedo $w$ can be assumed to be $1$ (implying that all particle extinction is due to scattering).
Eq.~\ref{eq:hapke} can then be simplified and becomes \citep{Hap_1993,Hap_2001}: 
\begin{equation}
\label{eq:hapke_rc}
\Gamma(\gamma) =  \frac{r_c(\rm sample)}{r_c(\rm standard)} = \frac{1-\gamma^2}{(1+2\gamma\mu_0)(1+2\gamma\mu)},
\end{equation}
Eq.~\ref{eq:hapke_rc} can be solved for $\gamma$:
\begin{equation}
\label{eq:hapke_gamma}
\gamma = \frac{  \sqrt{ \Gamma^2 \left( \mu^2 - \mu_0^2 \right)^2 + \Gamma \left(   4 \mu \mu_0 - 1 \right) + 1 } -  \Gamma \left( \mu + \mu_0 \right)  }{4 \mu \mu_0 \Gamma + 1}.
\end{equation} 
From $\gamma$, the single scattering albedo for each end-member is then immediately derived (Eq.~\ref{eq:gamma}).


Once the single scattering albedo for each end-member has been combined linearly, the bidirectional reflectance of the mixture is computed back using the same equations to obtain the composite spectrum. The best abundance of each end-member which fits at best the asteroid spectrum is then derived using a Levenberg-Marquardt optimization technique.

\subsubsection{Space weathering}
\label{sssec:Space_Weath}

Space weathering is a general process due to chemical and physical mechanisms that affect an airless body surface exposed to the space environment. It is the result of the exposure of an asteroid regolith to micro meteoritic impacts and heavy radiation (Solar wind and cosmic rays). It was observed that space weathering affects different bodies in different ways. In general, the effects of space weathering on bright, silicate-rich, asteroids tends to increase their spectral slope (reddening of the spectra), reduce the optical geometric albedo (darkening) and decrease the absorption bands depth \citep{Cha_1996,Hap_2001,Bru_2013,Bru_2014}. However, recent studies have shown that the effect of space weathering on dark asteroid surfaces in nearly opposite. The reflectance spectrum tends to become bluer and brighter than fresh materials \citep{Lan_2017}. It is believed that these effects are due to the progressive implantation of nanophase metallic iron particles (${\rm npFe^0}$) into regolith grains, as the effect of micrometeoritic impacts and solar wind sputtering.

\citet{Hap_2001} proposed a model to take into account the optical constants of iron within the host material to modify a spectrum in the framework of his Hapke reflectance and scattering theory. This model is based on the idea of computing the absorption coefficient of the host material (laboratory spectra) ($\alpha_{\rm h}$) and of  ${\rm npFe^0}$ particles ($\alpha_{\rm Fe}$). The total absorption coefficient ($\alpha$) of the space-weathered material is considered to be simply given by the sum of the two: 
\begin{equation}
\alpha = \alpha_{\rm h} + \alpha_{\rm Fe}.
\end{equation} 

In this work, we made use of this model to modify the spectra of each of our chosen end-members, to make them more similar to what we expect to be the case of space-weathered material on the surfaces of celestial objects. The first step is the computation of the single scattering albedo of each end-member using the procedure already explained above. Then, $\alpha_{\rm h}$ is determined as: 
\begin{equation}
\alpha_{\rm h} = \frac{1}{D}\ln{ \left[ S_i + \frac{\left( 1- S_e \right) \left( 1 - S_i \right) }{w-S_e}\right] } 
\end{equation}  
where $D$ is the effective size of the particles in the media,  $n$ is the refractive index of the end member, and $S_e$ is the Fresnel reflection coefficient of the particle surface averaged over all angles of incidence for light incident from outside the particle, while $S_i$ is the same, but for light incident from inside the particle. They are given by:
\begin{equation}
S_e = \frac{\left( n-1 \right)^2}{ \left( n+1 \right)} +0.05
\end{equation}  
and,
\begin{equation}
S_i = 1.014 - \frac{4}{n\left( n+1 \right)^2},
\label{eq:S_i}
\end{equation}  

The above expressions are useful approximation of the true integral given by \citet{Hap_1993}. In the case of $S_i$, \citet{Hap_1993} used 1 instead of 1.014. However, \citet{Lucey_1998} found that in the range of refractive indexes n = 1.5 to 2,
Eq.~\ref{eq:S_i} is a better approximation. 

As for the contribution of ${\rm npFe^0}$, we have the relation:
\begin{equation}
\alpha_{\rm Fe} = \frac{36\pi}{\lambda}\phi z
\end{equation} 
where 
\begin{equation}
z = \frac{n_{\rm h}^3 n_{\rm Fe} k_{\rm Fe}}{\left( n_{\rm Fe}^2 - k_{\rm Fe}^2 + 2n_{\rm h}^2 \right)^2 + \left( 2 n_{\rm Fe} k_{\rm Fe} \right)^2}
\end{equation} 
and $\phi$ is the volume fraction of ${\rm npFe^0}$ that are embedded in the host material. $n$ and $k$ are the refractive index and absorption coefficient, respectively, of the host material (h) (end-member) and of iron (Fe).
 
If the iron particles are uniformly distributed $\phi = \rho_{\rm Fe}f/\rho_{\rm h}$ where $\rho_{\rm Fe}$ and $\rho_{\rm h}$ are the solid density of iron in the host material and the solid density of the host material respectively. Finally, $f$ is the bulk mass fraction of iron.  

\begin{figure}
\includegraphics[width=8.8cm]{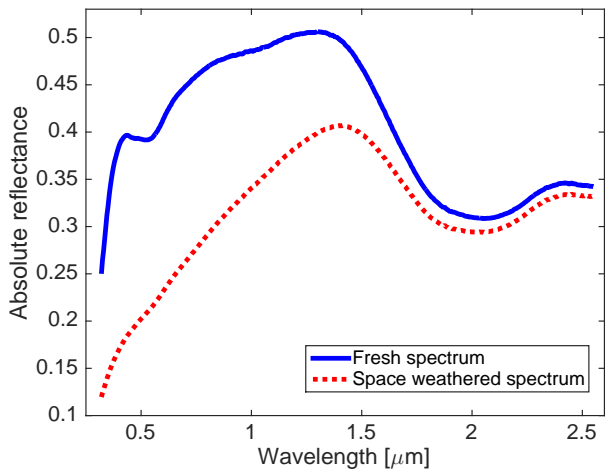}
\caption{Example of the computation of the space weathering effect (see text) on the reflectance spectrum of the fluffy-type A CAI considered in our analysis. }
\label{fig:SW_Ex}
\end{figure}

The last step is to compute the new bidirectional reflectance of the space-weathered material using the following relations: 
\begin{equation}
\Theta = \exp{\left(-\alpha D\right)}
\end{equation} 
and
\begin{equation}
w = S_e + \left( 1-S_e \right) \frac{1-S_i}{1-S_i\Theta}\Theta
\end{equation}
from which, by using Eqs. \ref{eq:hapke_rc} and \ref{eq:gamma}, one can compute the final space-weathered spectrum. An example is shown, in the case of the fluffy type A CAI, in Fig.~\ref{fig:SW_Ex}.

\subsubsection{End-members}
\label{sssec:End_Members}

We describe here the end-members used in our analysis. All end-member spectra were obtained from the NASA REflectance experiment LABoratory (RELAB) spectral database.

In our case, the classical mixture of olivine and pyroxene, cannot be assumed. We have therefore to assume the presence, of spinel bearing CAIs \citep{Bur_1992} like those found in CV3s meteorites to model the 2~${\rm \mu}$m band. Convincing arguments supporting this choice can be found in \citet{Bur_1992} and \citet{Sun_2008a}. A complete review of the mineralogy of CVs meteorite can be found in \citet{Clo_2012}. 

\begin{figure}
\includegraphics[width=8.8cm]{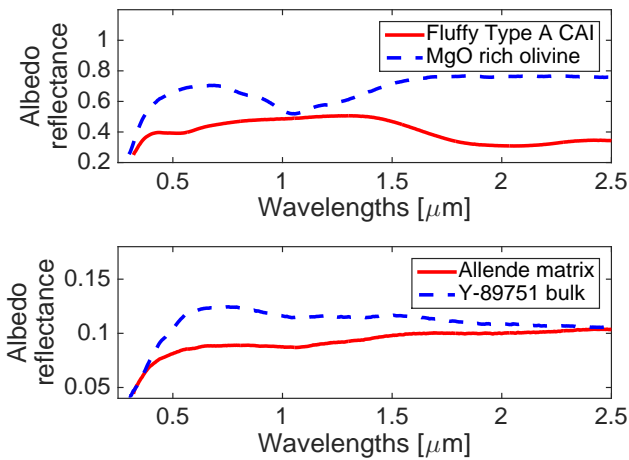}
\caption{Spectra of all the end-members used to model the spectra of the asteroids studied in our analysis.}
\label{fig:end-members}
\end{figure}

We took into consideration different types of CAIs, that are known to produce large absorptions at wavelengths around 2~${\rm \mu}$m. In addition, we assumed the presence of the olivine mineral, one of the most important silicates found in many meteorites, as well as two example of the meteorites in which CAIs are found as inclusions. In particular, we chose the matrix of the well-known Allende and the Y-86751 meteorites. Note that we are aware that other possible materials could be taken into consideration. Moreover, we did some preliminary tests in which we included also pyroxene among the end-members, but these tests did not give satisfactory results. In what follows, we give a brief description of our selected end-members, while Fig.~\ref{fig:end-members} shows the spectrum of each of them in terms of absolute reflectance.
 
{\it Calcium Aluminium rich Inclusion} -  Three distinct types of CAIs (A, B and C) classified on the basis of petrography and geochemistry, are known to exist. The types B and C show evidence of melting by transient heating events before accretion. As opposite, fluffy type A CAIs (FTA) do not show any evidence of melting. FTAs are found in all chrondritic meteorites, whereas types B and C are found only in CV3 meteorites. 
FTAs and B type CAIs are both dominated by an absorption feature around 2~${\rm \mu}$m, but this absorption is much stronger in the case of FTAs. This stronger absorption feature is due to a much higher concentration of FeO in the aluminous spinel present in these CAIs.

In this study we have considered the same spectra of CAIs, three FTAs and three B-type CAIs, already analyzed by \citet{Sun_2008a} (RELAB: TM-TJM-001 to 005, and 007). The three FTA samples contain from $3$ to $14\%$ in weight mass of FeO while type B CAIs only a fraction of percent. The strength of the 2~${\rm \mu}$m absorption band is directly correlated to the abundance in FeO \citep{Sun_2008a}. 
Considering that abundant FeO is excluded from CAI minerals during condensation, and that we see that CAI-types showing highest percentage of FeO also show an abundance of alteration phases, the FeO present in FTAs and B-type CAIs should come from post-accretion enrichment.

In our procedure, we always made use as end-member the FTAs sample showing the highest percentage of FeO. Type B CAI was always found to result in poorer fitting than using FTAs. This result was already noticed by \citep{Sun_2008a}.

\textit{MgO rich olivine} - Olivine is an important component of many meteorites, and is most abundant in chondrites. Differences in abundance and composition of olivines is an important criterion for the classification of meteorites \citep{Mas_1963}. 

Olivine is a magnesium-iron silicate. The most general formula is $ \left[ {\rm Mg}^{+2}, {\rm Fe}^{+2} \right] {\rm SiO}_4$. The end-members are called forsterite (Fo) (${\rm Mg}_2 {\rm SiO}_4$) and fayalite (Fa) (${\rm Fe}_2{\rm SiO}_4$). Olivine is usually described by the relative fraction of Fo and Fa (${\rm Fo}_x$ and ${\rm Fa}_x$). Forsterite is olivine with ${\rm Fo}_x$ between $100\%$ and $90\%$, whereas fayalite is olivine with ${\rm Fo}_x$ between $0\%$ and $10\%$. The spectrum of olivine shows a broad 1~${\rm\mu}$m absorption band which slightly depends upon ${\rm Fo}_x$ content.

The real part of the refractive index of olivine is highly dependent on the iron content. \citet{Lucey_1998} determined that the refractive index of olivine satisfies the relation
$n = 1.827 - 0.192 {\rm Fo}_x$. This value of $n$ appears to be constant over the whole interval of visible wavelengths.

Olivine is already present in the matrix of the meteoritic component that we are considering as end-members (see next paragraph). However, the composition of olivine is varying with respect to their degree of alteration. Unaltered olivine should be Mg-rich (forsterite) and the fraction of Fe will then increase as a function of the alteration. The olivine present in the considered meteoritic components possess moderate to large amount of Fe. On the other hand, some of our asteroids show less alteration than Allende or Y-86751 and should contain more MgO-rich olivine. As a consequence, we are considering forsterite olivine as end-member (RELAB: PO-CMP-076), which has a refractive index $n = 1.635$. The presence of forsterite in our model can then be seen as an indicator of post-accretion alteration.

\textit{CV3 meteorites} - CAIs are one of the inclusions found in a matrix of material that constitutes the bulk composition of CV meteorites. The matrix must therefore be taken into account as one of the important constituents of any possible asteroidal composition. 

In this work, we used two different CV3 meteorites. First, we considered the matrix of the Allende meteorite from which CAI were removed \citep{Sun_2008a} (RELAB: MT-TJM-071). We also used spectral measurement of the Y-86751 meteorite (RELAB: MP-TXH-009). 

The composition of the matrix of the Allende meteorite, the largest carbonaceous chondrite found on Earth, and one of the most studied examples of primitive meteorites, was measured by \citet{Bla_2004}. They found that it is composed of more than $80\%$ of olivine. The Allende matrix is also found to be pyroxene-poor with only $5.9\%$ of enstatite. The refractive index for the Allende matrix (after removing CAI inclusions) is unknown. \citet{Zub_2015} estimated a value between 1.68 to 1.83 for the Allende meteorite using the polarimetric inversion angle as a proxy. They found a value of 1.7 by fitting the light-scattering response from Allende meteorite particles. In this work we used $n = 1.7$ for the Allende matrix.

The Y-86751 meteorite is of the same type than Allende and possess the same bulk composition \citep{Pal_1993}. Characterization using optical microscope to measure transmitted and reflected light shows flow texture possibly due to aqueous alteration \citep{Gyo_2011}.  It is known to contain CAIs where the spinel is more FeO-rich than Allende (18-$25\%$ \citep{Mur_1994} instead of 4 to $14\%$). Its matrix also contain fine-grained aluminous spinel \citep{Mur_1994}.


\subsubsection{Optimization procedure}

The matlab \textit{fmincon}\footnote{More information about this function can be found on the Mathwork website: https://nl.mathworks.com/help/optim/ug/fmincon.html} function was used as optimization procedure. This function allows the user to use constrained values of the optimized parameters. In our case, we set constraint on the end-member abundances so their sum will always be equal to 1.

Typical reduced chi-square optimization function was used. This function can be weighted in order to give priority to certain wavelengths. This is useful in the case of doubtful visible and NIR merging. In our case, only the near-infrared region, from to $0.82$ to 2.5~${\rm \mu}$m, was considered to constrain the fitting procedure. This prevent the featureless visible part of the spectrum to play to significant role in case of plausible wrong visible and NIR merging. 
\section{Results}
\label{sec:Res_Disc}
In this section we present the results of our analysis of polarimetric and spectroscopic data using the methods described in the previous Section. 

Table \ref{tab:Param} lists the 43 objects analysed in this work taking profit of the new spectroscopic and/or polarimetric data obtained in our observing campaigns. This table also summarizes some physical properties of the objects, and indicate family membership if any. Bold entries correspond to value directly determined in this work.
Albedo values are taken from the NEOWISE (Wide-field Infrared Survey Explorer) catalogue \citep{Mai_2016}, and can be affected by noticeable uncertainties, as suggested by \citet{Cel_2015a}, and also by the fact that in some cases we have for the same object more than one albedo estimate, showing noticeable differences. This suggests that the nominal albedo uncertainties listed in the Table \ref{tab:Param} can be in many cases fairly optimistic.
\begin{table*}
\centering
\begin{tabular}{| l | lll | l | l | l | l | }
\hline
Asteroid				&	Tholen	&	SMASS		&	DM	&	Barbarian	&	D	(NEOWISE)				&	$p_V$ (NEOWISE)	&	 Family	\\
						&				&						&						&					&	[km]				&										&			\\
\hline 
(12)~Victoria			&	S 			&	L				&	\textbf{D}		&	N				&	$115.1 \pm 1.2$  			&	$0.16 \pm 0.03$ &			\\
(122)~Gerda			&	ST		&	L				&	\textbf{S}		&	\textbf{N}	&	$70.7 \pm 0.9$  	&	$0.25 \pm 0.04$			&			\\
(172)~Baucis			&	S			&	L				&	\textbf{L}		&	Y				&	$63.5 \pm 3.1$ 			&	$0.13 \pm 0.02$			&			\\
(234)~Barbara			&	S			&	Ld				&	L					&	Y				&	$45.5 \pm 0.2$				&	$0.20 \pm 0.03$	&			\\
(236)~Honoria			&	S			&	L				&	L					&	Y				&	$77.7 \pm 1.2$				&	$0.16 \pm 0.02$	&			\\
(387)~Aquitania			&	S			&	L				&	L					&	Y				&	$97.3 \pm 3.4$			&	$0.20 \pm 0.03$					&			\\          
(402)~Chloe		&	S			&	K				&	L					&	Y				&	$55.4 \pm 1.7$				&	$0.16 \pm 0.03$	&		\\ 
(458)~Hercynia			&	S			&	L				&	\textbf{L}		&	Y				&	$36.7 \pm 0.4$		&	$0.43 \pm 0.07$	&	\\
(460)~Scania  			&				&	K				&	L					&					&	$19.7 \pm 0.1$		&	$0.26 \pm 0.06$	&			\\
(478)~Tergestre			&	S			&	L				&						&	N				&	$80.7 \pm 1.0$				&	$0.17 \pm 0.04$	&			\\
(599)~Luisa			&	S			&	K				&	L					&	Y				&	$70.2 \pm 0.5$		&	$0.12 \pm 0.03$			&		\\
(606)~Brangane			&	TSD		&	K				&	L					&	\textbf{Y}	&	$35.0 \pm 0.2$		&	$0.10 \pm 0.01$		&	606		\\
(611)~Valeria   			&	S			&	L				&	\textbf{L}		&	\textbf{Y}		&	$57.5 \pm 0.2$		&	$0.12 \pm 0.01$			&		\\
(642)~Clara   			&	S			&	L				&						&					&	$38.2 \pm 5.3$				&	$0.11 \pm 0.02$					&		\\
(679)~Pax   		&	I		&	K			&	L				&	Y			&	$63.0 \pm 0.4, 63.9 \pm 0.2 $	&	$0.11 \pm 0.01, 0.10 \pm 0.02$		&		\\
(729)~Watsonia				&	STGD	&	L				&	L					&	Y				&	$50.0 \pm 0.4$				&	$0.13 \pm 0.01$	& 729	\\
(753)~Tiflis				&	S			&	L				&	\textbf{S}		&	\textbf{N}	&	$20.9 \pm 0.8$			&	$0.33 \pm 0.05$			& 	\\
(824)~Anastasia				&	S			&	L				&	L					&	\textbf{Y}	&		$32.5 \pm 0.3$		&	$0.11 \pm 0.03$		&			\\
(908)~Buda				&				&	L				&	D					&	\textbf{N}	&	$30.8 \pm 0.5$				&	$0.09 \pm 0.01$		&		\\
(980)~Anacostia			&	SU		&	L				&	L					&	Y				&	$74.7 \pm 0.6$				&	$0.23 \pm 0.06$	&			\\
(1040)~Klumpkea			&				&					&						&					&	$22.3 \pm 0.2$				&	$0.24 \pm 0.04$	& 1400	\\
(1284)~Latvia			&	T			&	L				&						&	\textbf{Y}	&	$41.1 \pm 0.5$		&	$0.08 \pm 0.02$	&			\\
(1372)~Haremari			&				&	L				&	\textbf{L}		&	Y 	&	$26.5 \pm 0.3$	&	$0.04 \pm 0.01$	&	729	\\ 
(1406)~Kommpa			&				&	Ld				&	D					&	\textbf{N}	&	$24.2 \pm 0.4$	&	$0.17 \pm 0.05$	&			\\
(1702)~Kalahari			&	D		&	L			&					&		&	$34.6 \pm 0.1$, $32.7 \pm 0.2$	&	$0.06 \pm 0.01$, $0.06 \pm 0.01$&\\
(2085)~Henan				&				&	L				&	L					&	\textbf{Y}		&	$13.4 \pm 0.1$	&	$0.30 \pm 0.06$	& 2085 \\
(2354)~Lavrov					&				&	L				&	\textbf{L}		&					&	$13.3 \pm 0.1$	&	$0.23 \pm 0.05$	& 2085	\\
(2448)~Sholokhov						&				&	L				&	L					&	\textbf{N}	&	$38.5 \pm 0.2$	 &	$0.16 \pm 0.02$	&			\\
(2732)~Witt				&				&	A				&	L					&					&	$11.0 \pm 0.3$				&	$0.30 \pm 0.02$					&	\\
(3269)~Vibert Douglas						&				&					&						&				&	$11.7 \pm 0.2$	&	$0.14 \pm 0.03$	&	729 \\
(4917)~Yurilvovia						&				&	Ld				&	\textbf{L}		&					&	$8.0 \pm 2.4$				&	$0.21 \pm 0.21$			&			\\
(8250)~Cornell 				&				&					&						&					&	$9.0 \pm 0.3$				&	$0.34 \pm 0.08	$		&	1400	\\
(15552)~Sandashounkan				&				&					&						&				&	$7.6 \pm 0.1$	&	$0.37 \pm 0.04$ &1400	\\
(19369)~1997 YO		&			&			&			&			&	$14.3 \pm 0.1$, $13.6 \pm 0.1$	&	$0.16 \pm 0.03$, $0.18 \pm 0.02$	&1400	\\
(26219)~1997 WO21				&				&					&						&					&	$7.6 \pm 0.2$		&	$0.22 \pm 0.03$		& 1400	\\
(67255)~2000 ET109				&				&					&						&					&	$6.6 \pm 0.1$				&	$0.28 \pm 0.02$		& 1400	\\
\hline \hline
(673)~Edda						&	S	&	S			&	L		&												& 	$37.6 \pm 0.4$        	& $0.09 \pm 0.03$&				\\
(3734)~Waland					&		&	Ld			&	L		&												&	$9.0 \pm 0.2$		&	$0.20 \pm 0.05$	&				\\
(3844)~Lujiaxi 					& 	 	&	L			&	L		&													&	$15.0 \pm 0.7$			&	$0.17 \pm 0.03$&		2085 \\
(4737)~Kiladze					&		&	L			&	L		&													&	$8.8 \pm 0.1$			&	$0.15 \pm 0.02$&			\\
(5840)~Raybrown				&		& Ld			&	L		&													&	$9.7 \pm 0.1$			&	$0.22 \pm 0.04$&		\\
(7763)~Crabeels				&		& L 			&	L		&													&	$7.8 \pm 0.1$			&	$0.22 \pm 0.02$&			\\
\hline
\end{tabular}
\caption{List of the targets observed (spectroscopy and/or polarimetry) during the different campaigns (upper part). Some targets which were not observed by us but discussed in this work were added in the lower part. Each bold entries mean that this is a result determined in this work. The first column corresponds to the number and name of the considered asteroid. The columns Tholen \citep{Tholen84}, SMASS \citep{Bus_Bin,Mot_2008} and DM \citep{Dem_2009,Bus_2009} stand for the taxonomic class in these 3 taxonomies. The Barbarian column indicates whether the asteroid is considered as a Barbarian \citep{b5,b6,b31}. D (NEOWISE) and $p_V$ (NEOWISE) correspond to respectively the diameter and the geometric albedo as given by the NEOWISE catalog \citep{Mai_2016}. Finally, the Family column indicates the number of the parent member of the family in which the asteroid is classified (606 for the Brangane, 729 for the Watsonia, 1400 for the Tirela/Klumpkea and 2085 for the Henan family). }
\label{tab:Param}
\end{table*} 

\subsection{Spectroscopy}
\subsubsection{Spectral classification}
\label{ssec:Spec_Class}
Using IRTF, we have obtained new NIR spectra between $0.82$ and 2.45~${\rm \mu}$m for 14 objects. For 9 of them, the visible part of the spectrum is available in the SMASS database and were merged together to produce the visible + NIR spectra. For each of them we derived a taxonomic classification according to the criteria used by \citet{Dem_2009}. Six objects are found to be (DM) L-class. The remaining three objects do not belong to the (DM) L class. (12)~Victoria is a (DM) D-type, while (122)~Gerda and (753)~Tiflis are (DM) S-types. The information about the derived taxonomical classes of the asteroids observed in this work is included in Table \ref{tab:Param} (bold entries). The other five observed asteroids for which no visible spectrum is available seem to be compatible with the (DM) L-class. However, no definitive classification can be made in the absence of the visible part of the spectrum.


\subsubsection{Space weathering}

\begin{figure*}
\includegraphics[width=18cm]{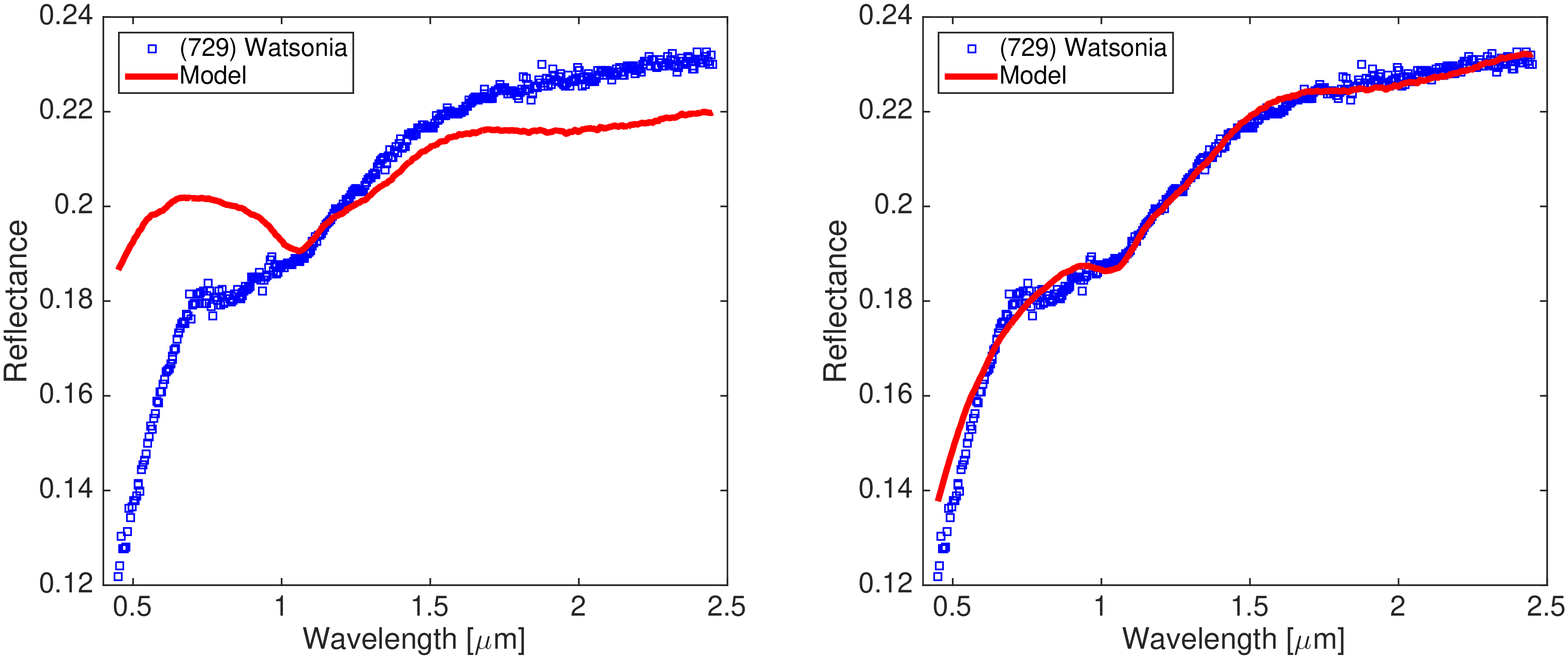}
\caption{Example of the fits performed for asteroid (729) Watsonia using fresh end-members (left panel) and weathered end-members (right panel).}
\label{fig:A_SL_SW}
\end{figure*}

Using space-weathered end-member spectra allowed us to model simultaneously the visible (even though the visible region is not taken into account in the optimization procedure) and near-infrared regions of the asteroid spectra. However, in applying our space weathering correction, some information about the optical properties of the end-members are needed. These properties are the real part of the refractive index $n$ and the effective size of the end-members particles ($D$). We choose the values $n = 1.635$ for the Mg-rich olivine \citep{Lucey_1998}, and $n = 1.7$ for the meteoritic component (see Section \ref{sssec:End_Members}). In the case of the spinel bearing CAIs, we used the values computed of $n$ with respect to wavelength derived by \citet{Hos_2008}. However, \citet{Hos_2008} derived an $n$ value valid only up to 1.4~${\rm \mu}$m, we considered $n$ to be constant over a wider interval of wavelengths. According to \citep{Gun_2013} the average size of regolith particles may be dependent upon the size D of the asteroid, ranging from $10$ to 100~${\rm \mu}$m for large asteroids (D $\textgreater 100$ km) and from millimeter to centimeter for asteroids smaller than 100 km. However, the finest fraction of regolith particles is responsible for the principal optical effect of the space weathering \citep{Pie_1993}. We then choose the effective size of the particles to be equal to 25~${\rm \mu}$m. On the other hand, this model gives as a result the mass fraction of nano-phase iron particle implanted on the regolith particles (see $f$ column of Table \ref{Tab:Fit_result2}). This parameter can be interpreted in terms of the extent at which an asteroid spectrum can be affected by space weathering. We show in Fig.~\ref{fig:A_SL_SW} an example of the best fit obtained for asteroid (729)~Watsonia when using fresh (left panel) and weathered (right panel) end-members.

\subsubsection{Spectral fitting}
\label{ssec:Spec_Fit}

\begin{table*}
\centering
\begin{tabular}{| c | cccccc | cccccc |}
\hline
										& \multicolumn{6}{ c |}{Allende matrix} & \multicolumn{6}{ c |}{Y-86751 bulk} \\
Asteroid							& FTA	&  Olivine	& Allende & $f$ & $p_V$ & $\chi^2$ & FTA	&  Olivine	& Y-86751 & $f$ &  $p_V$ &  $\chi^2$	\\
\hline
(172)~Baucis						&  $29$	&	$32$	&	$39$	 & 0.078& 0.15 &	2.96 &	$17$	&	$29$		&	$53$	 & 0.115& 0.13 &2.45		\\
(234)~Barbara					&  $25$		&	$62$	&	$13$	 & 0.039& 0.24 & 3.26 &	$20$	&	$65$		&	$15$	 & 0.047& 0.24 & 3.05	\\ 
(236)~Honoria					&  $12$		&	$29$	&	$58$	 & 0.061& 0.09 & 3.82	&	$0$	&	$51$		&	$49$	 & 0.093& 0.15 &3.79		\\
(387)~Aquitania					& $31$	&	$35$	&	$34$	& 0.031& 0.18 & 5.25	&	$19$	&	$34$		&	$47$	 & 0.066& 0.16 &3.56	\\
(402)~Chloe						&  $32$	&	$17$		&	$50$	& 0.012& 0.16 & 5.18 	&	$6$	&	$0$		&	$94$	 & 0.106& 0.10 &2.56		\\
(458)~Hercynia					&  $32$	&	$14$		&	$53$ & 0.019& 0.13 & 3.29	&	$9$	&	$1$		&	$90$	 & 0.121& 0.09 &1.86		\\ 
(599)~Luisa						&	$24$	&	$29$		&	$47$	& 0.074	& 0.15 & 3.29	&	$8$	&	$30$		&	$61$	 & 0.121& 0.13 &2.32			\\
(611)~Valeria					&	$3$	&	$11$		&	$85$	& 0.085	& 0.19 & 1.60	&	$0$	&	$68$		&	$32$	 & 0.080& 0.08 &3.09		\\
(673)~Edda						&	$34$	& $0$		& $66$	&0.014	& 0.12  & 2.29	&	$10$	&	$3$		&	$87$	 &	0.100& 0.10	& 2.10		\\
(679)~Pax							&	$26$	&	$20$		&	$53$	& 0.036& 0.14 & 4.12	&	$0$	&	$0$		&	$100$& 0.186& 0.08 &1.24		\\
(729)~Watsonia					&  $6$	&	$26$		& 	$68$	& 0.083	& 0.10 & 2.58	&	$0$	&	$61$		&	$39$	 & 0.092& 0.17 &3.64	\\ 
(824)~Anastasia					&  $45$	&	$9$		& 	$46$	& $0.000$	& $ 0.19$ & $9.33$	&	$10$	&	$0$		&	 90	 & 0.030 & 0.12 & 4.82	\\ 
(980)~Anacostia				&  $48$	&	$28$		&$23$ 	& 0.042& 0.17 & 5.20 	&	$36$	&	$0$		&	$64$	 & 0.106& 0.11 &3.64	\\     
(1372)~Haremari				&  $13$	&	$56$		& 	$31$ & 0.085& 0.12 & 2.44	& 	$0$	&	$44$		&	$56$	 & 0.137& 0.12 &1.85			\\   
(2085)~Henan					&	$36$	&	$6$		&	$58$	&	0.000		& 0.13 & 1.88	&	$10$	&	$0$		&	$90$	 &	0.075& 0.10 &1.49		 \\
(2354)~Lavrov					&	$41$	&	$35$		&	$25$ & 0.028& 0.18 & 1.47	&	$26$	&	$20$		&	$54$	 & 0.065&  0.14 &1.35		\\	
(2732)~Witt						&	$27$	&	$56$		&	$17$	&	0.045&	0.22	&	2.47	&	$22$	&	$58$		&	$20$	 &	0.058&	0.21 &	2.17		\\
(3734)~Waland					&	$40$	&	$16$		&	$44$	&	0.017&	0.15	&	1.75	&	$17$	&	$0$		&	$83$	&	0.088&	0.10	&	1.50	\\	
(3844)~Lujiaxi					&	$51$	&	$12$		&	$37$	&	0.000		& 0.19 & 1.78	&	$28$	&	$0$		&	$72$	 & 0.031& 0.13 & 1.32			\\
(4737)~Kiladze					&	$49$ & $0$		&	$51$	&	0.020	&	0.14	&	2.46	&	$29$	&	$0$		&	$71$	&	0.073&	0.12		&	1.92	\\
(4917)~Yurilvovia				&	$20$	&	$45$		&	$35$	&	0.061&	0.15 &	8.10	&	$5$	&	$32$		&	$62$	&	0.140& 0.11	&	6.05	\\
(5840)~Raybrown				&	$39$	&	$30$		&	$31$	&	0.010&	0.18	&	3.82	&	$22$	&	$0$		&	$78$	&	0.075&	0.10	&	3.04	\\
(7763)~Crabeels				&	$49$ & $11$		&	$39$	&	0.022	&	0.18	&	1.16	&	$32$	&	$0$		&	$68$	&	0.071&	0.13		&	1.09	\\
(8250)~Cornell					&	$23$	&	$76$		&	$1$	& 0.027 & &1.17	&	$23$	&	$76$		&	$1$	 & 0.028& & 1.17		\\
(15552)~Sandashounkan	&	$21$	&	$70$		&	$9$	& 0.040	& & 2.21	&	$18$	&	$73$		&	$9$	 & 0.043& &2.28		\\
(19369)~1997 YO				&	$8$	&	$0$		&	$92$	& 0.103	& & 1.13	&	$2$	&	$67$		&	$31$	 & 0.079& &1.96		\\
(26219)~1997 WO21			&	$18$	&	$77$		&	$5$	& 0.017	& & 1.06	& 	$19$	&	$77$		&	$5$	 & 0.019& & 1.07		\\
(67255)~2000 ET109		&	$22$	&	$45$		&	$33$	& 0.062	& & 1.51	&	$13$	&	$57$		&	$30$	 & 0.075 &	&1.69	\\
\hline

\end{tabular}
\caption{Result of the Hapke fitting procedure of reflectance spectra. For each object, identified by its asteroid number, we give the relative abundances of each of the three considered end-members in percent. The column ($f$) gives the fraction of npFe$^0$. The last column (Albedo) gives the reflectance at 0.55~${\rm \mu}$m of the fitted asteroid spectrum (no value when the visible part of the spectrum is missing).}
\label{Tab:Fit_result2}
\end{table*} 

The results of the spectral fitting procedure explained in section \ref{ssec:Fit_Spectra} are shown in Table \ref{Tab:Fit_result2}. All fits are shown in \ref{app:fit}. The red continuous lines correspond to the results using the Allende matrix while the blue discontinuous lines are for the results obtained with the spectrum of the Y-86751 meteorite. For each fit, the plot of the residuals is shown below. In the majority of the cases, the residuals are smaller using Y-86751 than using the Allende matrix. All the spectra were normalized to unity at 0.55~${\rm \mu}$m, except those for which no visible counterpart was available. In those cases, the spectra were normalized to unity at 1~${\rm \mu}$m. For 3 of those, we have plotted the available Sloan Digital Sky Survey (SDSS) \citep{Ive_2002} measurements.   


\citet{Sun_2008a} modelled the spectra of (234)~Barbara, (387)~Aquitania, and (980)~Anacostia using a slightly different approach. They found for these asteroids high CAI  abundances never observed in meteorite samples, equal to $22\%$, $25\%$ and $39\%$ of spinel bearing CAIs, respectively. In their analysis, they used four different end-members. Fluffy type A CAI, CAI free Allende matrix, MgO rich olivine and the spectrum of (2448)~Sholokhov to simulate the typical slope of a (DM) L-class asteroid. 
In our analysis, we find relatively similar values for the same three asteroids. Our values of CAIs abundances are respectively $25\%$, $31\%$, and $48\%$, when using the Allende matrix and $20\%$, $19\%$, and $36\%$ using Y-86751.



\begin{figure*}
\includegraphics[width=18cm]{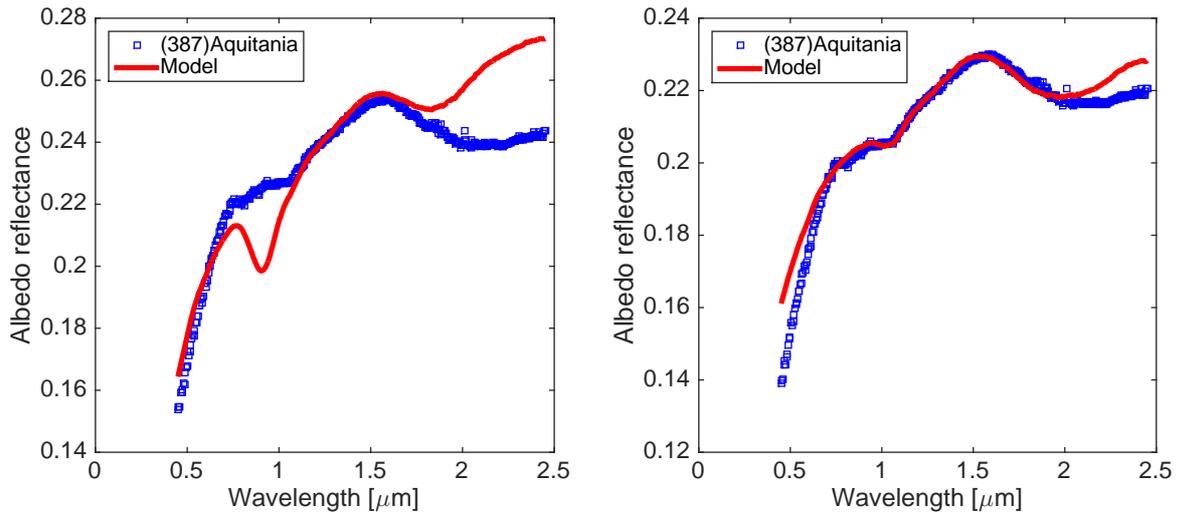}
\caption{Best-fit models for the reflectance spectrum of asteroid (387)~Aquitania. The left panel shows the best result obtained assuming the presence of pyroxene without CAIs. The right panel shows the opposite case, namely CAIs and no pyroxene.}
\label{fig:Pyr_CAI}
\end{figure*}

As it was already mentioned in \citet{Sun_2008a}, only the fluffy type A CAI seems to be able to explain the observed spectra. Among the three different FTAs spectra adopted as possible end-members, the one showing the highest fraction of FeO ($\sim14\%$) within spinel was always found to give a better fit. In no case, using B-type CAIs led to satisfactory fits of the spectra.
We did some attempts to consider also pyroxene as a possible end-member (see Fig. \ref{fig:Pyr_CAI}). Even though in very few cases, very small amount of pyroxene ($1\%$ or below) could be added as a possible component, we found that pyroxene, in general terms, does not help to improve the fits of our spectra.

The composition of L-type asteroids seems to show some diversity as we observe different kind of behaviour. First, we notice that there are only two asteroids for which the Allende matrix provide a better fit (base on the $\chi^2$). From these, (611)~Valeria is the only one which is convincing. For the case of (729)~Watsonia,  the fit using the Allende matrix show a shallow absorption band around 2~${\rm \mu}$m which is not present in the asteroid spectrum. In the following, we will always discuss the result associated to Y-86751.

Only four spectra exhibit a positive slope from 1.5 to 2.5~${\rm \mu}$m (236, 611, 729, 1372). These spectra are also characterized by the total or nearly absence, of a 2 $\rm \mu$m absorption band. As a result, our model does not need the addition of CAIs, but includes a large part of meteoritic component. Since the meteorite possesses by itself some CAIs, we can still consider that these asteroids are not CAIs free. We note that (729) and (1372) belong to the Watsonia family. Another member of the Watsonia family, (599)~Luisa, also shows a low fraction of CAIs and shows a shallow absorption band around 2~${\rm \mu}$m associated with a null slope. We note that the visible part of the spectrum of (599)~Luisa is not properly modelled.  

Two asteroids possess very large amount of CAIs ($>30\%$). From those, the spectrum of (980) is relatively well fitted in both the visible and NIR regions. However, the fit could still be improved for the 2~{$\rm \mu$}m absorption band which is shifted toward shorter wavelength in our modeling. The visible part of (7763) is not properly fitted, and also the fit of the 2 $\rm \mu$m region is not ideal.

For four of them, the meteorite Y-86751 is almost the unique component ($>90\%$) associated with no (or almost no) olivine and a few percent of CAIs. These spectra are characterized by a monotonically decreasing slope after 2~${\rm \mu}$m, a high value of space weathering and low albedo. Two of them, (402) and (679), have a poor agreement in the visible region while the other two (458) and (2085) show a good agreement. Even though these asteroids do not show significant CAIs, their presence in the Y-86751 composition provides more or less $10\%$ of CAIs.  On the other hand, Y-86751 and Allende are almost absent for three asteroids ($<10\%$). All these three asteroids (8250), (15552), and (26219) belong the the Tirela family. The spectrum (only NIR) of these asteroids is well modelled with the highest fraction of olivine found in our sample and moderate fraction of CAIs. 

The asteroid (824)~Anastasia is a particular case. The model failed when using the Allende meteorite and produce a quite poor fit when using the Y-86751 bulk. The model provide 90\% of meteoritic associated with 10\% of CAIs. Although the fit is quite poor the Y-86751 bulk seems able to reproduce some very specific features such as a small absorption band around 1.1 ${\rm \mu}$m. The spectrum of (824)~Anastasia presents also a very steep slope in the visible which can be only very approximately reproduced. Although it seems that our model is missing some end-member to satisfactorily reproduce the spectrum of Anastasia, we also clearly see that the Y-86751 bulk seems to be the major component.

Our model failed in reproducing the spectrum of (2448) and (2732). The composition for these three asteroids should then be different than the rest of the L-class. In the case of (2448) this result agrees with the fact that this asteroid does not show a large inversion angle. We note that this asteroid was previously classified as A-type in the SMASS taxonomy.

\subsection{Polarimetry}

\subsubsection{Phase-polarization curve}
\label{ssec:Phase_Pol_Curv}
In this work, $109$ polarimetric measurements of 32 individual asteroids are presented. All the observations were acquired using a standard Johnson $V$ filter. The individual measurements $P_{\rm r}$  are listed in \ref{app:pola}. 

Phase-polarization curves were built by using data from the literature to complement our new data. Literature data are available at the PDS web site\footnote{Planetary Data System. The data are available at the URL address http://pds.jpl.nasa.gov/ (files maintained by D.F. Lupishko and I.N. Belskaya)}, while others were taken from some recent papers \citep{Gil_2014}. The obtained phase-polarization curves were fitted using Eq.~\ref{eq:Pr2} whenever enough data are available, using the techniques explained in Section~\ref{ssec:Phase_Pol}. 

\begin{figure}
\includegraphics[width=9cm]{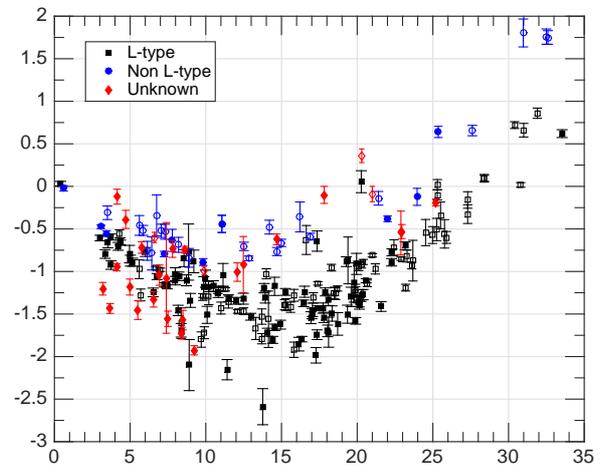}
\caption{All polarimetric measurement of asteroids studied in this work. All full symbols correspond to data obtained by us while empty symbols correspond to measurements retrieved form literature. The black squares stand for (DM) L-type , the blue circles stand for asteroids which do not belong to the (DM) L-class, and the red diamonds stand for asteroids for which the (DM) classification is unknown.}
\label{fig:All_Pola}
\end{figure}

Fig.~\ref{fig:All_Pola} presents all the polarimetric measurements available for all asteroids studied in this work. In this figure and for all the following figures presenting polarimetric measurements, full symbols represent data obtained by us while empty symbols correspond to data retrieved in the literature. In Fig.~\ref{fig:All_Pola}, asteroids were subdivided into three groups based on their (DM) taxonomy. Those belonging to the (DM) L-class are plotted as black squares. Their phase-polarization curve is characterized by $P_{\rm min}$ around -1.5\% occurring between 15 to 20 degrees of phase angle and an inversion angle between 25 to $30^{\circ}$. Asteroids which do not belong to the L-types are shown in blue circles. They display different phase-polarization curves which are not compatible with the one displayed by L-class. Those for which the (DM) classification is unknown are displayed as red diamonds. Some of these measurements are compatible with the phase-polarization curves of L-class objects, whereas others are not.

\begin{figure*}
\includegraphics[width=19cm]{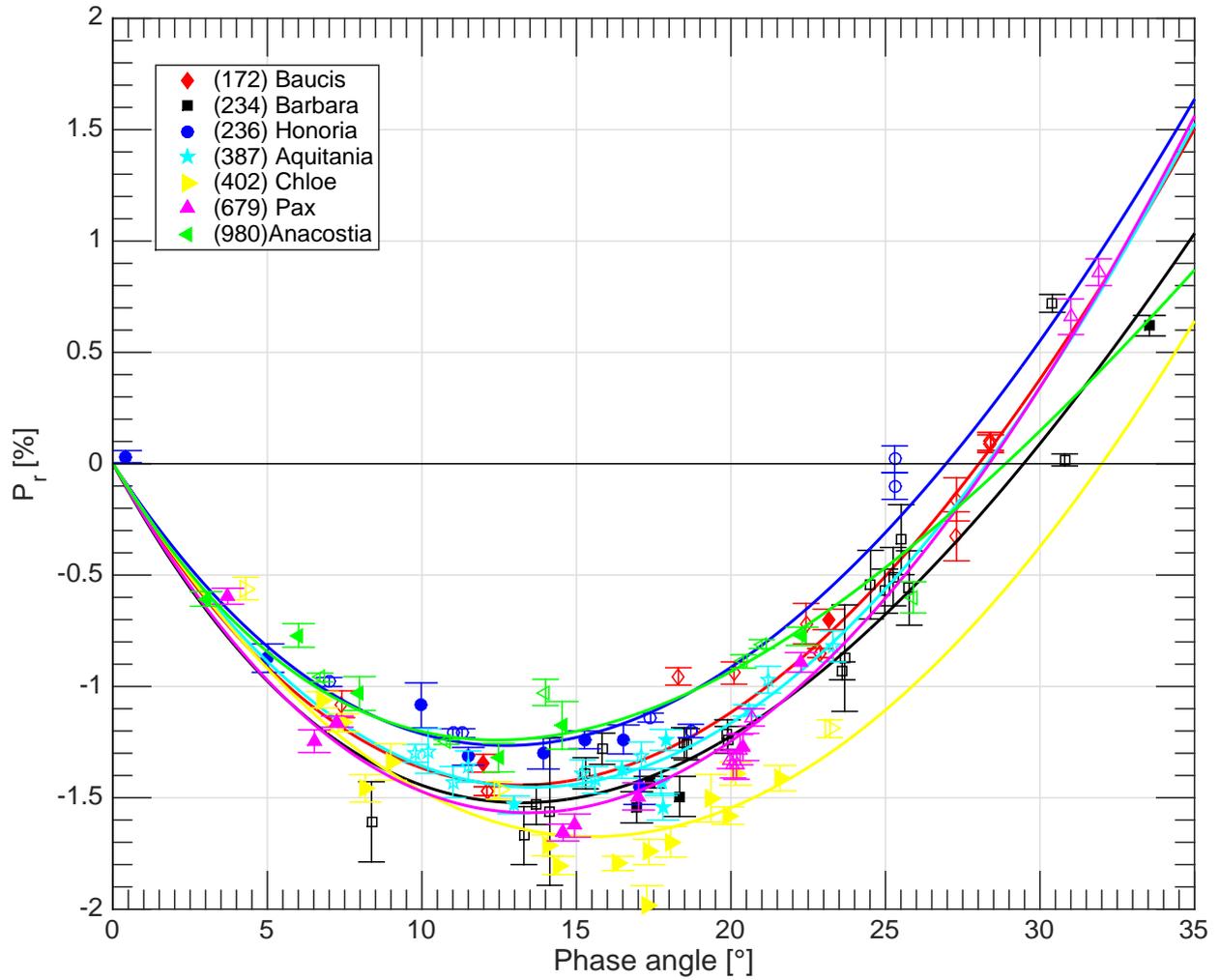}
\caption{Phase-polarization curves derived for Barbarian asteroids using new polarimetric data presented in this work. These asteroids are the same as those of Fig.~\ref{fig:CAI_Inv}. Full symbols refer to data obtained in this work while open symbols refer to data obtained in the literature.}
\label{fig:AI_Inv_pola}
\end{figure*}

Among these 32 observed asteroids, 11 of them (172, 234, 236, 387, 402, 458, 599, 679, 729, 980, and 1372) were already known to be Barbarians \citep{b5,b6,b31,Gil_2011,Gil_2014,Bag_2015,Dev_2017a}. Our measurements allowed us to improve the coverage of the phase-polarization. For seven of them, the phase angle coverage is sufficient to calculate a phase-polarization curve. The measurements and derived phase-polarization curves for these asteroid are presented in Fig.~\ref{fig:AI_Inv_pola}. The derived polarimetric parameters are summarized in Table~\ref{tab:Pol_info}. From these, (172), (234), (236), (387), and (980) had been previously modelled \citep{b6,Cel_2015a,Bel_2017}.  Our results are in agreement with previously published phase-polarization modellings. The only major difference concerns (234)~Barbara for which the inversion angle is found to be $2^{\circ}$ lower by \citet{b6} and \citet{Bel_2017}. This is mainly due to one measurement at $\alpha = 30.4^{\circ}$ where $P_{\rm r} = 0.72$ \citep{b6} which is not compatible with two more recent measurements at $30.8^{\circ}$ and $33.54 ^{\circ}$ which provide $P_{\rm r}$ equal to $0.017 \pm 0.027$ and $0.62 \pm 0.046$, respectively. More observations at high phase angle of (234)~Barbara are needed to see if these contradictions between several observations are due to measurement errors or to intrinsic variation of the polarization degree measured in different observing circumstances.
\begin{table}
\begin{tabular}{c c c c}
\hline
Asteroid   & $\alpha_{\rm inv}$ & $\alpha(P_{\rm min})$ & $P_{\rm min}$ \\
\hline
(12)~Victoria  		   & $23.0 \pm 0.2$  &  $11.3 \pm 0.1$   & $-0.81 \pm 0.01$           \\ 
(122)~Gerda        &  $18.4 \pm 2.3$ &   $8.2 \pm 0.5 $ & $ -0.73 \pm 0.05 $       \\
(172)~Baucis        & $28.1 \pm 0.1$  & $12.8 \pm 0.2$     & $-1.42 \pm 0.02 $           \\
(234)~Barbara         &  $29.4 \pm 0.1$ & $12.5 \pm 0.4$   & $-1.57 \pm 0.04 $         \\ 
(236)~Honoria         & $26.6 \pm 0.2$    & $13.0 \pm 0.1$ & $-1.27 \pm 0.01 $        \\
(387)~Aquitania        & $28.2 \pm 0.5$  &  $13.7 \pm 0.2$  & $-1.46 \pm 0.02 $            \\
(402)~Chloe         & $32.0 \pm 0.4$  & $15.7 \pm 0.2$     &$-1.68 \pm 0.02$            \\
(679)~Pax         &  $28.2\pm0.1$ &  $13.7 \pm 0.1$    & $-1.59 \pm 0.02$          \\
(980)~Anacostia         &  $28.6 \pm 0.5$ &    $12.5 \pm 0.2$  &$ -1.24 \pm 0.01$             \\
(1284)~	Latvia	&	$26.1 \pm 0.4$	          &	                          	&			\\
\hline
\end{tabular}
\caption{Summary of the phase-polarization curve parameters for some asteroids studied in this work. $\alpha_{\rm inv}$ is the inversion angle, $\alpha(P_{\rm min})$ is the phase angle in the negative polarization branch where linear polarization reaches its largest (negative) value $P_{\rm min}$, which is listed in the last column.}
\label{tab:Pol_info}
\end{table} 

For some targets, our measurements are the first polarimetric observations. The new data allow us, in some cases, to obtain for the first time an estimate of the inversion angle, leading us to decide whether they are Barbarians or not (we include this information in Table~\ref{tab:Param}).

\subsubsection{Identification of new Barbarians}

The new polarimetric measurements presented in this work allow us to identify some new Barbarian asteroids.

{\it (606)~Brangane}. For this asteroid, we have two polarimetric measurements at phase
angles around $20^{\circ}$. The corresponding $P_{\rm r}$ values are around $-1.2\%$ which is a clear diagnostic of a Barbarian behaviour.

{\it (611)~Valeria}. For this asteroid, we have two polarimetric measurements at similar phase angles around $20^{\circ}$. The corresponding $P_{\rm r}$ values are around $-0.8\%$ which is a clear diagnostic of a Barbarian behaviour.

{\it (824)~Anastasia}. For this asteroid, we have five polarimetric measurements at phase angles ranging from $3.35^{\circ}$ to $18.11^{\circ}$ . The observed $P_{\rm r}$ values at $\alpha = 18.11^{\circ}$ is $-1.71 \pm 0.18\%$ which is an indication of a Barbarian behaviour.

{\it (1284)~Latvia}. Four polarimetric measurements are available at medium and high phase angles ($\alpha = 9.23^{\circ}$, $22.89^{\circ}$, $22.95^{\circ}$, and $25.20^{\circ}$) with $P_{\rm r}$ values respectively equal to $-1.93$, $-0.55$, $-0.53$, and $-0.19$. The three measurements at phase angles higher than $22^{\circ}$, although one of them has a large error bar, nicely indicate an inversion angle around $26^{\circ}$ which is clearly Barbarian-like.

{\it (1372)~Haremari}. We have only one single measurement for this asteroid at a phase angle of $19.96^{\circ}$ for which the value of $P_{\rm r}$ is $-0.911\pm0.124\%$. As for (606)~Brangane, this negative value at a phase angle as high as $19.96^{\circ}$ is by itself a strong evidence of its Barbarian nature. Haremari belongs to the dynamical family of Watsonia. This single available measurement for Haremari is in agreement with the (yet fairly noisy) phase-polarization curve of Watsonia, the largest remnant of this family. This is the first family identified among the Barbarians \citep{Cel_2014}. 

In addition to the above-mentioned objects, we have also one polarimetric observation of (2085)~Henan, for which we find
$P_{\rm r} = -1.920 \pm 0.090$ at a phase angle of $15.82^{\circ}$. This observation is still far from the inversion angle, but the polarization is strongly negative. This makes (2085)~Henan a reasonable Barbarian candidate, to be confirmed by future measurements.

\begin{figure}
\includegraphics[width=9cm]{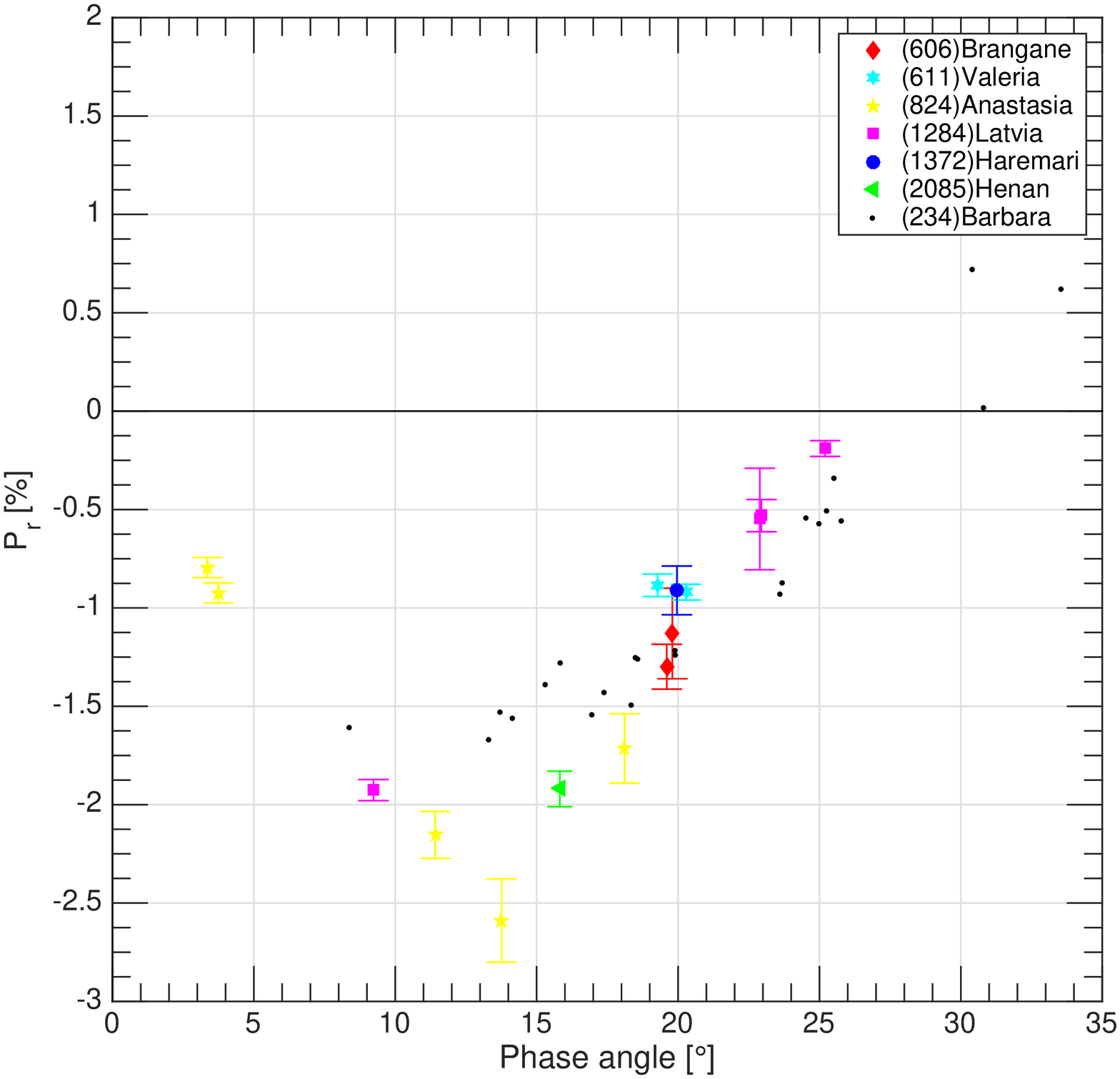}
\caption{Polarimetric data for (606)~Brangane (red diamonds), (824)~Anastasia (yellow pentagrams), (1284)~Latvia (magenta squares), (1372)~Haremari (large blue disk), and (2085)~Henan (green triangle). Polarimetric data of (234)~Barbara are also shown as small black points for a comparison.}
\label{fig:new_Barb}
\end{figure}

Fig.~\ref{fig:new_Barb} shows the polarimetric data for the asteroids discussed in this Section. The phase - polarization curve of (234)~Barbara is also shown for a comparison.

\section{Discussions}

\label{Sec:Inter_Disc}

\subsection{The abundance of CAIs}

In our sample of asteroids, we are observing a wide range of CAIs abundance from almost 0\% (including the CAIs present inside the Y-86751 bulk) to more than 36\%. This wide range of CAIs suggest that all the L-type asteroids do not have formed either in the same time and/or in the same location in the Solar nebula. This support the hypotheses that CAIs were spread inhomogeneously in the Solar nebula after their formation close to the Sun.

\subsection{Relation between L-class (DM) and Barbarians}

We already saw in Fig.~\ref{fig:All_Pola} that L-class asteroids display a distinct phase-polarization characterized by the deep and very wide negative polarization branch typical of Barbarians. 

The first Barbarian asteroids were discovered when asteroid taxonomy was based only on reflectance spectra limited to the visual region. We have already mentioned that, according to the SMASS taxonomy, these first Barbarians belonged to the L-, Ld- and K-classes. Later, it has been found that known Barbarians belong to the (DM) L taxonomic class, defined by taking into account also the NIR region of the spectrum.

\begin{figure}
\includegraphics[width=8.8cm]{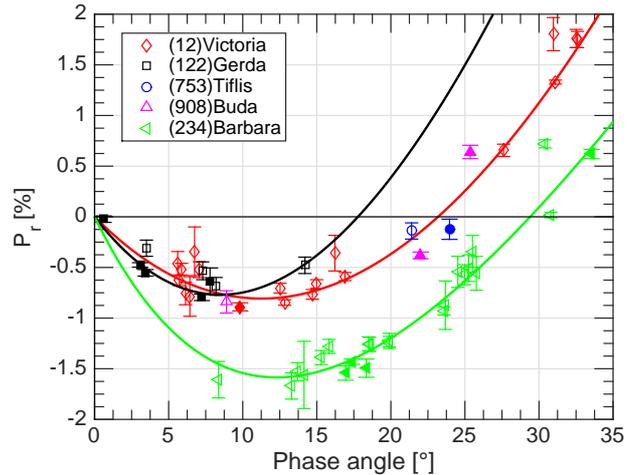}
\caption{Polarimetric data for (12)~Victoria (red diamonds), (122)~Gerda (black squares), (753)~Tiflis (blue circles), (908)~Buda (magenta triangles), and (234)~Barbara (green left-oriented triangles). The four first asteroids are (SMASS) L-type, but are not L-type in the (DM) taxonomy. (234)~Barbara is an L-type in the (DM) taxonomy and a Barbarian. It is displayed here as reference for the typical behaviour of (DM) L-type /Barbarian asteroids.} 
\label{fig:Non_barb_pola}
\end{figure}

Some asteroids belonging to the (SMASS) L-class have been included in our sample, even if there was no evidence that they belong to the (DM) L-class, to get a more definitive evidence that the Barbarian polarimetric behaviour is indeed uniquely associated with the (DM) L-class. These objects are (12)~Victoria, (122)~Gerda, (753)~Tiflis. New spectroscopic measurements in the NIR presented in this work allow us to conclude that they belong to the (DM) D- (Victoria) and (DM) S-class (Gerda and Tiflis). Our polarimetric measurements for these three asteroids confirm that they do not exhibit a large inversion angle and certainly are not Barbarians (see Fig.~\ref{fig:Non_barb_pola}).

Another asteroid in our sample, (908)~Buda, was also classified as (SMASS) L-class, but it was later reclassified as a (DM) D-class \citep{Dem_2009}. Our two polarimetric observations of this asteroid at high phase angle ($P_{\rm r} = -0.381 \pm 0.033$ at $\alpha = 22.01^{\circ}$ and $P_{\rm r} = 0.640 \pm 0.166$ at $\alpha = 25.36^{\circ}$) suggest an inversion angle around $23^{\circ}$. This value is fairly high in general terms, but still too low to be considered as clearly diagnostic of a Barbarian asteroid, until new data will confirm or rule out this hypothesis.

Asteroid (1406)~Komppa was already found to belong to the (DM) D-class (MIT-UH-IRTF survey), but it had been previously classified as a (SMASS) Ld asteroid. Our unique polarimetric measurements indicate a  $P_{\rm r} = -0.44 \pm 0.10$ at a phase angle of $11.06^{\circ}$. This phase angle is too low to draw any conclusion about its inversion angle, but the degree of linear polarization seems to be exceedingly low to be considered as a likely Barbarian candidate.

Some asteroids, (478)~Tergestre and (1702)~Kalahari are found to belong to the (SMASS) L-class, but no NIR spectra are available to determine the (DM) taxonomic classification. (478)~Tergestre does not possess a large inversion angle and most probably is not a (DM) L-type asteroid. In the case of (1702)~Kalahari, our polarimetric observations were taken at too low phase angles to draw any reliable conclusion.

On the other hand, asteroids (606), (824), (1372), and (with some more uncertainty) (2085), are found in this work to be Barbarians, and all of them are found to belong to the (DM) L-class. Based on these results, and on the fact that no Barbarian identified so far belongs to taxonomical classes other than the (DM) L-class, while only one peculiar (DM) L-class asteroid (2448)~Sholokhov (see Sec.~\ref{ssec:Spec_Fit} for a discussion about this peculiar case), we can safely and definitively confirm the existence of a biunivocal relation.





\subsection{Aqueous alteration}

We found evidence in our modelling procedure that the meteoritic
sample Y-89751 was almost always providing better result than using
the matrix of Allende. As already discussed, these two meteorites are of similar type and similar bulk composition. Y-89751 contains CAI inclusions with more FeO-rich spinels (18-25\% compared to 4-14\% for Allende). Neither Allende nor Y-89751 present hydrated minerals on their matrices, but according to \citet{Gyo_2011}, the material in the groundmass around the chondrules of Y-89751 may show a flow texture due to aqueous alteration.
  
The possible presence of aqueous alteration for L-type asteroids was already suggested by \citet{Sun_2008a} to explain the apparent
absence of igneous differentiation that would have destroyed the
observed FTAs \citep{Gri_1989}. On the other hand, \citet{Riv_1998}
observed a 3~${\rm \mu}$m feature on the spectrum of (387)~Aquitania which is
characteristic of aqueous alteration. 

However, the improvement of the modelling with Y-89751 is not a proof of the presence of hydration on L-type asteroids. The improvement could as well be due to slight modifications in composition, FeO enrichment of the spinel and/or different preparation of the measured laboratory samples. Indeed, most of the differences observed between Allende and Y-89751 arise in the visible part of the spectrum and around the 2~${\rm \mu}$m absorption band. We do not see the typical 0.7~${\rm \mu}$m absorption feature associated to hydrated silicates neither in the spectrum of Y-89751 nor in that of Allende or any L-type studied in this work, and the 2~${\rm \mu}$m band is not diagnostic of aqueous alteration. Only a spectroscopic survey of L-types around 3~${\rm \mu}$m would assess the aqueous alteration of L-types.

\subsection{Interpretation of the high polarimetric inversion angle of Barbarian asteroids}
\label{ssec:CAI_Barb}
In this section, different possibilities are given to explain the large polarimetric inversion angle of Barbarian asteroids.

\subsubsection{Regolith size}
In a recent paper, \citet{paper2} have shown an updated version (see Fig.~9 of the above-mentioned paper) of a classical plot \citep[see. for instance,][]{Dol_1989} showing the distribution of the asteroids in the space of the polarimetric parameters $P_{\rm min}$ versus inversion angle. The authors display that the Barbarians show a well distinct behaviour with respect to ``regular'' asteroids, since they occupy a region of the $\alpha_{\rm inv}$ - $P_{\rm min}$ space corresponding to the domain occupied, according to laboratory experiments, by very finely divided silicaceous powders and thin lunar fines, whereas regular asteroids are usually found in a region of the plot corresponding to pulverized rocks and meteorites with coarser grains, having  sizes between 30~${\rm \mu}$m and 300~${\rm \mu}$m. \citet{paper2} also noted that regular asteroids tend to group together, in the above-mentioned space, according to the albedo. In turn, the albedo is known to be a parameter that is related both to composition and regolith properties \citep[see also, in this respect,][]{Vesta}.

This polarimetric property suggests therefore that the Barbarian
behaviour is related to anomalous surface regolith properties, in
particular by regolith particle sizes much smaller than
usual. \citet{paper2} also noted that another non-Barbarian asteroid,
(21)~Lutetia visited by the \textit{Rosetta} probe, is also found in a
location close to that occupied by Barbarians in the $(\alpha_{\rm{inv}},P_{\rm{min}})$ space. They noted that this asteroid is unusual in several respects, and there are reasons to believe that Lutetia is a primitive asteroid \citep{Cor_2011,Sie_2011}.
The \textit{Rosetta} instruments VIRTIS and MIRO also found evidence of very low thermal inertia \citep{Gul_2012,Oro_2012}.
This is in general agreement with polarimetric data suggesting that Lutetia's surface could be rich in fine dust. 

Surfaces rich of fine dust could also be interpreted in terms of age, through the cumulative effect of a long exposure to moderate impacts unable to totally disrupt the body, but more than sufficient to finely pulverize its surface. This is a possible interpretation deserving further scrutiny, because we have already seen that the likely composition and also possibly the slow rotations of Barbarians could suggest that these asteroids could be extremely old.

However, the size of regolith is believed to be dependent on the asteroid size. The fact that Barbarian asteroids are also found among small members of dynamical family is contradictory with a relation between asteroid regolith size and Barbarian behaviour.

\subsubsection{High abundance of spinel bearing-CAIs}

\begin{figure}
\includegraphics[width=8.8cm]{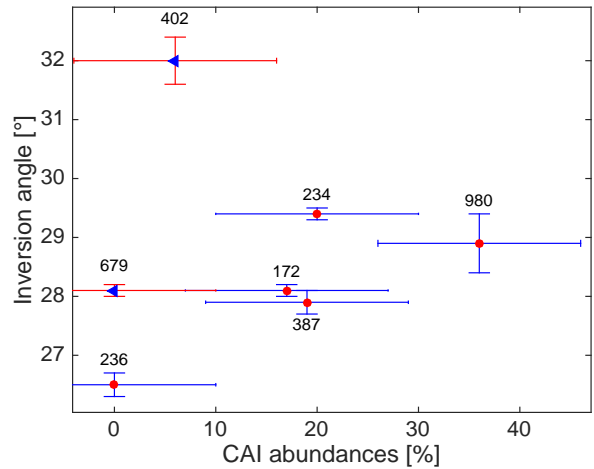}
\caption{Plot of the polarimetric inversion angles for 7 Barbarian asteroids studied in this work, as a function of the derived CAI abundances obtained from fitting their reflectance spectra.}
\label{fig:CAI_Inv}
\end{figure}

Fig.~\ref{fig:CAI_Inv} shows a graph of the relative abundance of
fluffy type A CAIs (obtained by our analysis) against the inversion
angle of the phase-polarization curve for some Barbarian asteroids for
which we have decent phase-polarization curves. The asteroids (402)
and (679) have been represented using different symbol (blue
triangles). Indeed, as already discussed in Sec.~\ref{ssec:Fit_Spectra} these two objects seem to be different with respect to the other known Barbarians. Based on our model, they are the only ones (out of the 7) to be almost only composed of meteoritic component ($100\%$ for (679) and $96\%$ for (402)). These asteroids also possess very high values of space weathering. We also note that the visible part of their spectrum is not well fitted in both cases.  

With the exception of (402) and (679), a correlation between the
modelled abundance of CAIs and the inversion angle seems to be
apparent, in spite of the fairly large uncertainties in the
determinations of CAI abundances. The interpretation of such a
correlation is not trivial, however. 
  
The results suggest that the polarimetric inversion angle of Barbarians tends to increase with increasing abundance of CAIs. According to current understanding, the only active phase of CAIs in determining the  2~${\rm \mu}$m absorption band in the reflectance spectrum is spinel. Moreover, there are reasons to believe that the strength of the absorption band is determined primarily by the spinel FeO content. So, when we plot our resulting abundances of the fluffy A-type CAI component in our modelled mineralogic compounds, we are also indirectly dealing with the FeO content of the spinel assumed to be present in the modelled CAI component.

One should note that even if (236) seems to not possess any CAIs, its spectrum is still modelled using meteoritic material which possess CAIs inclusions in which spinel has as much as $25\%$ of FeO. Taking into account an abundance of CAIs of $\sim10\%$, we can assume a CAI abundance of $\sim 5\%$ for (236).

The most optically active compound inside CAIs is the FeO-bearing spinel. This material has a high refractive index which is highly dependent on the wavelength in the visible range \citep{Hos_2008}. Because olivine, and also the meteoritic components (which are mainly composed of olivine) all have a refractive index mostly constant in the visible, an indirect proof for the presence of FeO-bearing spinel would be a wavelength dependence of the polarimetric properties measured at different visible wavelengths \citep{Zub_2015}. Most polarimetric observations of asteroids have been historically done in $V$-band, only, but there are a few exceptions regarding asteroids particularly bright and/or interesting. Among them, there is (234)~Barbara.
  
\begin{figure}
\includegraphics[width=8.8cm]{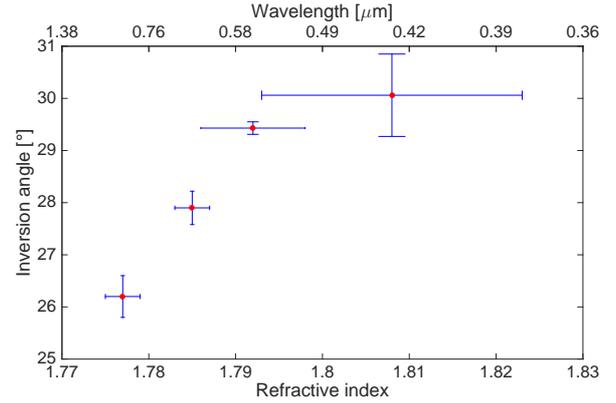}
\caption{Inversion angle of (234)~Barbara as a function of the refractive index of spinel and wavelength.}
\label{fig:Barb_refind}
\end{figure}

Fig.~\ref{fig:Barb_refind} shows our computations of the inversion angle of the asteroid (234)~Barbara using data taken in the Johnson-Cousin $B$, $V$, $R$ and $I$ standard filters. In doing this plot, we are using some $BRI$ data that are still unpublished and were obtained in the past mostly at the CASLEO observatory (San Juan, Argentina). In producing this plot, we chose to plot in the horizontal axis not the wavelength (which would be the direct observable, being based on the known properties of the standard photometric filters), but the value of the refractive index of spinel taken at the effective wavelength of each filter. As indication, the corresponding wavelengths are indicated in the upper horizontal axis. The value of the inversion angle is shown on the vertical axis. One can notice that, in spite of all uncertainties, there is a clear trend of increasing inversion angle with increasing refractive index. The error bars for the refractive index in Fig.~\ref{fig:Barb_refind} have been estimated based on the FWHM of each filters and the wavelength dependence of the refractive index. The errors are bigger at shorter wavelength since the refractive index is varying quickly in this region while it is almost constant for the I filter. One should take into account also that the sensitivity of most detectors decreases quickly in the blue spectral region, justifying the larger error bars for the inversion angle derived with the B filter. A similar variation of the inversion angle with respect to the wavelength was also found for the other Barbarian (599)~Luisa. One spectropolarimetric measurement was acquired by \citet{Bag_2015} at high phase angle ($26.9^{\circ}$). This measurement shows VRI polarization of respectively $-0.39$, $-0.30$, and $-0.16\%$. This confirms the strong correlation of inversion angle with decreasing wavelength. However, in the case of Luisa, this variation is smaller. Based on our phase-polarization curve, Barbara would have VRI polarization of respectively $-0.41$, $-0.18$, and $0.15\%$ at phase angle equal to $26.9^{\circ}$. One should note that the CAI abundance derived in this work for Luisa is mush lower than the one found for Barbara ($8$ and $20$\% respectively). This could be an explanation for the differences we observe between these two asteroids in spectropolarimetry.

The variations of the polarimetric inversion angle as a function of wavelength and of the derived abundance of CAIs suggest that the large inversion angles of Barbarian asteroids can be a consequence of a higher-than-normal and wavelength-dependent refraction index of the surface regolith, to be possibly interpreted as due to the presence of a high abundance of spinel-bearing minerals, fluffy A-type CAI being our preferred candidates to explain the available observational evidence.

\subsubsection{Space weathering}

Space weathering certainly affects the spectroscopic properties of the objects, but it is expected to affect also some polarimetric properties. The most direct effect of space weathering (in the case of S-type asteroids) is a darkening of the surface and it is known that polarimetry is highly sensitive to the albedo.

\begin{figure}
\includegraphics[width=8cm]{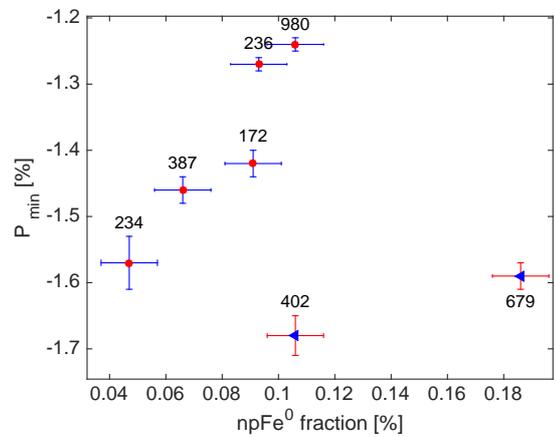}
\caption{Abundance of ${\rm npFe^0}$ (mass fraction) vs. $P_{\rm min}$}
\label{fig:SW_Pmin}
\end{figure}

Fig.~\ref{fig:SW_Pmin} shows an apparent relation between the derived amount of nano-phase iron particles needed to fit the reflectance spectra, and the extreme value of negative polarization, $P_{\rm min}$. This effect could possibly be interpreted as due to an increase of the imaginary part of the refractive index, according to \citet{Zub_2015}. Since  ${\rm npFe^0}$  have a high imaginary refractive index, this interpretation is consistent with our results. As it was already noted for the relation between the inversion angle and the abundance of CAIs, the asteroids (402) and (679) seems (blue triangles) to follow a different behaviour. 

\subsection{Geometric albedos}
\label{ssec:Alb_Det}

\begin{figure}
\includegraphics[width=8.8cm]{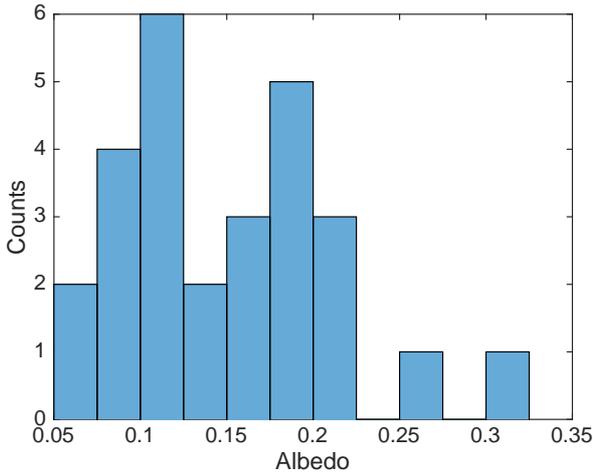}
\caption{Histogram of the albedo of L type asteroids studied in this work.} 
\label{fig:LType_Alb}
\end{figure}

Geometric albedos have been obtained for a large number of asteroids using the thermal radiometry technique applied to thermal IR data obtained by the WISE \citep{Mas_2011} and AKARI \citep{Usui_2012} satellites. The results tend to suggest that (DM) L-class objects, including also the new ones identified in this paper, have an albedo which appears to be bimodal. The distribution is peaking around $0.11$ and $0.18$ as seen in Fig.~\ref{fig:LType_Alb}. 

\begin{figure}
\includegraphics[width=8.8cm]{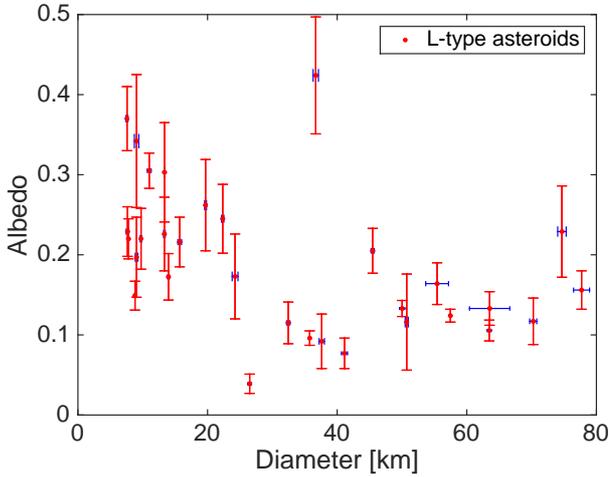}
\caption{Albedo of (DM) L-type asteroids with respect to their derived IRAS diameter.} 
\label{fig:ALB_VS_DIAM}
\end{figure}

One should note that small L-type asteroids tend to have a higher albedo than bigger ones. Fig.~\ref{fig:ALB_VS_DIAM} represents the albedo of L-type asteroids as a function of derived diameter by the NEOWISE survey. Actually, all asteroids with a size below $20$ km have an albedo higher than $0.15$ while asteroids bigger than this treshold tend to have albedo lower than $0.15$. This property is expected if space weathering act as a surface darkening process. 
Since the collisional lifetime decreases with the size of an asteroid \citep{Far_1992,Bin_2004}, smaller asteroids are expected to have a younger surface than larger ones. This hypothesis was strengthened by the observation of a size dependence of the spectral slope of S-type asteroids \citep{Gaf_1993,Car_2016}. \citet{Bin_2004} also observed a correlation between the size and the spectral slope of near earth asteroids. Since the space weathering is also expected to increase the slope of asteroid spectra, a relation between size and slope, and albedo should also be expected.

Families supposed to be populated by (SMASS) L-type asteroids, such as the Henan (2085) and Tirela (1400) families, possess high albedos (respectively $0.22 \pm 0.08$ and $0.28 \pm 0.11$). However, only three of these asteroids were considered in the histogram presented in Fig.~\ref{fig:LType_Alb}. It is then possible that more (DM) L-type high albedo asteroids will be identified in the near future. 

\begin{figure}
\includegraphics[width=8cm]{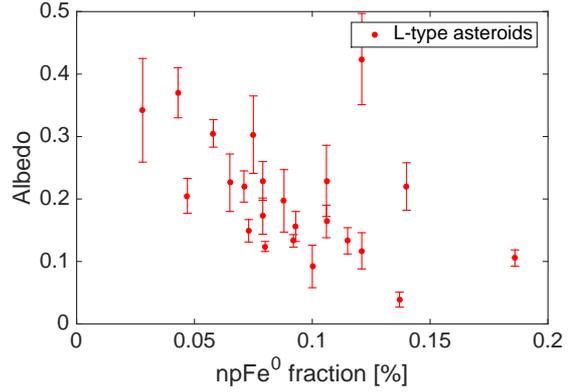}
\caption{Abundance of ${\rm npFe^0}$ (mass fraction) vs. the NEOWISE albedo for L-type asteroids.}
\label{fig:Alb_SpaceW}
\end{figure}

The darkening property of space-weathering can be seen in our modelling. Fig.~\ref{fig:Alb_SpaceW} shows the derived fraction of nanophase iron particle derived by our modelling procedure as a function of the NEOWISE albedo. This figure strongly suggests that the albedo decreases with the space-weathering in the case of L-type asteroids. However, according to \citet{Cel_2010} this property does not seem apply in the case of S-type asteroids.

\subsection{Asteroid families}
\label{ssec:Ast_Fam}
Asteroid families are groups of asteroids that share very similar orbital proper elements, and are interpreted as swarms of fragments issued from the disruption of single parent bodies.

The asteroids in our sample include some objects belonging to different dynamical families. \citet{Cel_2014} identified for the first time the existence of an asteroid family composed of Barbarian asteroids, namely the Watsonia family. Two other families are suspected to include Barbarian asteroids, namely the families of Henan and Tirela/Klumpkea. 
Most members of these families are too faint to be observed by means of ToPol. However, we present here some polarimetric measurements and a few NIR spectra of candidate family members.



\subsubsection{The Watsonia family} 
\label{sssec:Watsonia}
Asteroid (729)~Watsonia was identified as the largest member of a dynamical family by \citet{Nov_2011}. This is a high inclination family (proper inclination $\sim 30^{\circ}$) located at an heliocentric distance of $2.76$ AU. The parent member was found to be a Barbarian by \citet{Gil_2014}. By means of VLT polarimetric observations of a sample of Watsonia family members, \citet{Cel_2014} discovered the first known case of a family consisting of Barbarian asteroids.

\begin{figure}
\includegraphics[width=8.8cm]{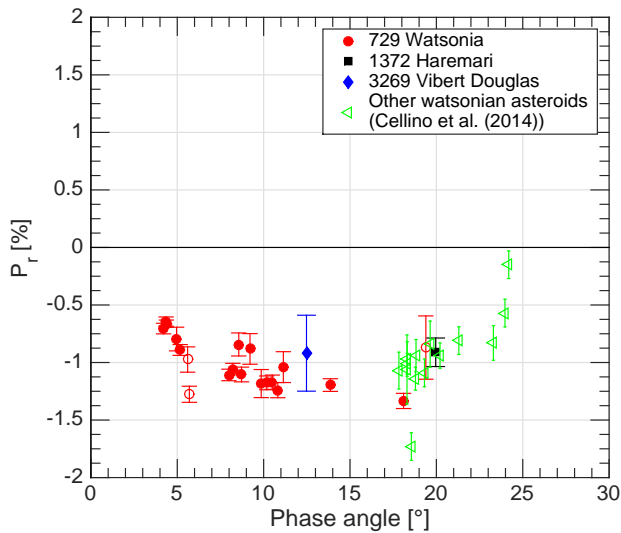}
\caption{Polarimetric data of asteroids belonging to the Watsonia family}
\label{fig:Watsonian_Pola}
\end{figure}

The polarimetric measurements presented in this work of the parent body (729)~Watsonia are shown in Fig.\ref{fig:Watsonian_Pola}, together with our measurement of the other largest asteroid in this family, (1372)~Haremari, and (3269)~Vibert-Douglas. Our data confirm that they seem to share the same polarimetric properties of Watsonia (see Fig.\ref{fig:Watsonian_Pola}). However, in the case of (3269) Vibert-Douglas the only one available measurement is insufficient to confirm a Barbarian behaviour.

\begin{figure}
\includegraphics[width=8.8cm]{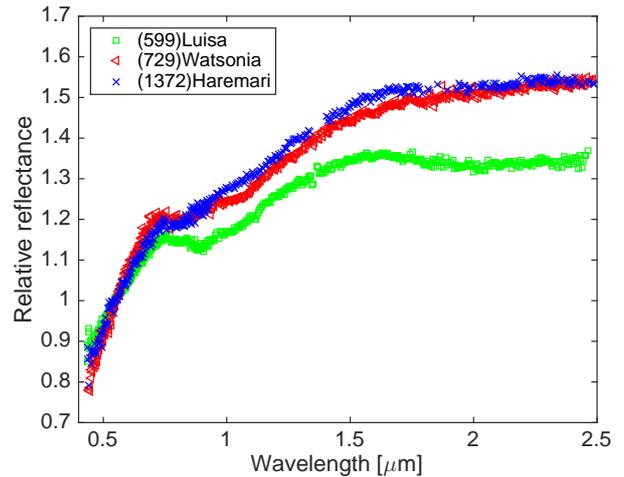}
\caption{Spectra of (599)~Luisa, (729)~Watsonia, and (1372)~Haremari
  normalized to one at 0.55~${\rm \mu}$m.}
\label{fig:Watsonian_Spec}
\end{figure}

The asteroid (599)~Luisa can also be considered as a member of the Watsonia family \citep{Cel_2015a}. A more ancient super-family including also the nearby Barbarians (387)~Aquitania and (980)~Anacostia is also suspected to exist \citep{Cel_2014}. Our polarimetric measurement of Luisa has been taken at a phase angle too low to conclude that Luisa is another confirmed Barbarian. However, \citet{Bag_2015} presented one spectropolarimetric measurement at high phase angle of this asteroid confirming its Barbarian nature. They exhibit similar spectra, (see Fig. \ref{fig:Watsonian_Spec}), with relatively low CAI abundances ($0\%$ for (729) and (1372), and $8\%$ for (599)). These results strengthen the hypothesis that these bodies are genetically related. 


Figs. \ref{fig:Watsonian_Pola} and \ref{fig:Watsonian_Spec} show the available polarimetric and spectroscopic data, respectively, for asteroids belonging to the Watsonia family.


\subsubsection{The Henan family} 
\label{sssec:Henan}
 \citet{Bro_2013} identified a family of 946 asteroids with (2085)~Henan as the largest member, but \citet{Mas_2013} did not classify this as a family because it is too dispersed and probably contaminated by many interlopers. 
This controversial family was the first one found to include some (SMASS) L-class asteroids by \citet{Bus_1999}.

As discussed above, we have only one polarimetric measurement of (2085)~Henan and one measurement of the second largest member, (2354)~Lavrov, the latter having been observed at a phase angle too small to draw any conclusion about its behaviour. The only one measurement of Henan, however, indicates a very high probability of being a Barbarian.

\begin{figure}
\includegraphics[width=8.8cm]{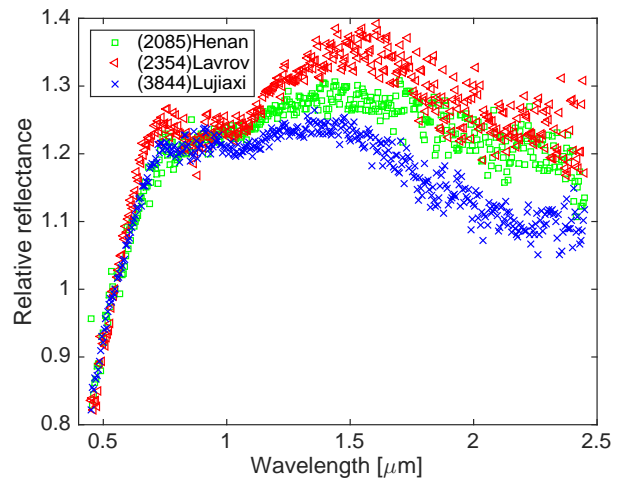}
\caption{Spectra of (2085)~Henan, (2354)~Lavrov, and (3844)~Lujiaxi normalized at
0.55~${\rm \mu}$m.} 
\label{fig:Henan_Spec}
\end{figure}

A few NIR spectra of Henan candidate members exist in the literature. Most of them are (DM) L-class and look similar, strengthening the possibility of a common origin. According to our modelling attempts, these objects are characterized by value of CAI ranging from 10 to 28 \%, and display moderate value of space weathering.  Fig.~\ref{fig:Henan_Spec} shows the spectra of (2085)~Henan, (2354)~Lavrov, and (3844)~Lujiaxi. A spectrum of (1858)~Lobachevski is also available, however, this asteroid was classified as an S-type and possesses an albedo of 0.37. This make it probably an interloper inside the Henan family. 

Some differences are noticed between the Henan and the Watsonia family. Henan family spectra exhibits a negative slope in the near-infrared region whereas members of the Watsonia family display positive slope. However, the visible spectral region shows identical behaviour.

\subsubsection{The Tirela/Klumpkea family} 
\label{sssec:Tirela}
The Tirela family was first identified by \citet{Nes_2005}. It is located at the edge of the outer belt  (proper semi-major
axis $a_p$ = 3.12 AU) and possesses high eccentricity and inclination ($e_p$ = 0.20 and $i_p = 16.8^{\circ}$). This family is characterized by high geometric albedo ($0.2-0.3$), whereas nearby asteroids in the same region have generally low albedo. This family was found to include (SMASS) L/Ld-class members by \citet{Mot_2008}. \citet{Mil_2014} found also a family in this region, but they assigned a different membership and called it the Klumpkea family.


We observed (1040)~Klumpkea in polarimetry, but only at low phase angles which cannot provide a diagnostic of Barbarian properties. In spectroscopy, five Tirela family members (8250), (15552), (19369), (26219), and (67255) were observed, but only in the NIR. However, for some of them, spectro-photometric data in the visible domain are available in the  SDSS database. Their NIR spectra are characterized by strong 2~${\rm \mu}$m absorption band which leads to high abundance of CAIs. Only (19369) is lacking a strong 2~${\rm \mu}$m absorption band. This asteroid also possesses lower albedo than all other Tirela/Klumpkea family asteroid observed in this work. We suspect this asteroid to be an interloper inside this family. For the other ones, the spectral modeling provide CAIs abundance ranging from 13 to 23 \% associated with almost no meteoritic component and high fraction of olivine.



\section{Conclusions and perspectives}
\label{sec:Conc_Persp}

Our comprehensive analysis of the evidence coming from polarimetric and spectroscopic data allows us to draw some robust conclusions as well as some more tentative interpretation attempts based on current observational evidence. 

The most robust result is the proven equivalence between the polarimetric Barbarian behaviour and the taxonomic classification as L-class objects according to the \citet{Dem_2009} taxonomy. This correlation between polarimetric and spectroscopic behaviour had been already suggested in the past, we show in this work a very convincing observational proof of that.

Another important result is that we confirm preliminary conclusions by \citet{Sun_2008a}, and we find that the spectra of (DM) L-class objects can be successfully modelled using primitive materials, including primarily CAIs, MgO-rich olivine and the mineral compounds forming CV3s meteorite. We tried two CV3 meteorites in our Hapke model. We obtained better results when using the CV3 displaying CAIs with more FeO-rich spinels and showed some possible clue of aqueous alteration (Y-89751). We could also rule out the presence of large amounts of pyroxene. Our fits of available reflectance spectra were generally good, both in the NIR and the visible spectral regions. 

An essential feature in our modelling exercises is that the presence of fluffy type A CAIs are needed to obtain acceptable fits of the reflectance spectra. We found evidence of a relation between the relative abundance of CAIs on the surface of these asteroids and the large polarimetric inversion angle which characterizes the Barbarian behaviour. Such a relation seems to be strengthened by the observed variation of the inversion angle of asteroid (234)~Barbara as a function of wavelength. This variation can be interpreted as due to the wavelength-dependent variation of the refractive index of the spinel mineral. 

Other possible explanations of the Barbarian behaviour, however, cannot be ruled out, including the possibility that Barbarians have surface regoliths formed by very thin particles, as suggested by \citet{paper2}. Of course, different possible explanations are not necessarily mutually exclusive. Instead, the high abundance of fluffy type A CAI suggest that Barbarian asteroids could be extremely old and primitive.

The important role played by space weathering processes was also stressed by the results of our investigations.  A tentative relation was found between the estimated abundance of nano phases iron believed to be characteristic outcomes of space weathering, and the extreme value of negative polarization $P_{\rm min}$.

Polarimetric and NIR reflectance spectra of a few members of dynamical families known to include L-class members were also obtained. We could confirm an L-classification for some of these family members. This is the first step of an investigation that deserves to be pursued making use of large telescopes. We plan also to extend our analysis in the future, by setting up laboratory activities, including polarimetric measurements of CAI material found in meteorite samples. These laboratory measurements will allow to definitely understand the polarimetric behaviour of CAIs and be able to provide more robust answers to the enigma represented by Barbarian asteroids. 

\section{Acknowledgements}
\label{sec:Ackn}

The authors wish to thank J. de Leon for her constructive review and remarks which improve the paper.

MD thanks the Li{\`e}ge University for their financial support during his scientific missions in Calern.

The Torino polarimeter was built at the INAF - Torino Astrophysical Observatory 
and funded by INAF in the framework of INAF PRIN 2009.

Part of the polarimetric data in this work have been obtained on the C2PU facility (Calern Observatory, O.C.A.).

Part of this work by MD was supported by the COST Action MP1104 ``Polarization as a tool to study the Solar System and beyond''.

This work is based on data collected with 2-m RCC telescope at Rozhen National Astronomical Observatory. 
The authors gratefully acknowledge observing grant support from the Institute of Astronomy and Rozhen National Astronomical Observatory, Bulgarian Academy of Sciences.

The near-infrared data were acquired by MD and PT as Remote Astronomer at the Infrared Telescope Facility, which is operated by the University of Hawaii under contract NNH14CK55B with the National Aeronautics and Space Administration.

Data were also obtained and made available by the The MIT-UH-IRTF Joint Campaign for NEO Reconnaissance. The IRTF is operated by the University of Hawaii under Cooperative Agreement no. NCC 5-538 with the National Aeronautics and Space Administration, Office of Space Science, Planetary Astronomy Program. The MIT component of this work is supported by NASA grant 09-NEOO009-0001, and by the National Science Foundation under Grants Nos. 0506716 and 0907766.

We also acknowledge the support from the French ''Programme Nationale de Plan\'etologie''. 

GB gratefully acknowledges observing grant support from the Institute of Astronomy and Rozhen National Astronomical Observatory, Bulgarian Academy of Sciences.

JL acknowledge support from the AYA2015-67772-R  (MINECO, Spain).

The asteroid diameters and albedos based on IRAS and NEOWISE observations were obtained from the Planetary Data System (PDS) .

\begin{appendix}

\section{Spectra observed in this work\label{app:spec}}

 \begin{figure*}
\centering
\begin{tabular}{|c|c|}
\hline
\includegraphics[width=7cm]{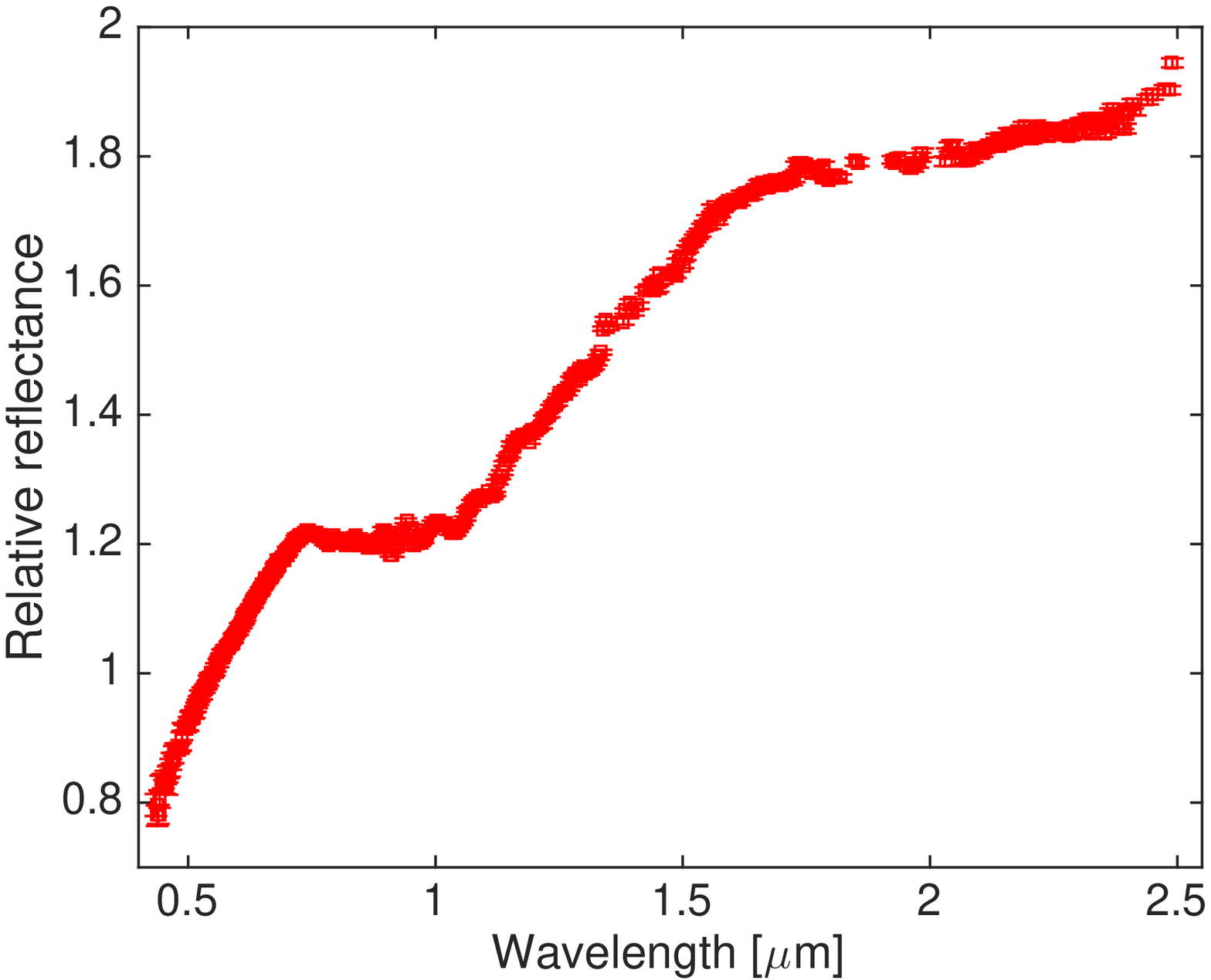}

&
{\includegraphics[width=7cm]{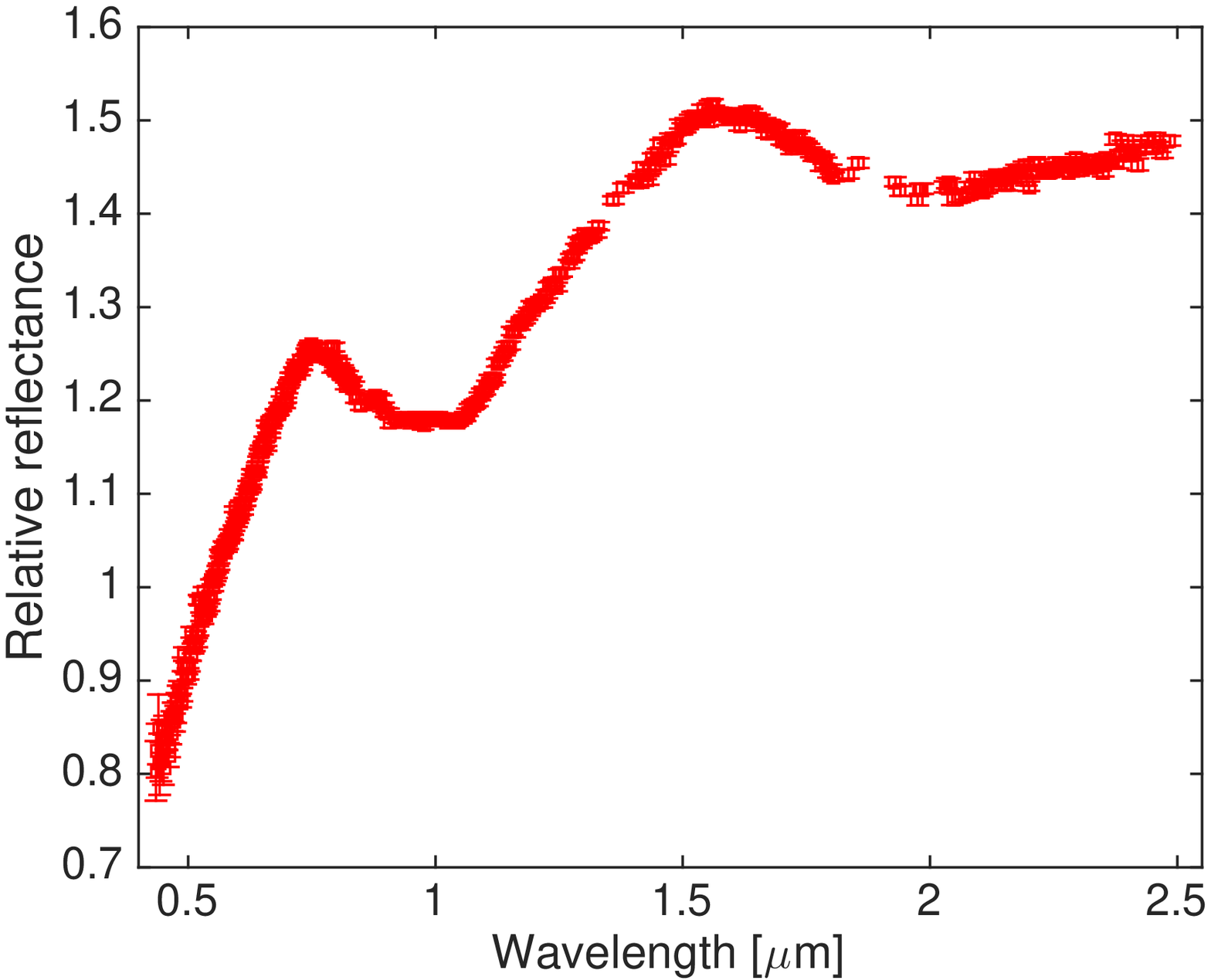}}

\\
     {(12)~Victoria}

&
     {(122)~Gerda}
     \\
     
\hline
{\includegraphics[width=7cm]{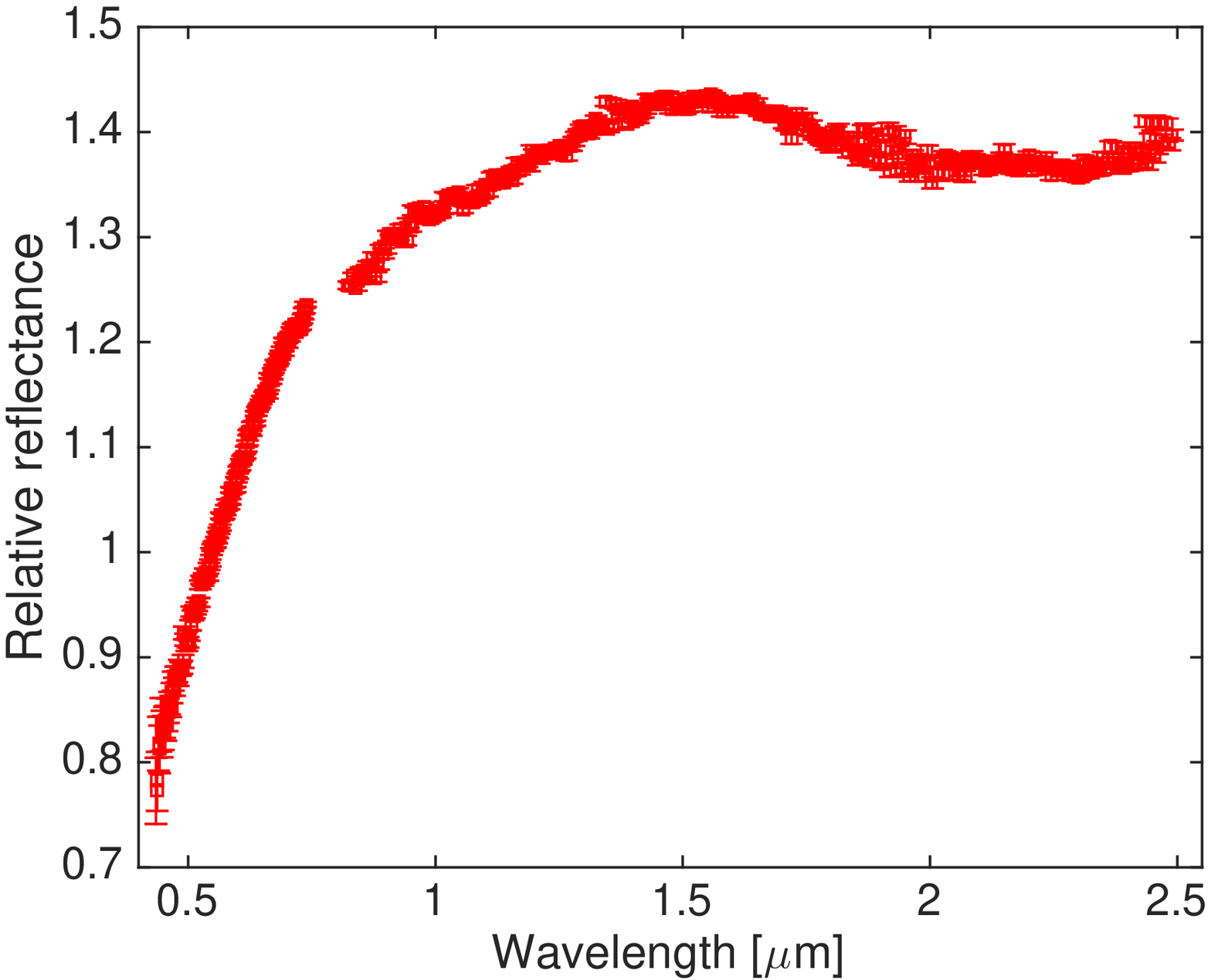}}

&
{\includegraphics[width=7cm]{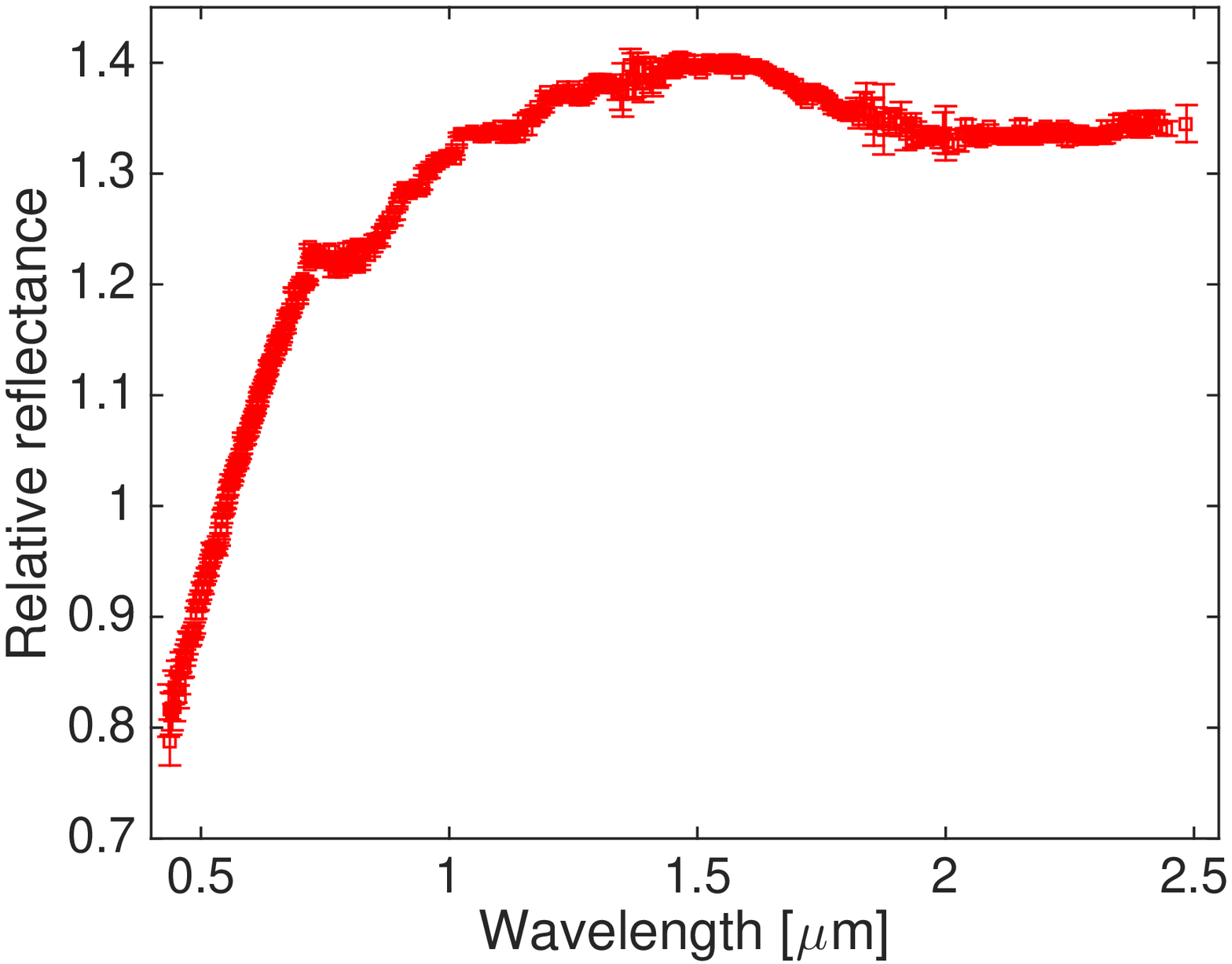}}

\\

     {(172)~Baucis}

&  {(458)~Hercynia}\\
\hline

\\
\hline
{\includegraphics[width=7cm]{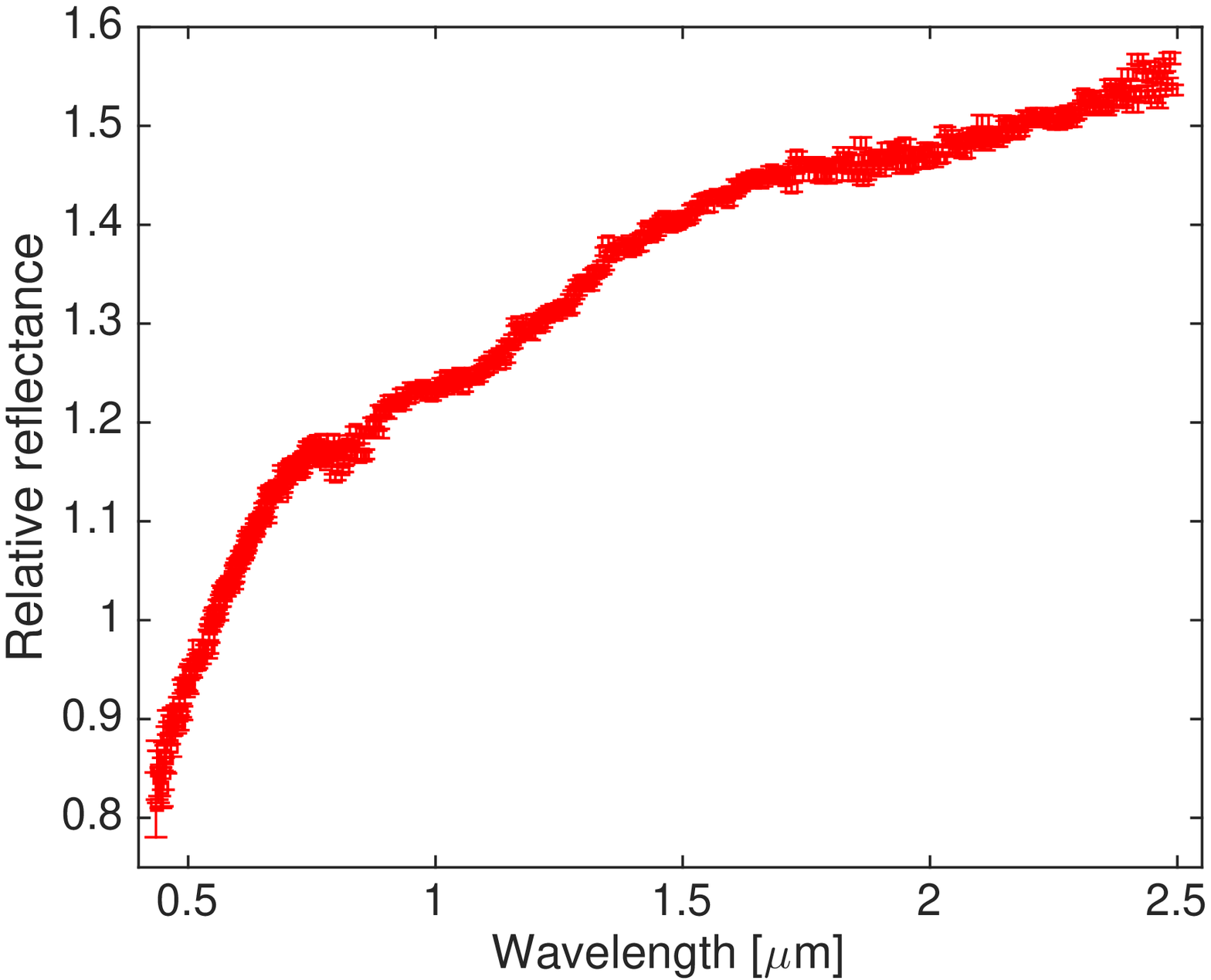}}

&

{\includegraphics[width=7cm]{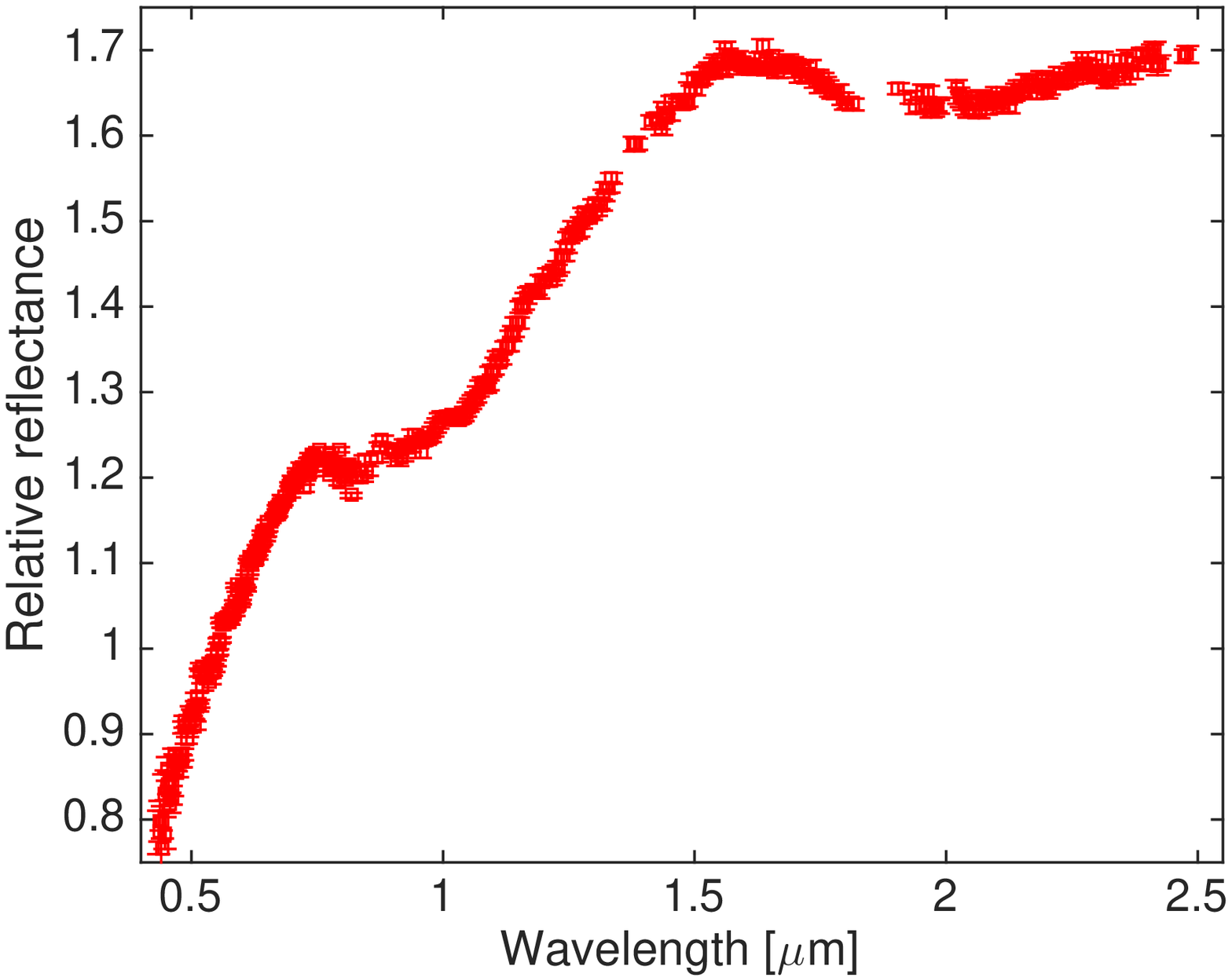}}

\\
     {(611)~Valeria}

& {(753)~Tiflis} \\
\hline
{\includegraphics[width=7cm]{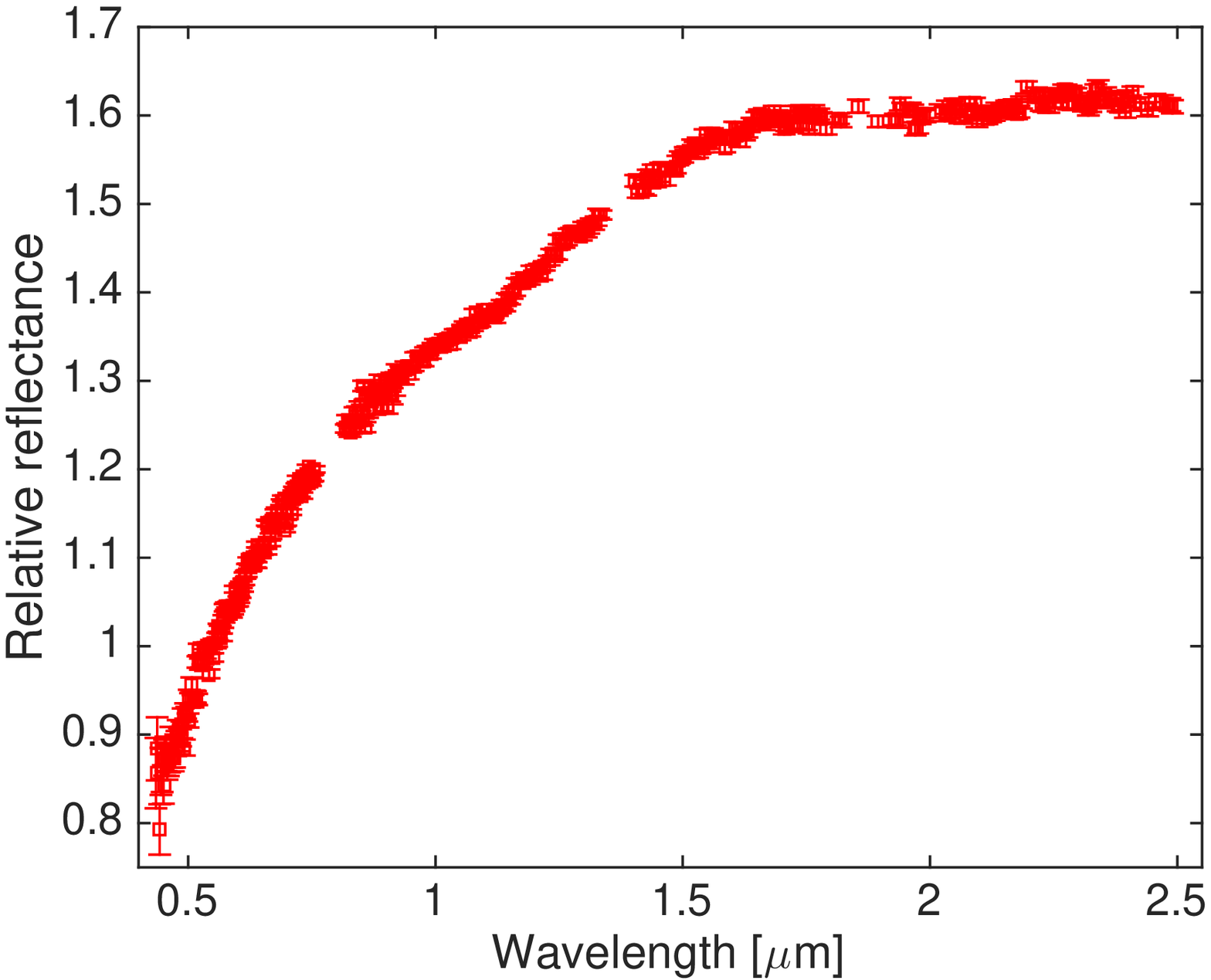}}
&
{\includegraphics[width=7cm]{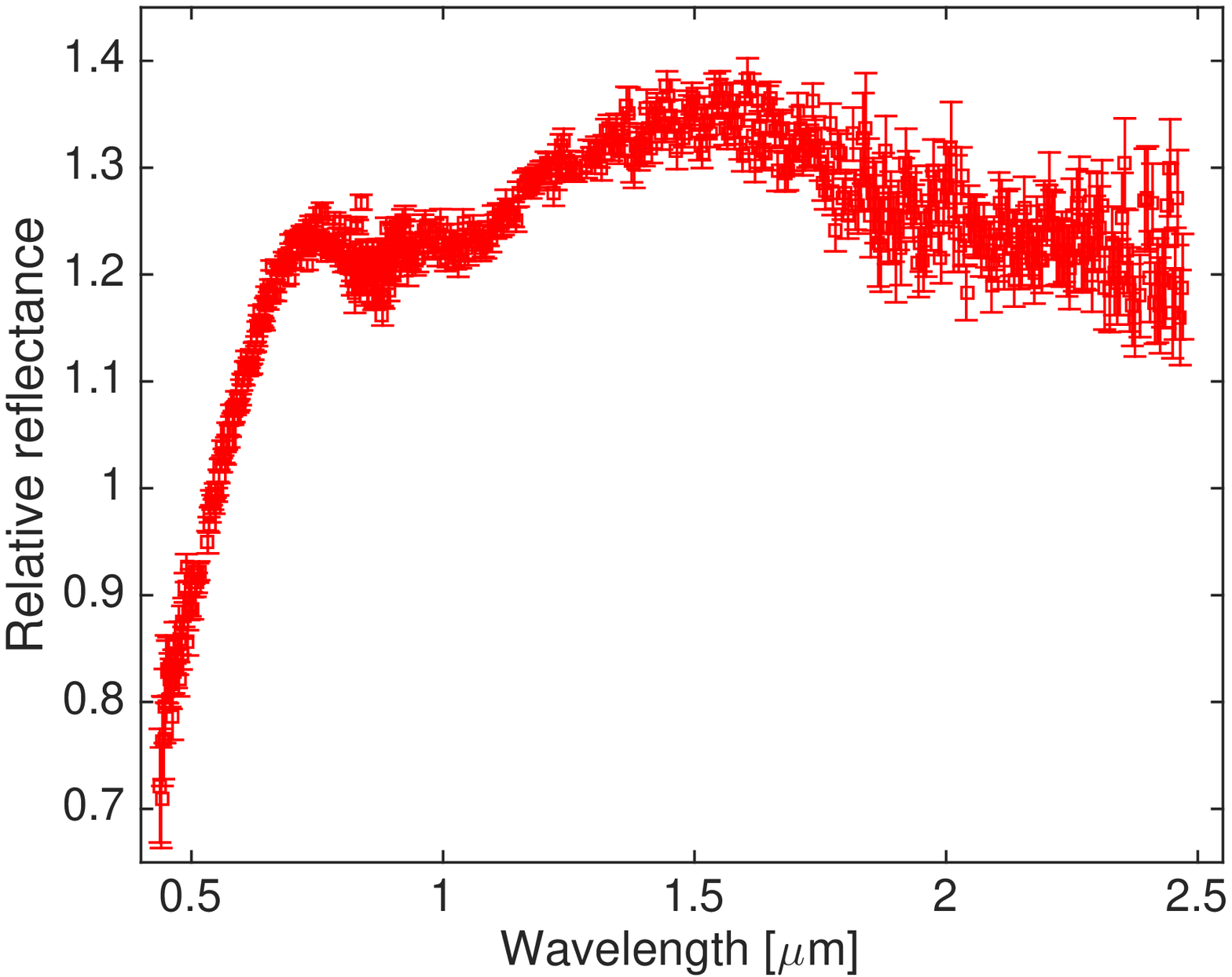}}

\\
     {(1372)~Haremari}

     & {(2354)~Lavrov	} \\
\hline

\end{tabular}
\caption{Spectra obtained during the two IRTF runs.}
\label{Spec1}
\end{figure*}

\begin{figure*}
\addtocounter{figure}{-1}
\centering
\begin{tabular}{|c|c|}
\hline
     {\includegraphics[width=7cm]{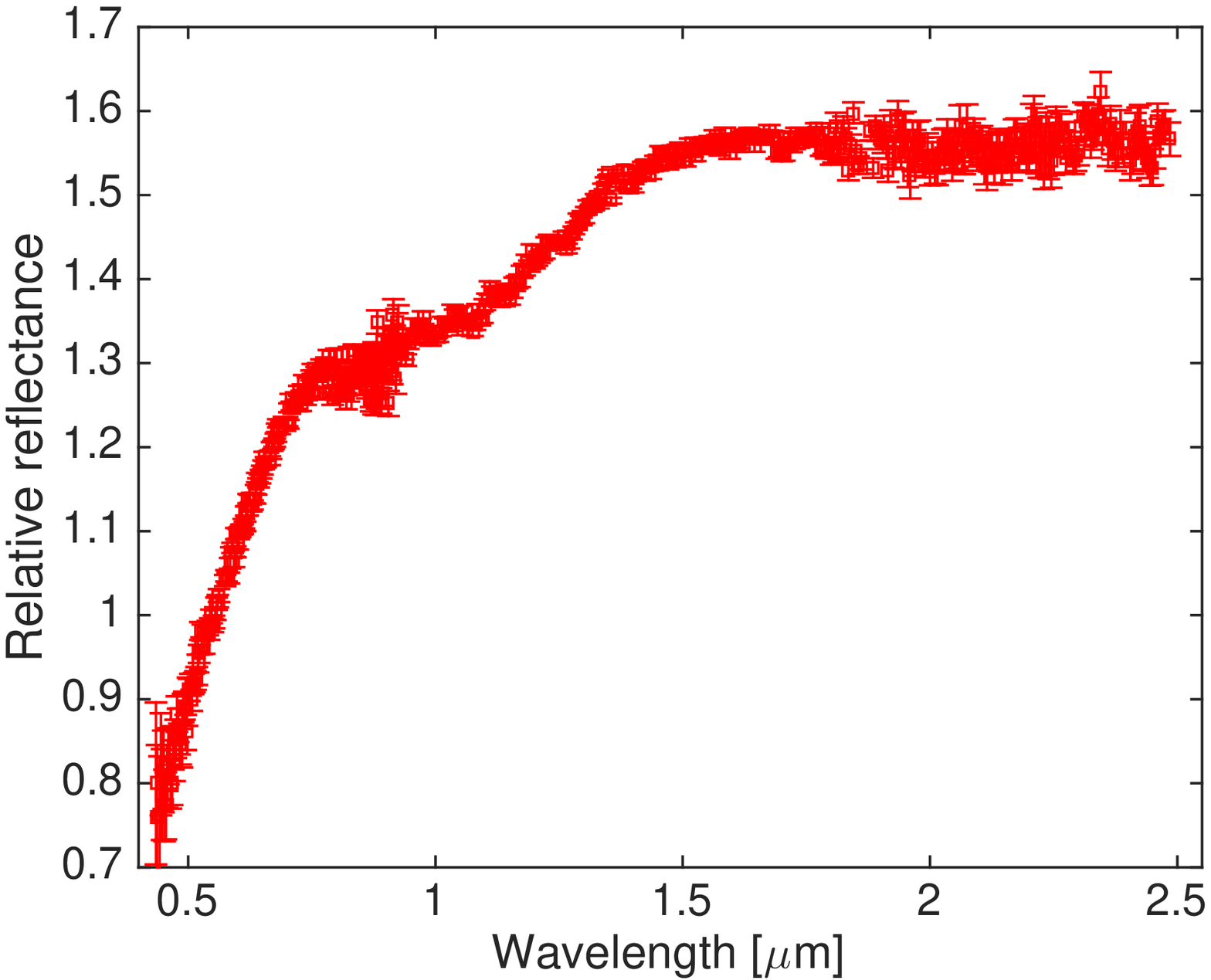}}

     &
{\includegraphics[width=7cm]{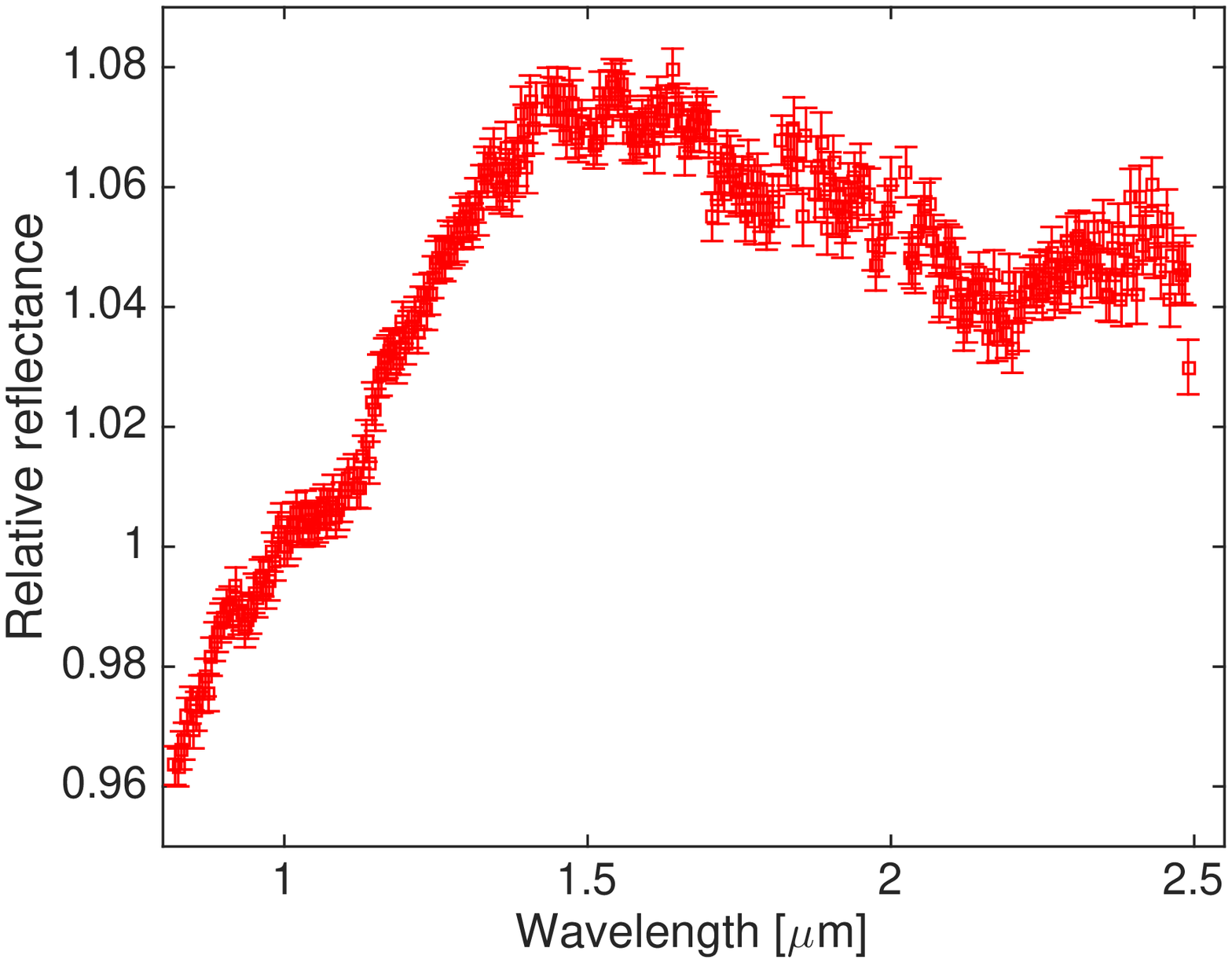}}
\\
     {(4917)Yurilvovia}
     
     &

       {(8250)~Cornell}

\\
\hline
{\includegraphics[width=7cm]{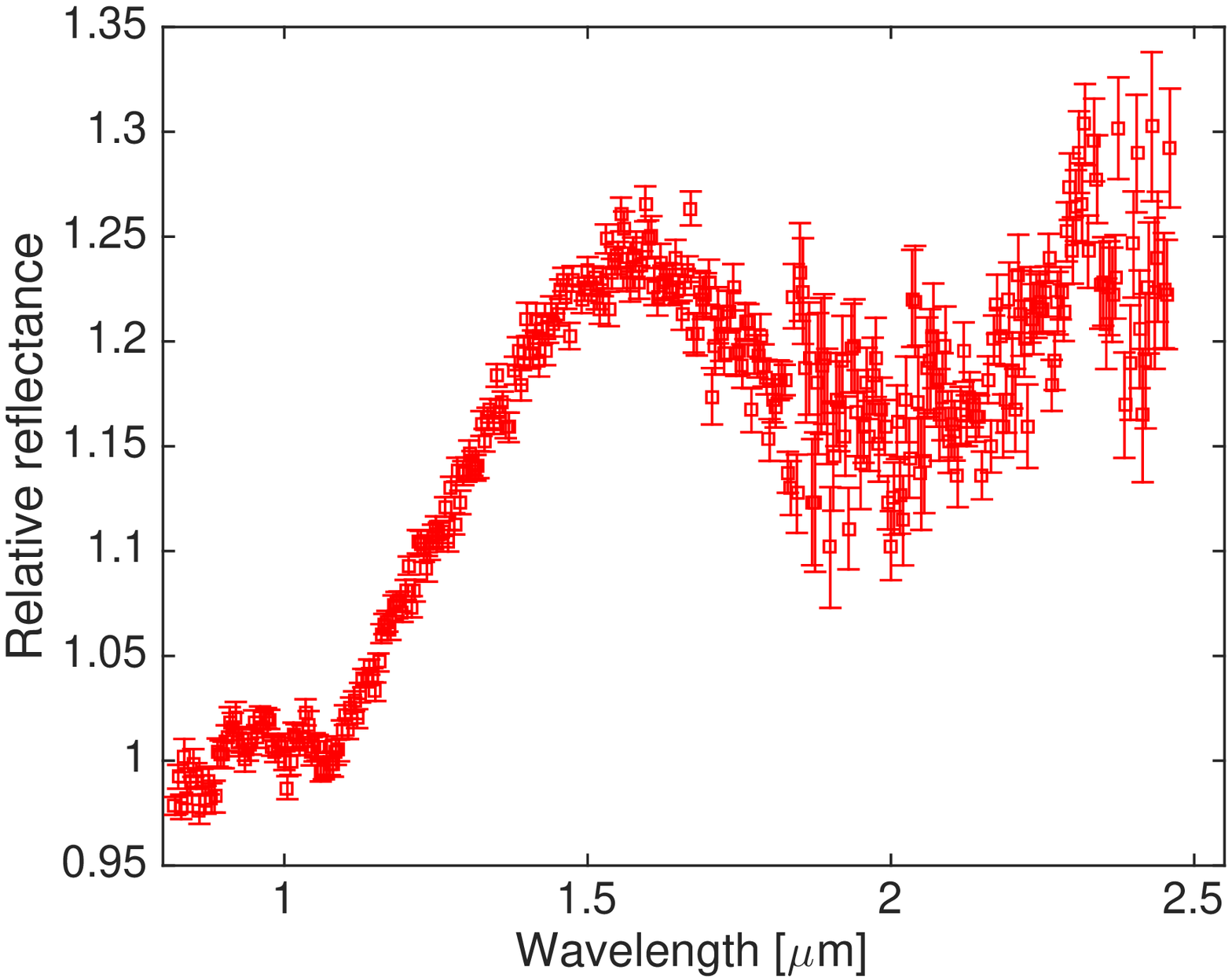}}

     &
     {\includegraphics[width=7cm]{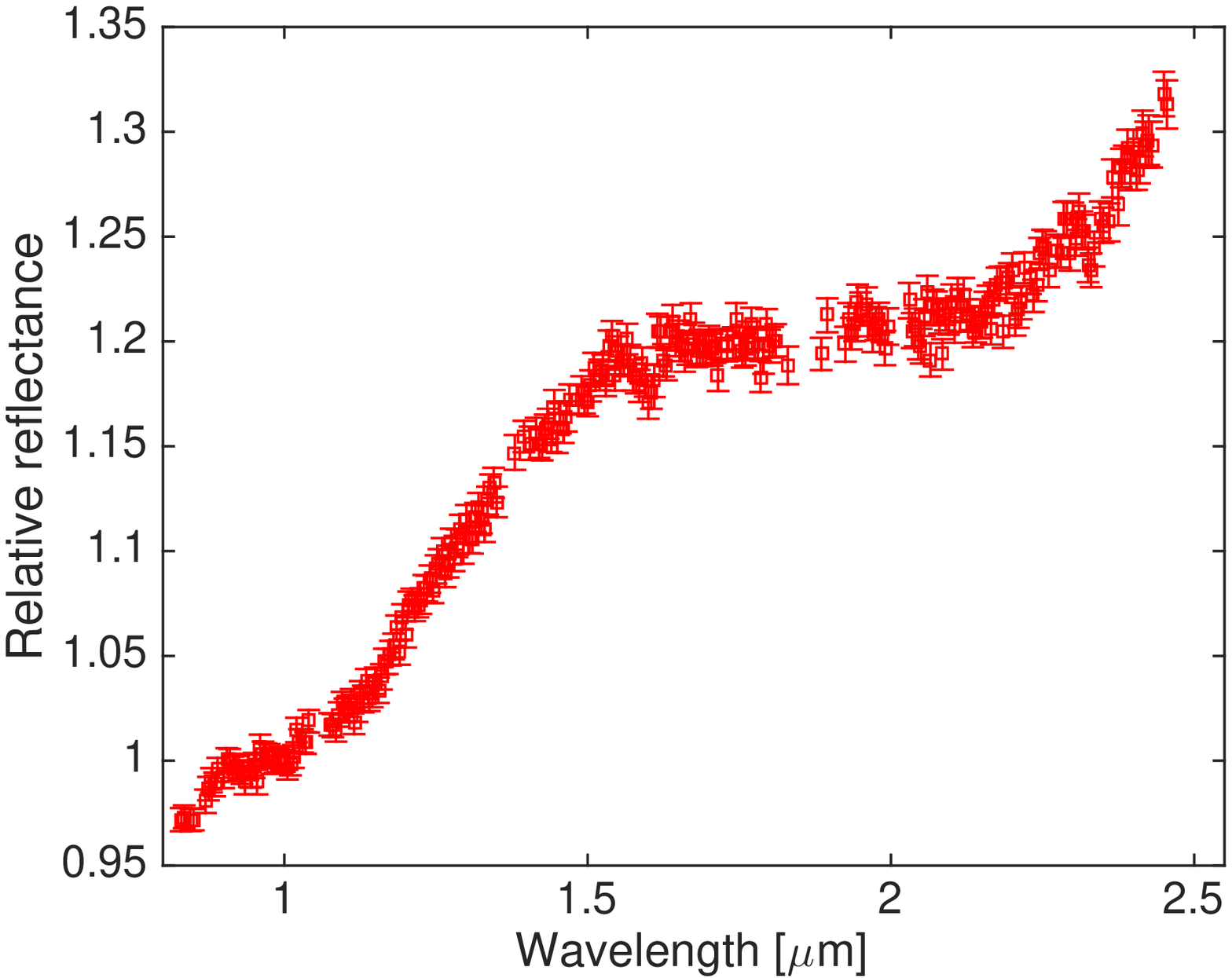}}
\\
     {(15552)~Sandashounkan}          

     &   
     {(19369)~1997 YO}

\\
\hline
     {\includegraphics[width=7cm]{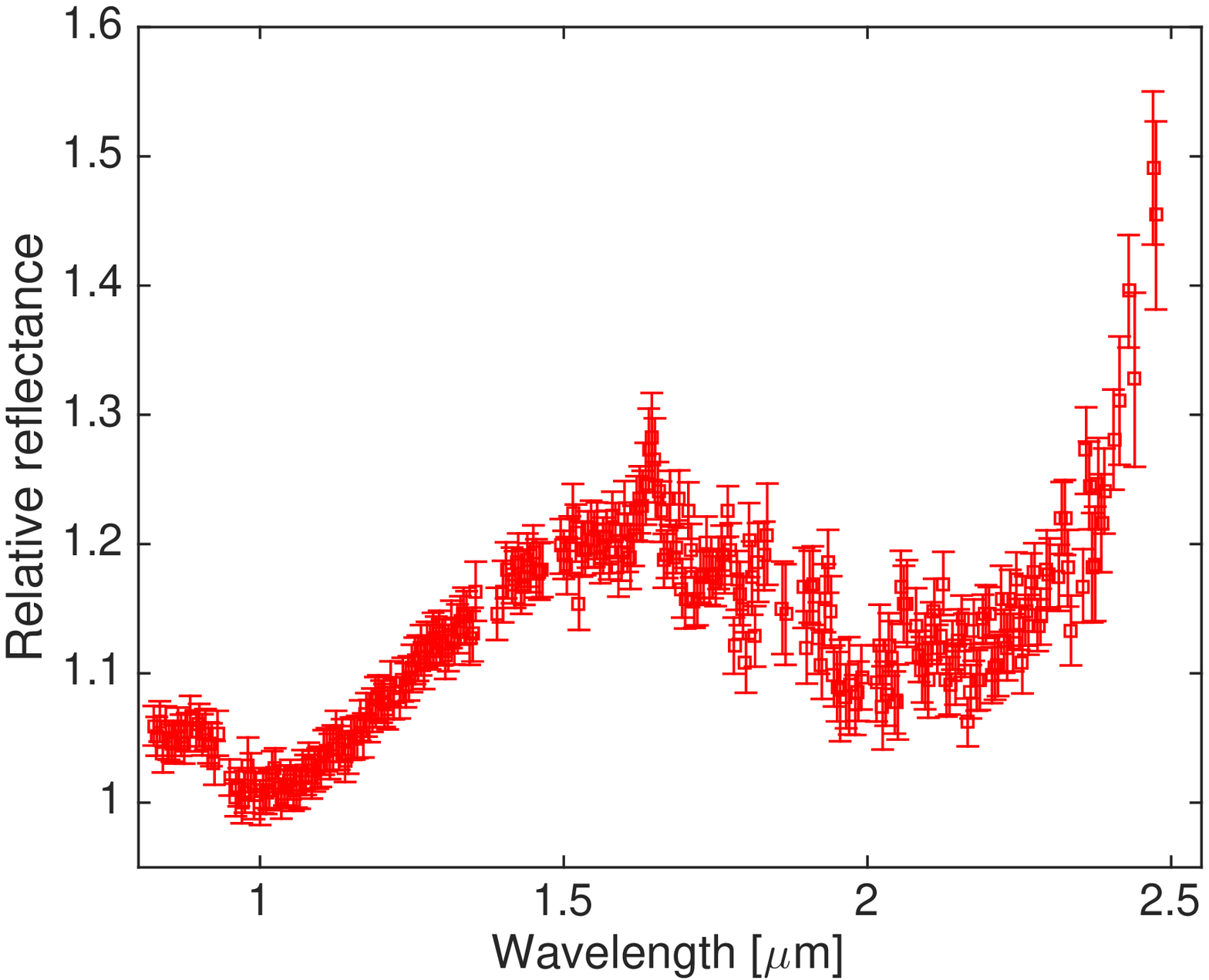}}
&
{\includegraphics[width=7cm]{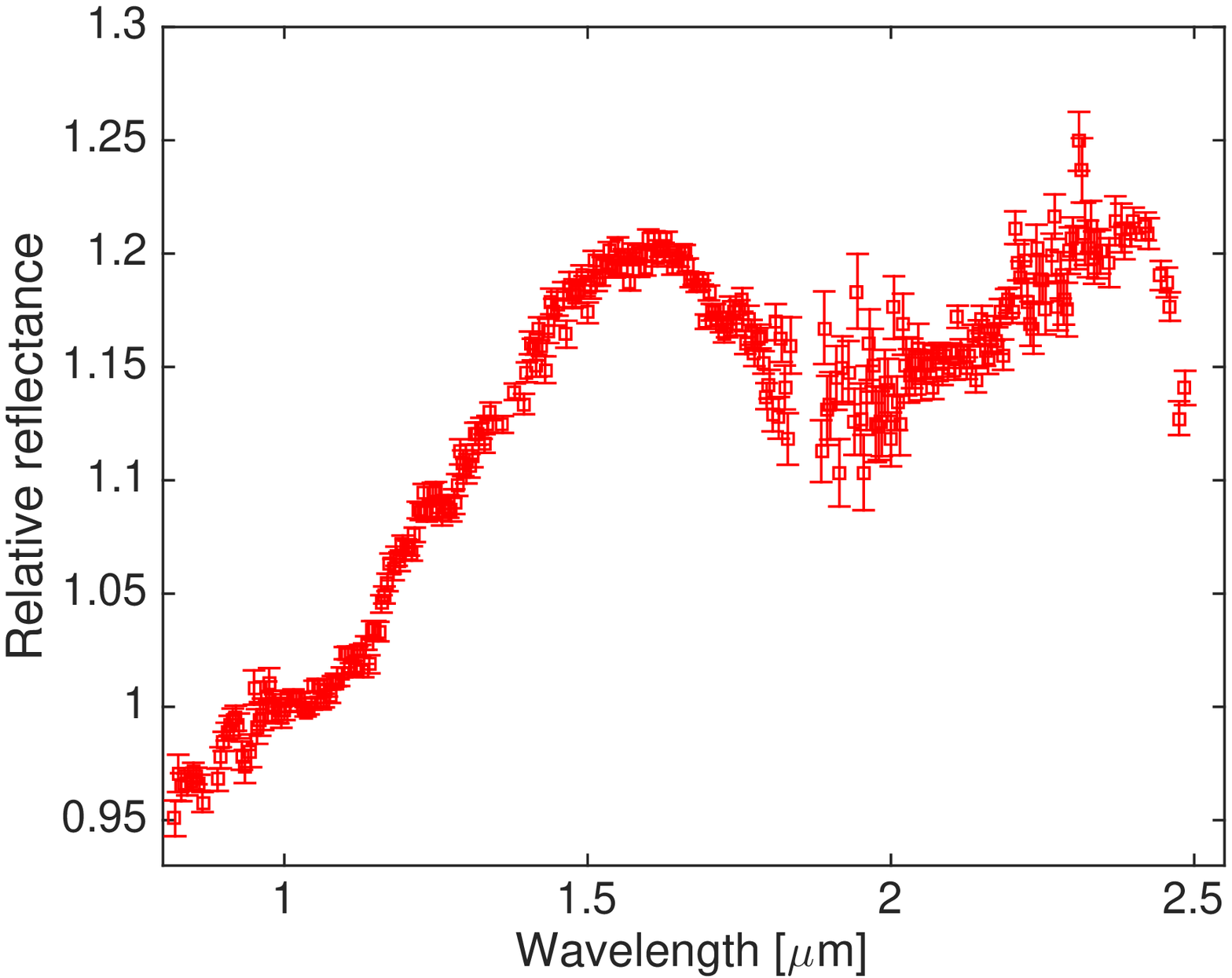}}
   
\\
  {(26219)~1997 WO21}

&
     {(67255)~2000 ET109}
\\
\hline

\end{tabular}
\caption{Spectra obtained during the two IRTF runs.}
\end{figure*}

\section{Results of spectra fitting presented in this work \label{app:fit}}

\begin{figure*}
\centering
\begin{tabular}{|c|c|}
\hline
\includegraphics[width=7cm]{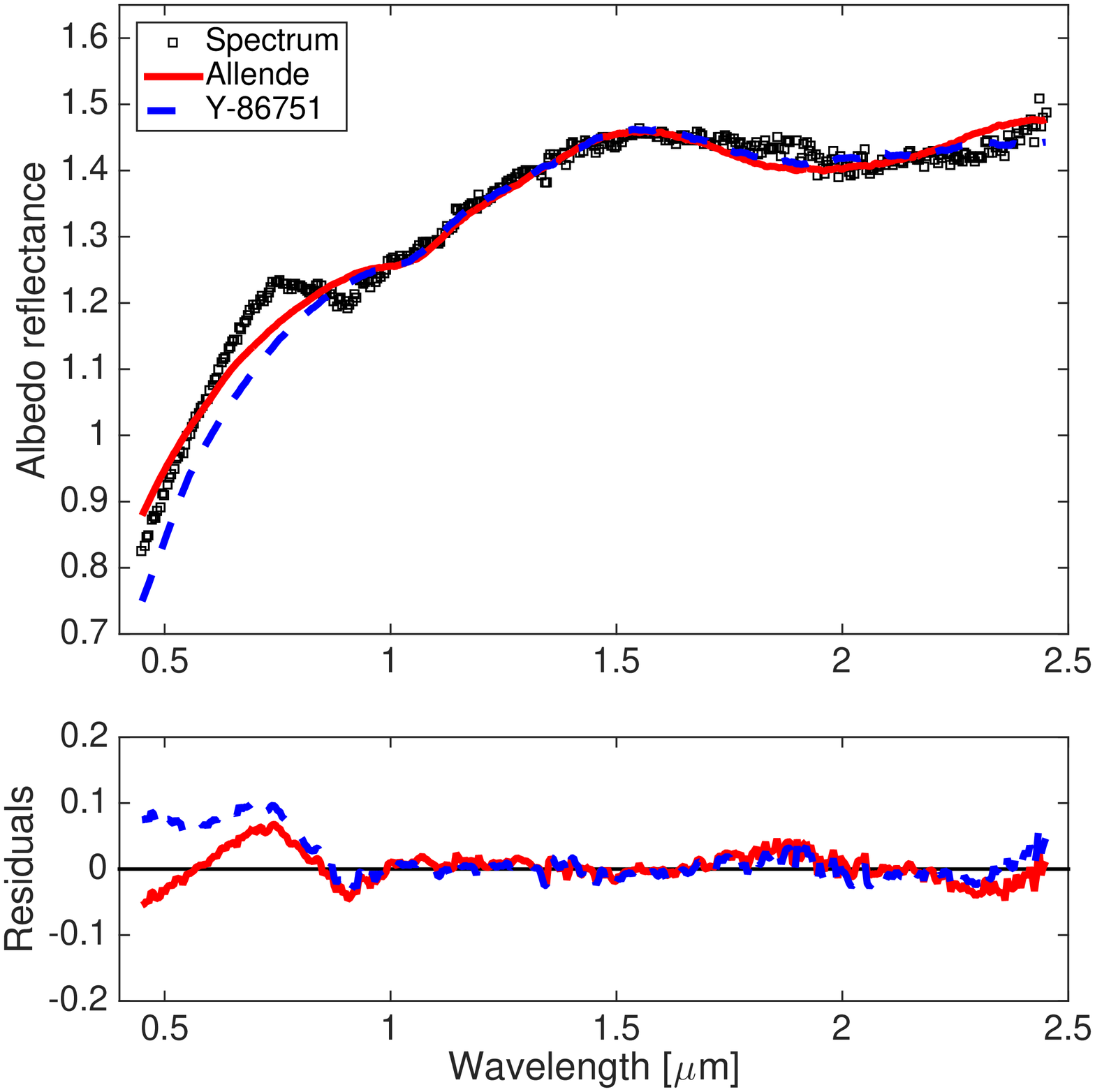}

&
{\includegraphics[width=7cm]{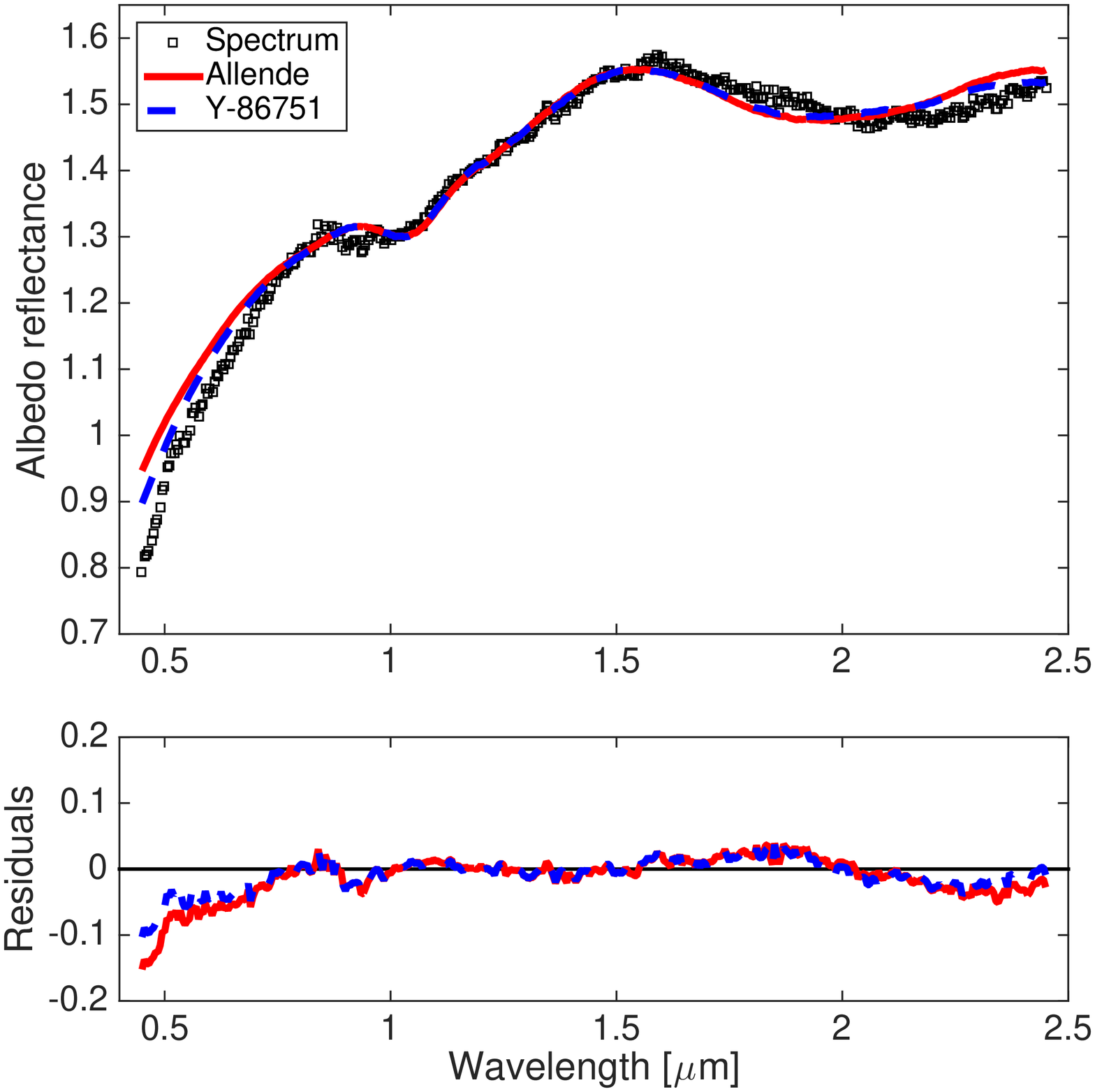}}

\\
     {(172)~Baucis}

&
     {(234)~Barbara}
     \\
     
\hline
{\includegraphics[width=7cm]{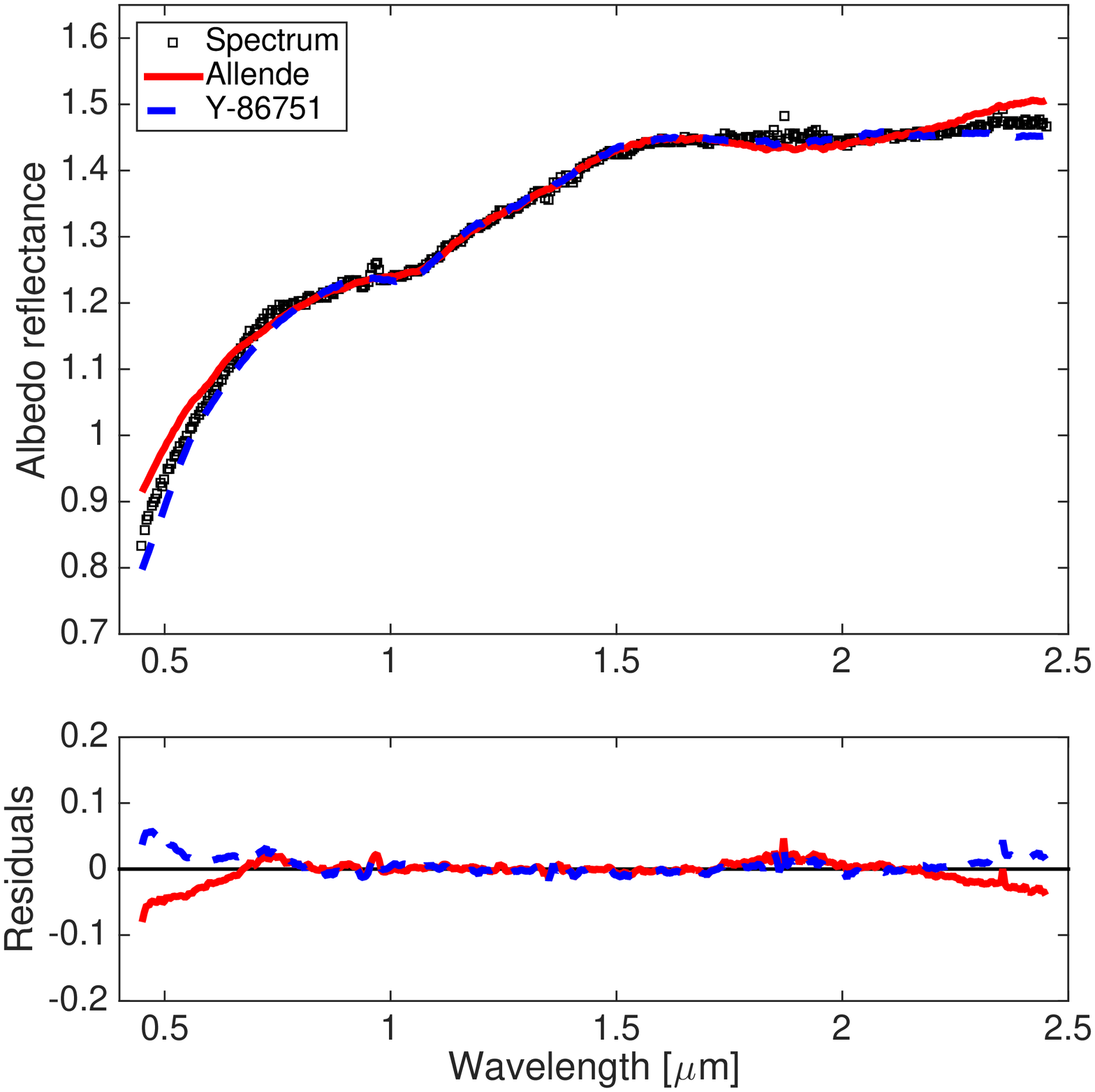}}

&
{\includegraphics[width=7cm]{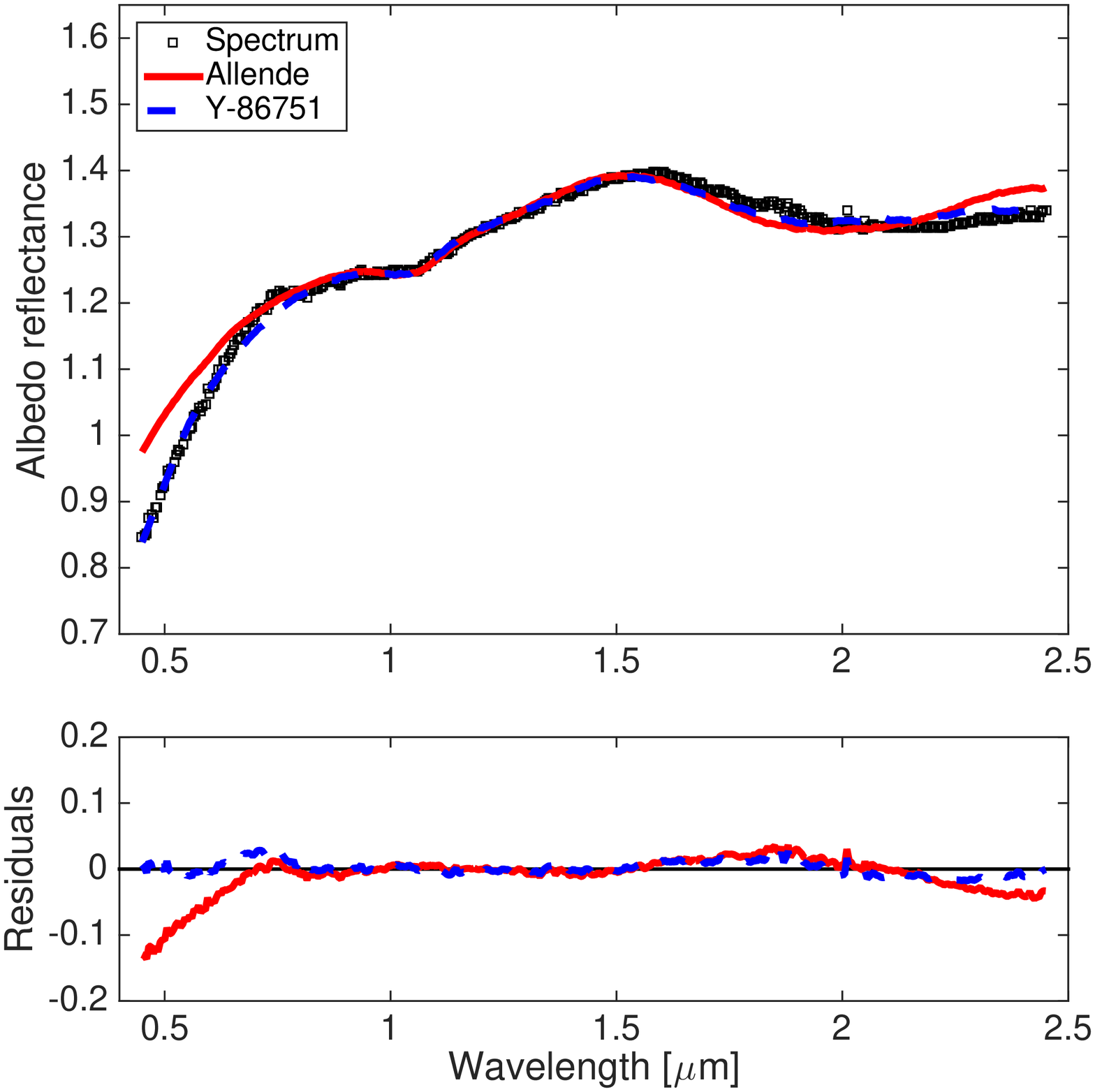}}

\\

     {(236)~Honoria}

&  {(387)~Aquitania}\\
\hline

{\includegraphics[width=7cm]{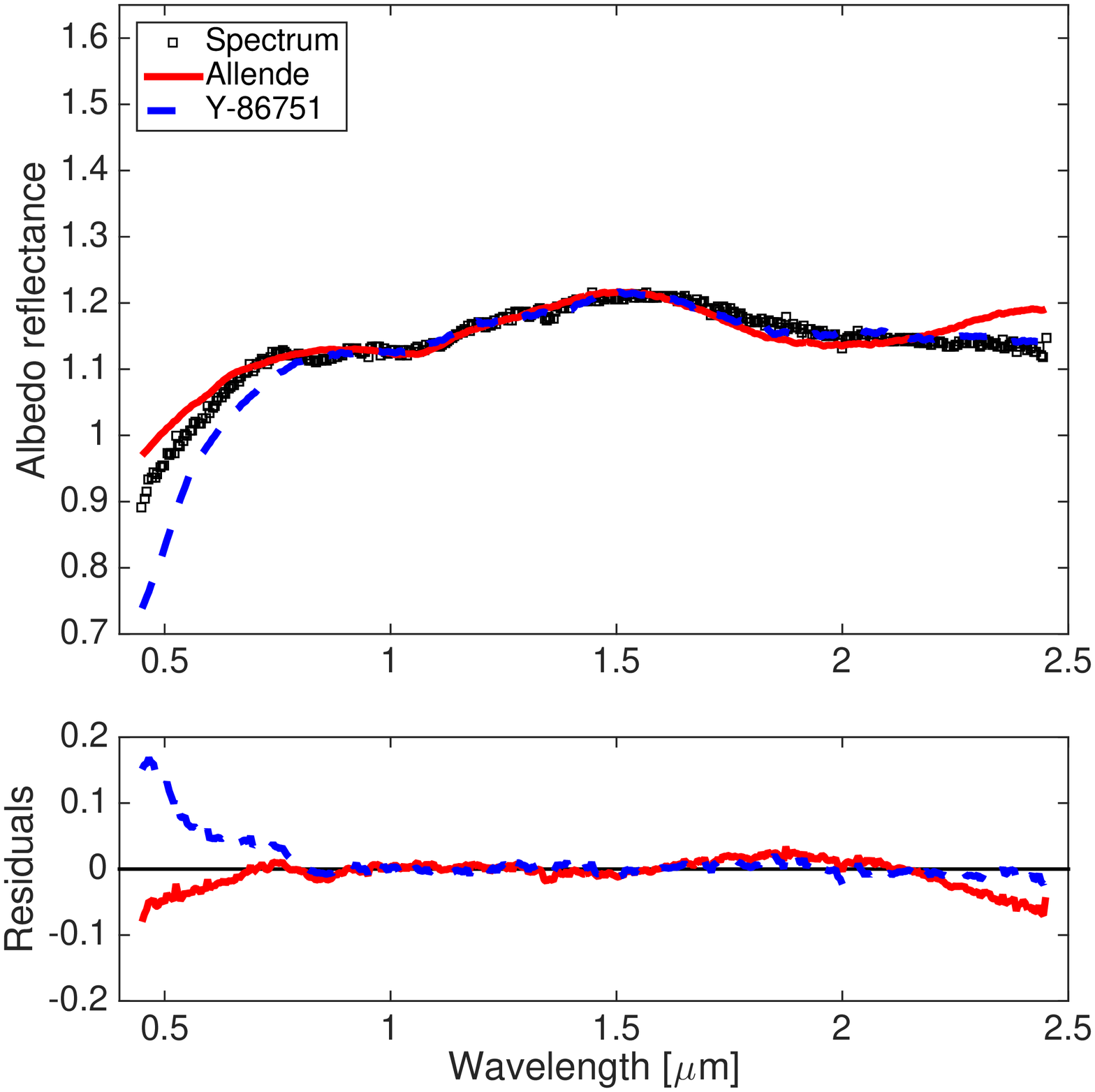}}

&
{\includegraphics[width=7cm]{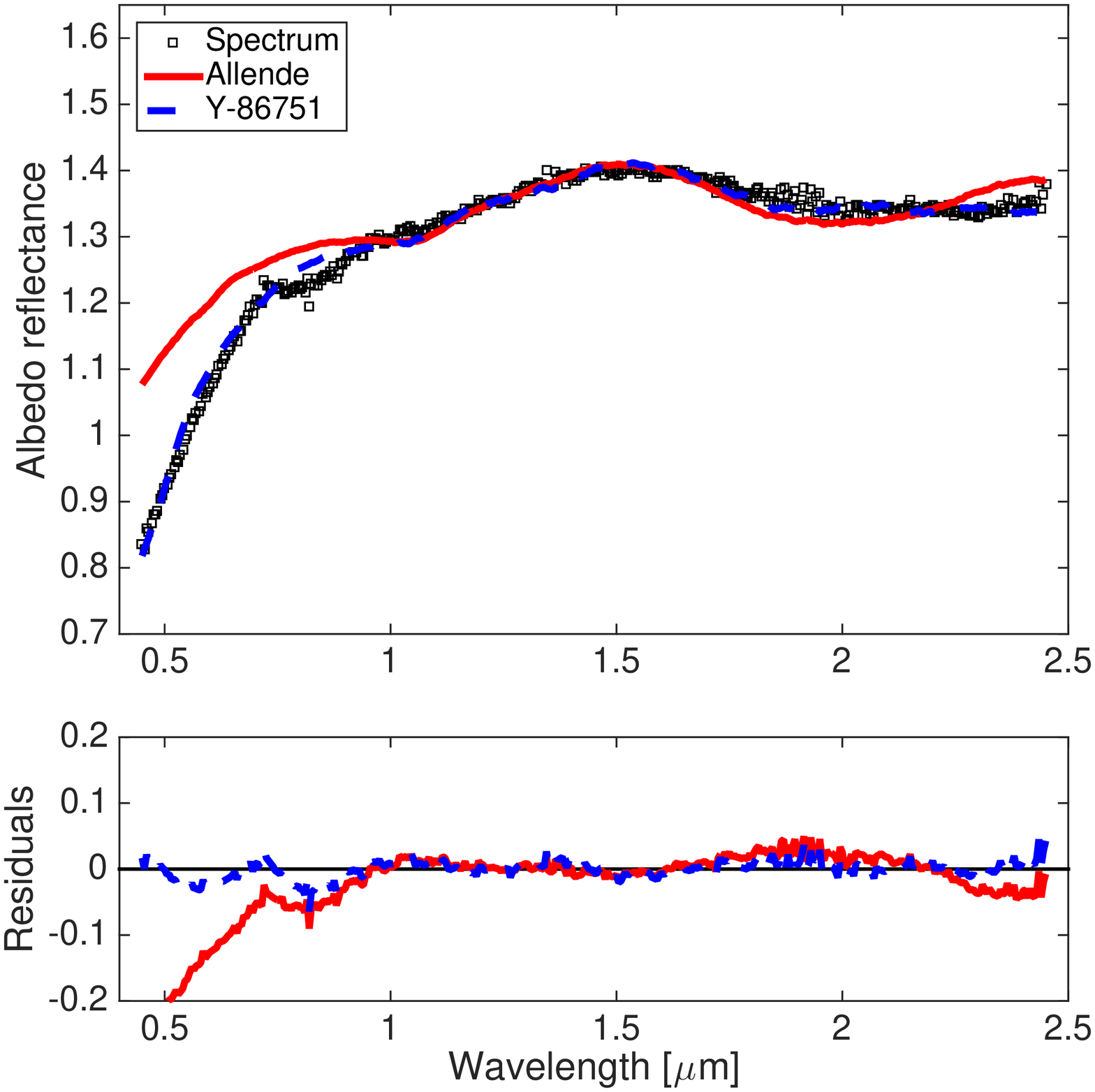}}

\\
     {(402)~Chloe}

& {(458)~Hercynia} \\
\hline
\end{tabular}
\caption{Spectra obtained during the two IRTF runs.}
\label{Res_Fit}
\end{figure*}

\begin{figure*}
\centering
\addtocounter{figure}{-1}
\begin{tabular}{|c|c|}
\hline

{\includegraphics[width=7cm]{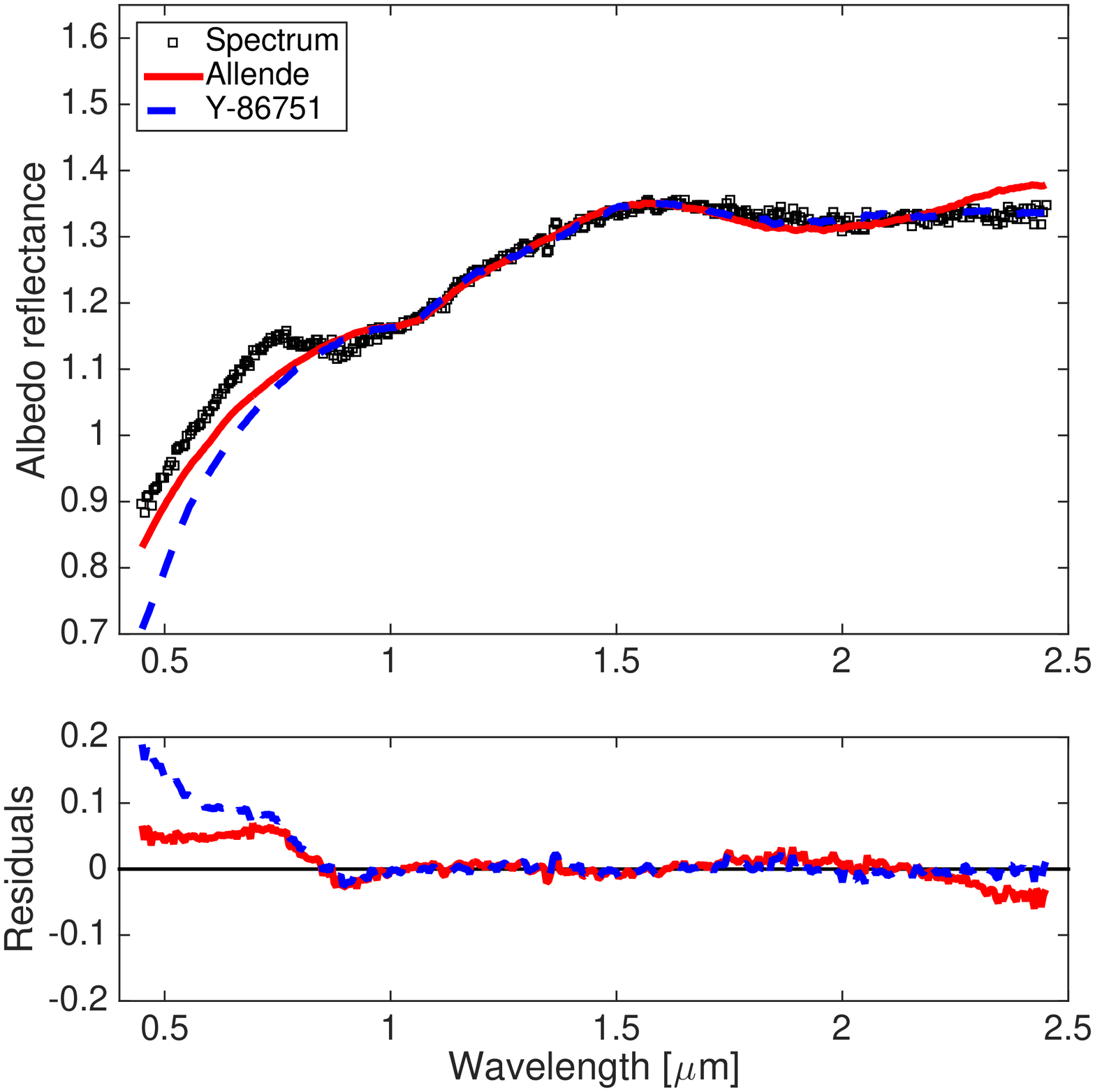}}

&
{\includegraphics[width=7cm]{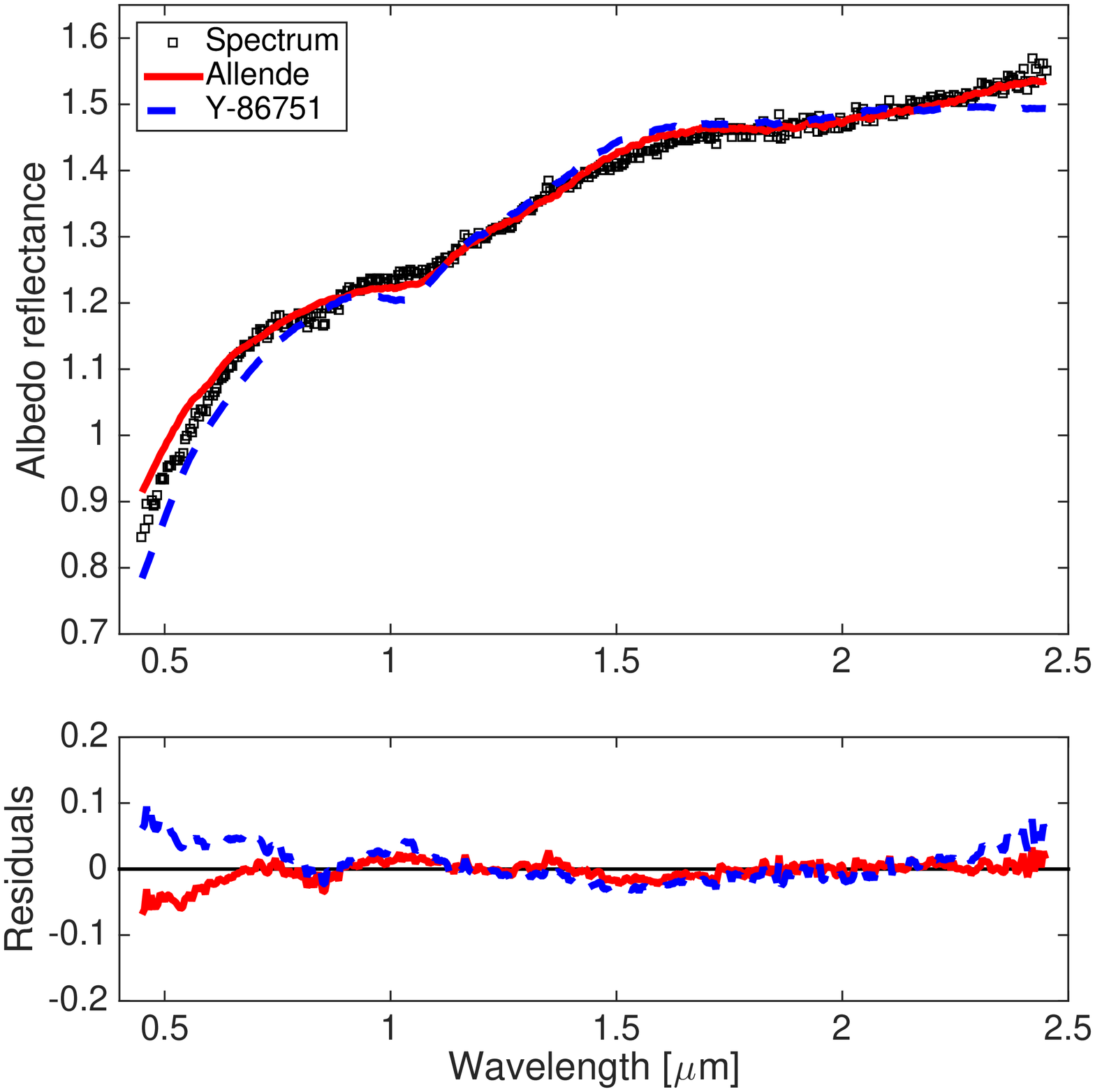}}

\\
     {(599)~Luisa}

& {(611)~Valeria} \\

\hline
 \includegraphics[width=7cm]{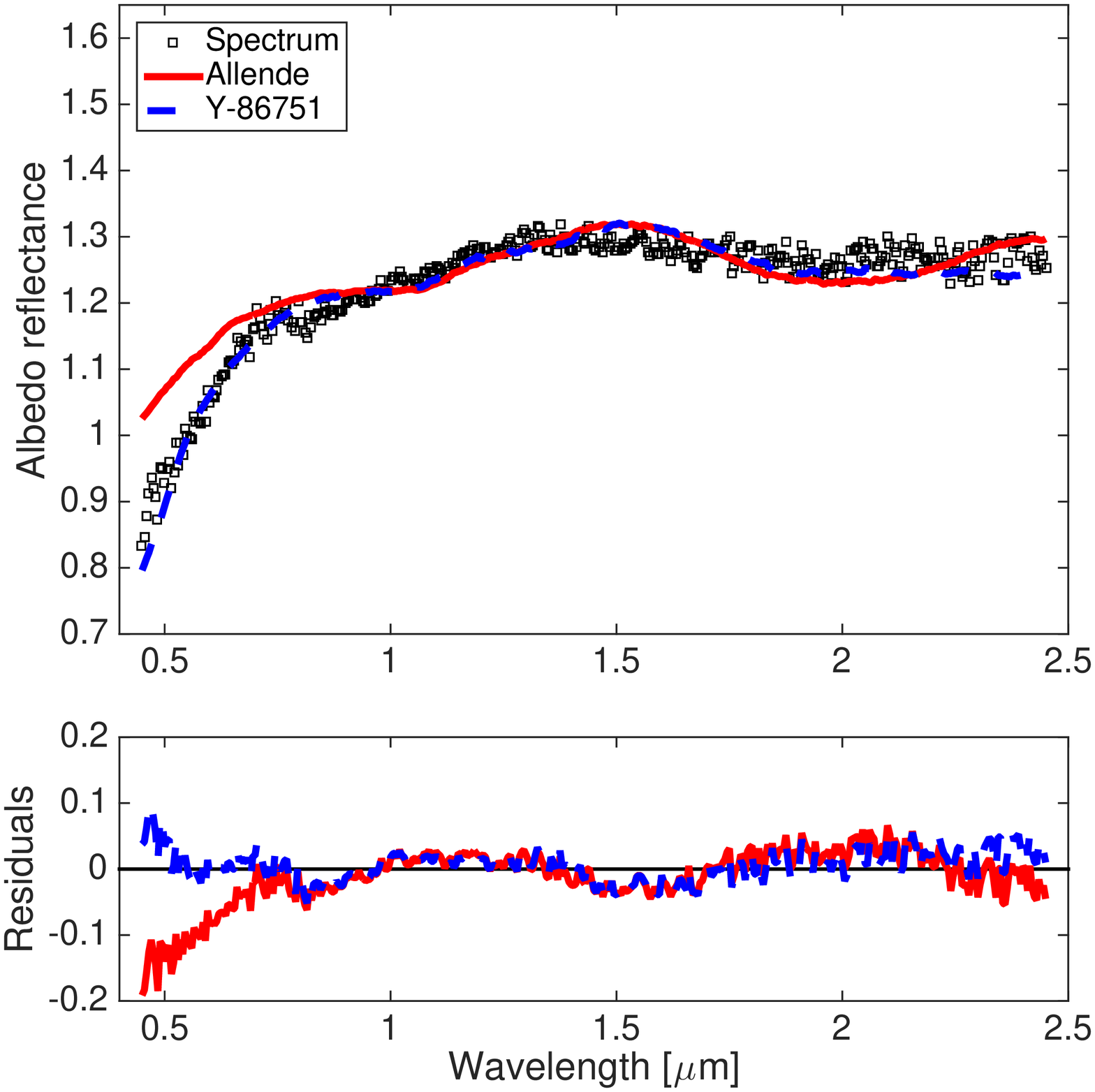}

&
\includegraphics[width=7cm]{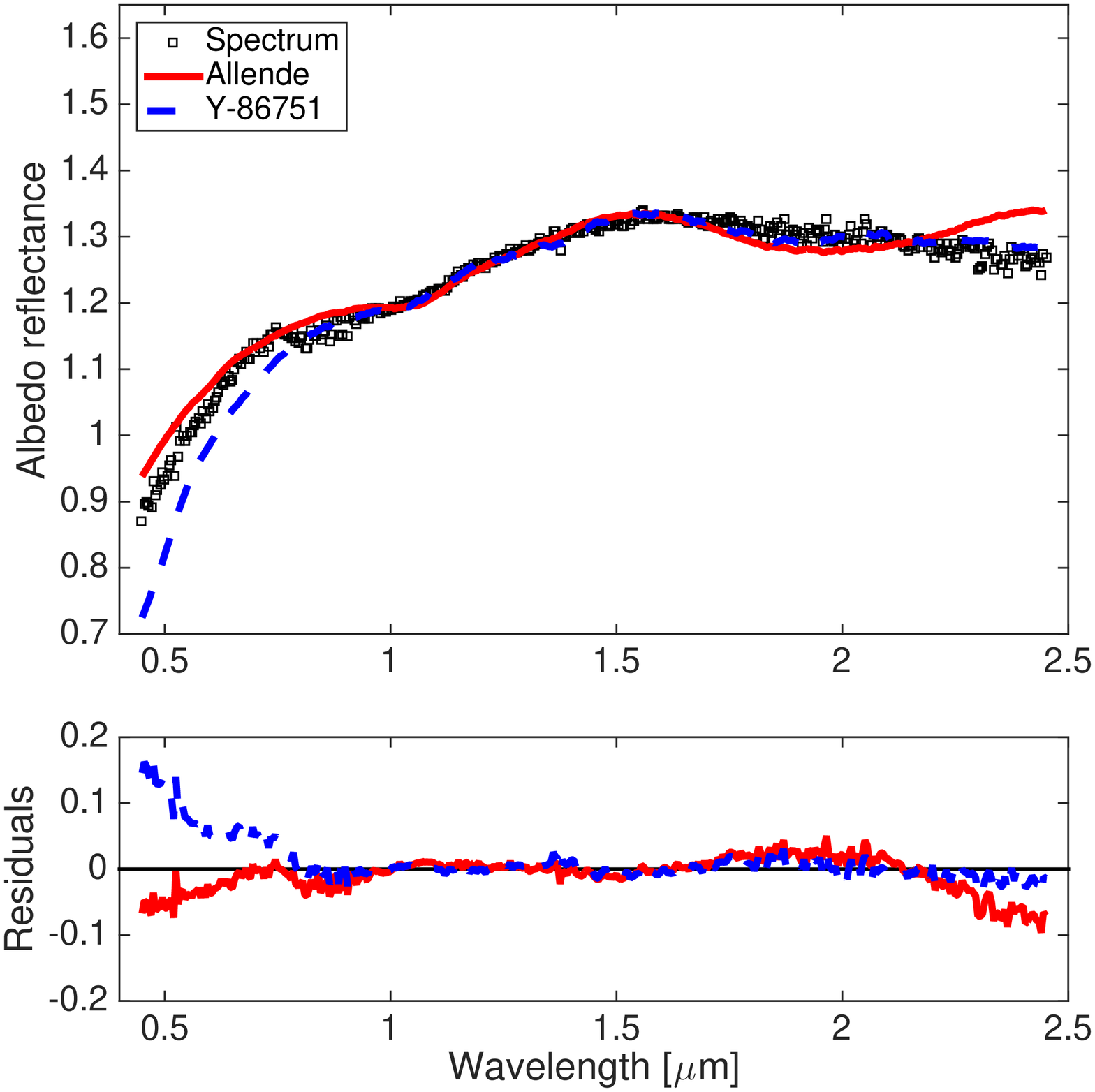}

\\
     {(673)~Edda}

&
     {(679)~Pax}
     \\
     
\hline
{\includegraphics[width=7cm]{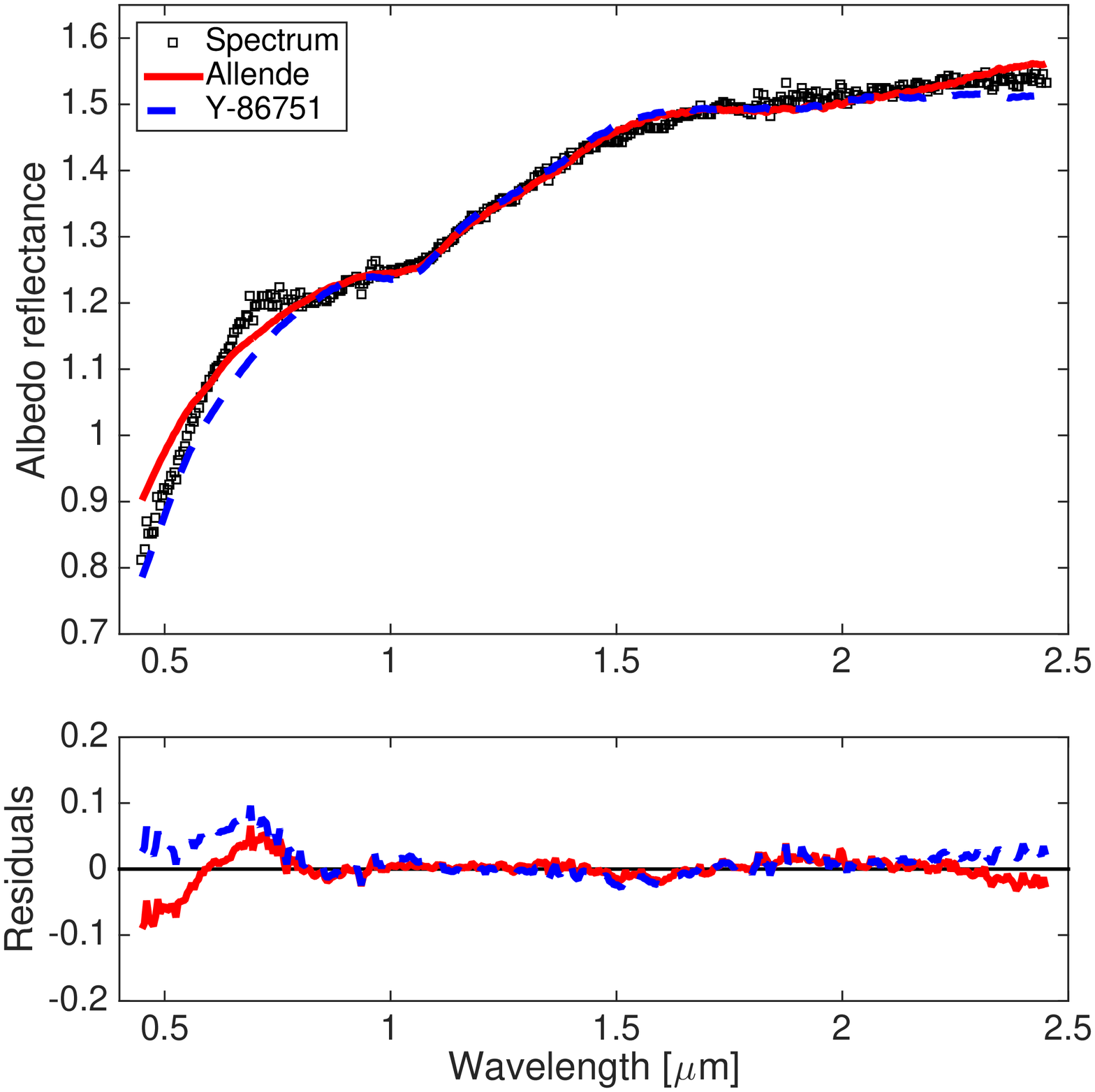}}

&
   {\includegraphics[width=7cm]{824_Fit_4.eps}}

\\
 {(729)~Watsonia}
& {(824)~Anastasia}\\
\hline
\end{tabular}
\caption{continued}
\end{figure*}

\begin{figure*}
\centering
\addtocounter{figure}{-1}
\begin{tabular}{|c|c|}
\hline
{\includegraphics[width=7cm]{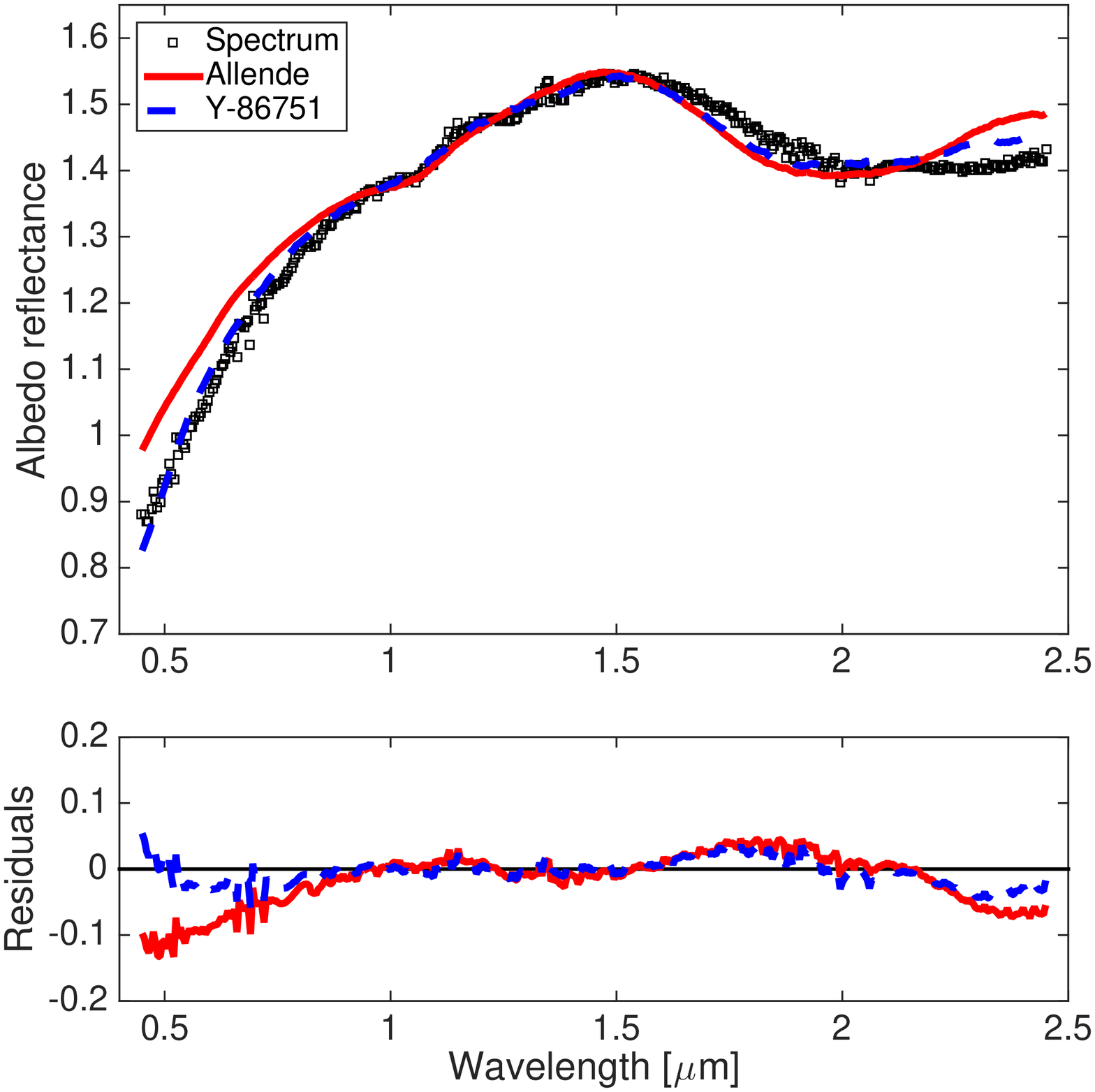}}

&
{\includegraphics[width=7cm]{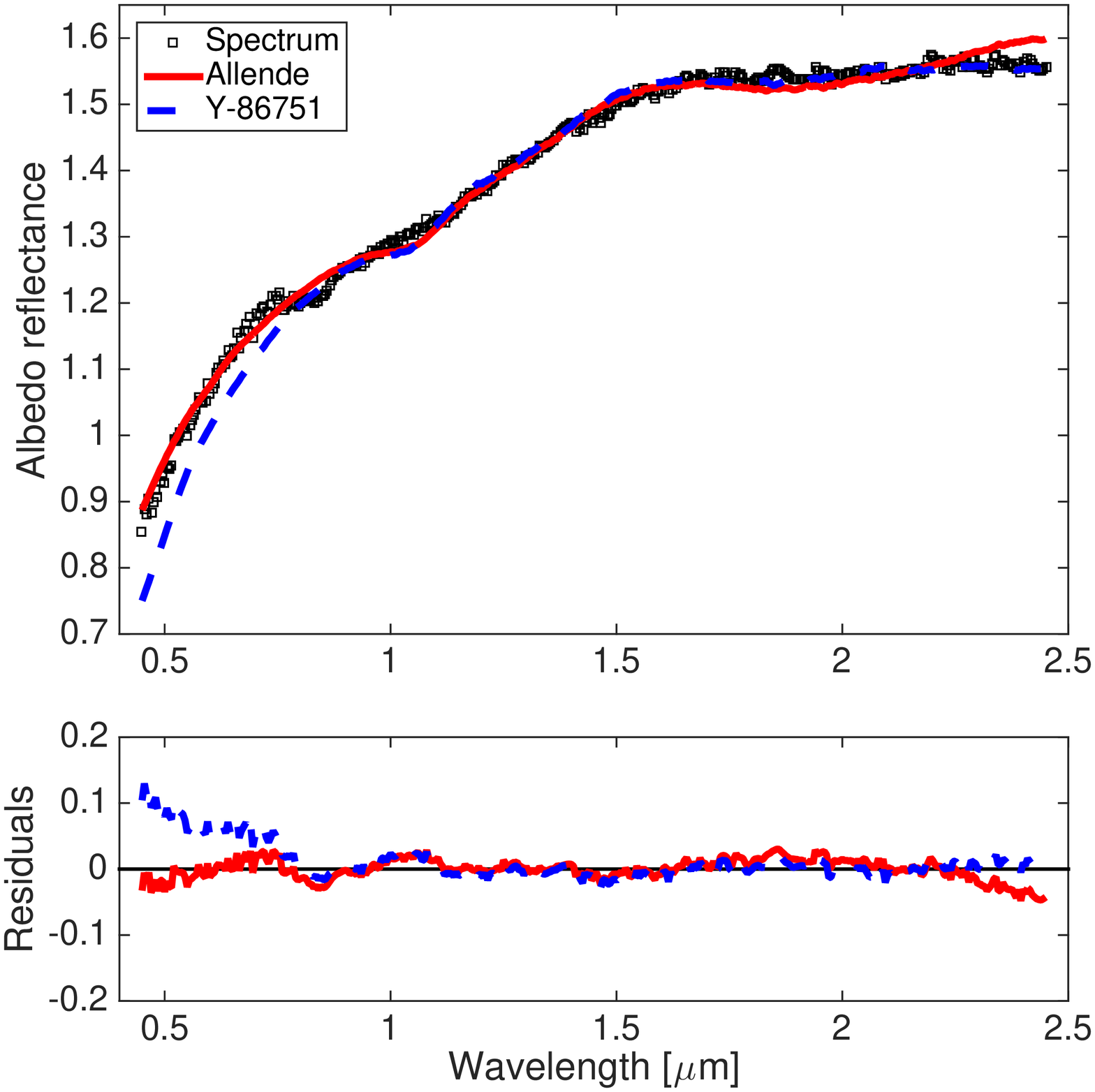}}

\\
     {(980)~Anacostia}

&     {(1372)~Haremari}  \\

\hline
{\includegraphics[width=7cm]{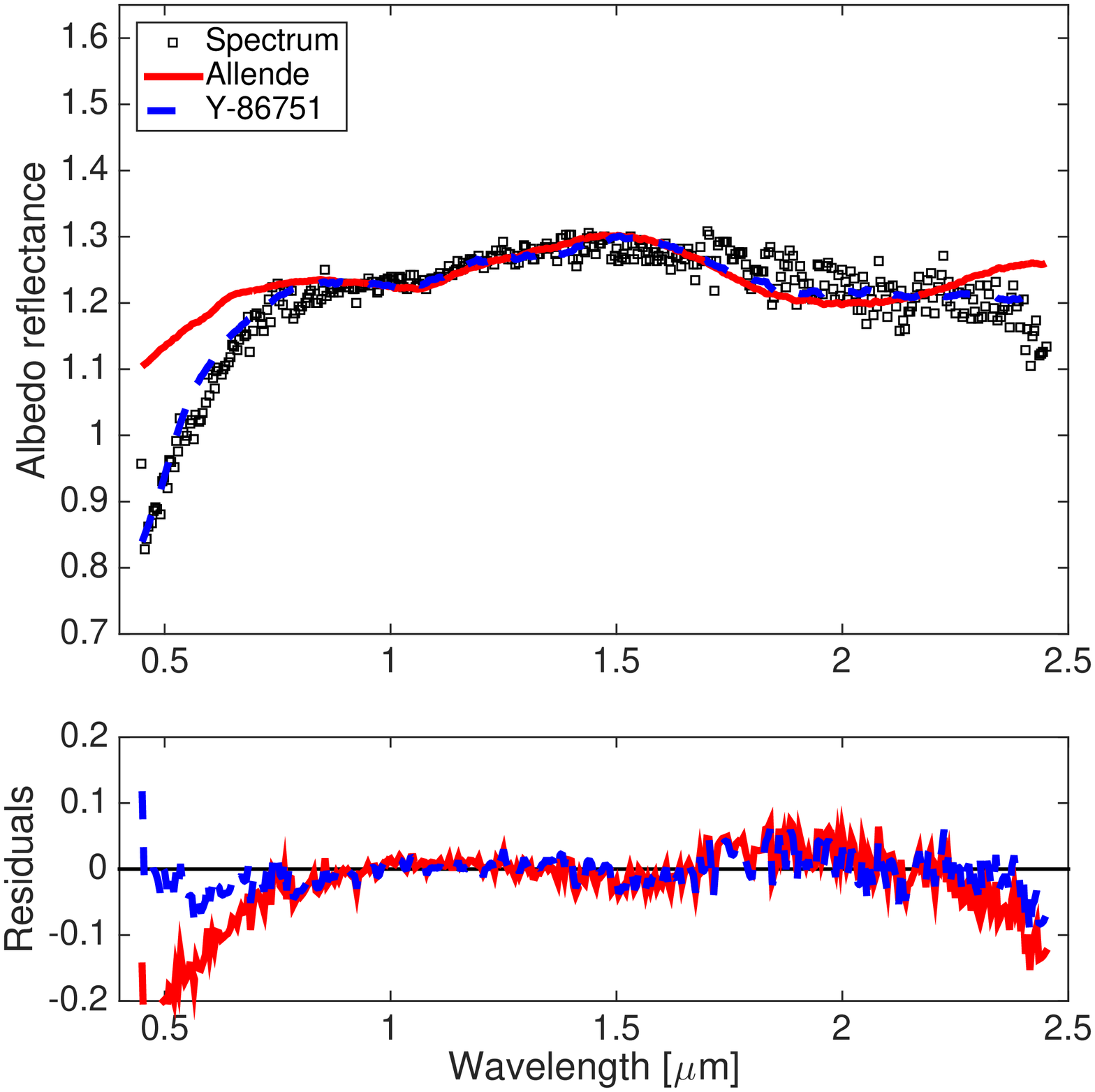}}

&

{\includegraphics[width=7cm]{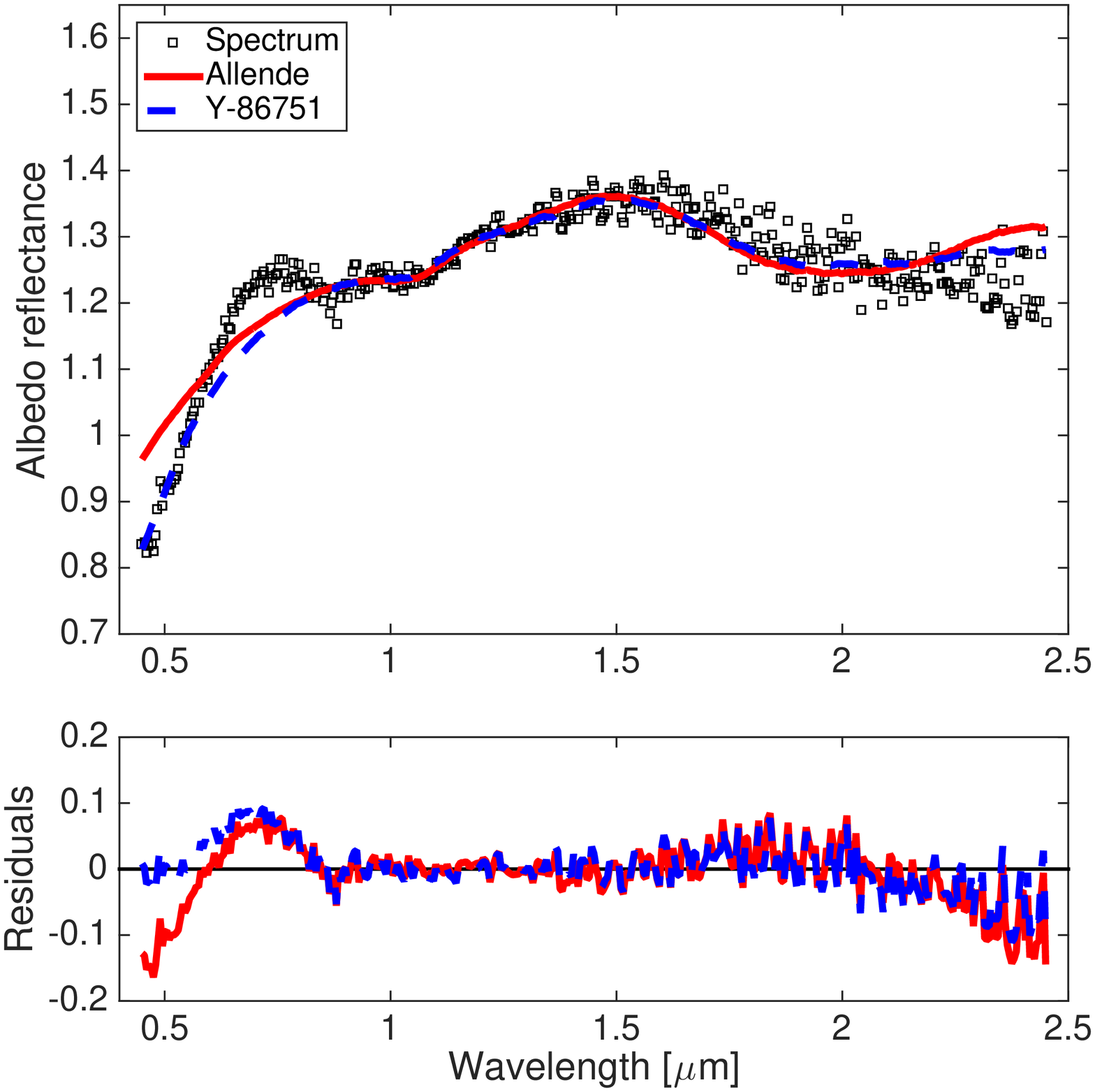}}

\\
     {(2085)~Henan}

&      {(2354)~Lavrov}  \\

\hline

 {\includegraphics[width=7cm]{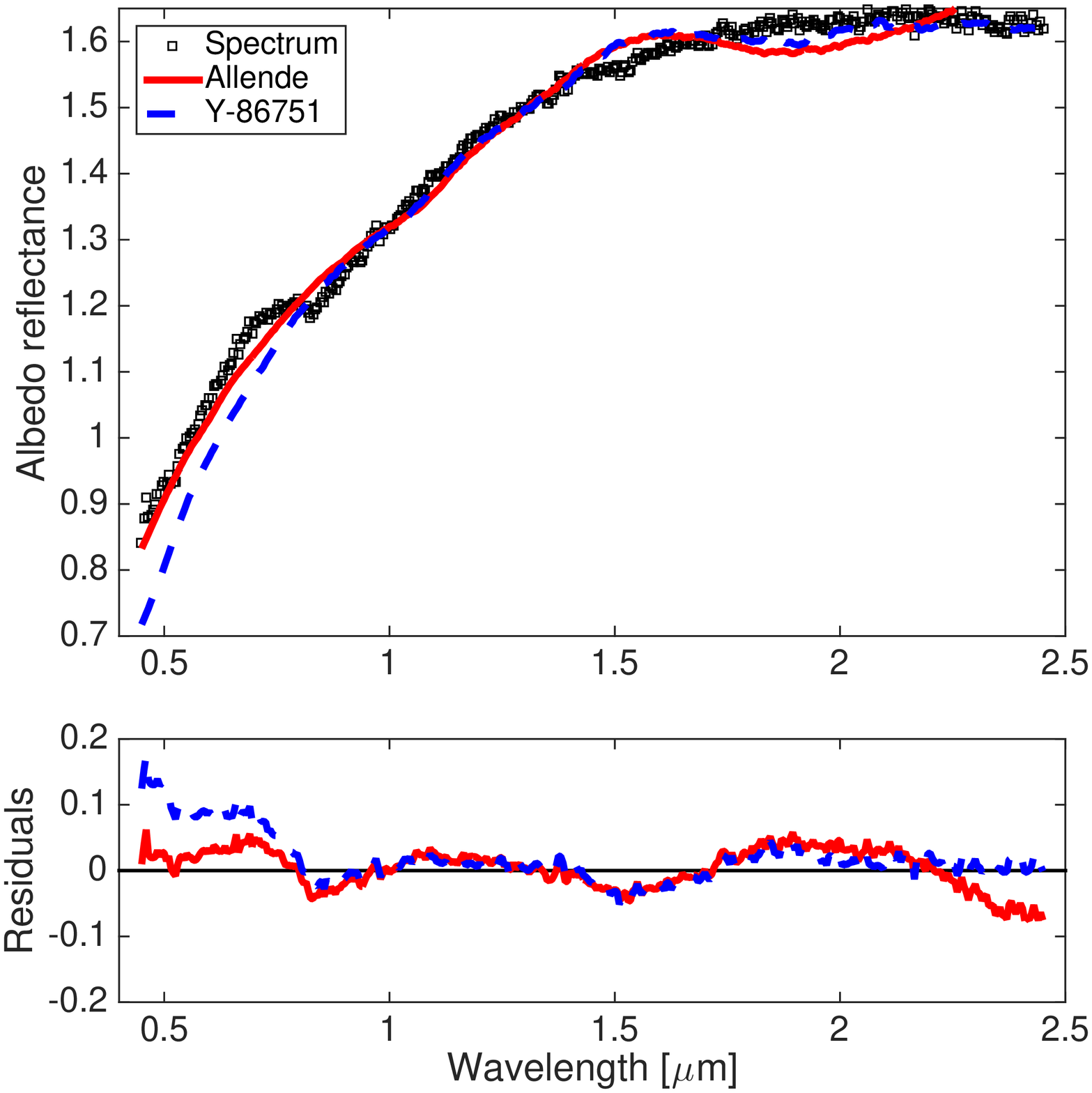}}

&
{\includegraphics[width=7cm]{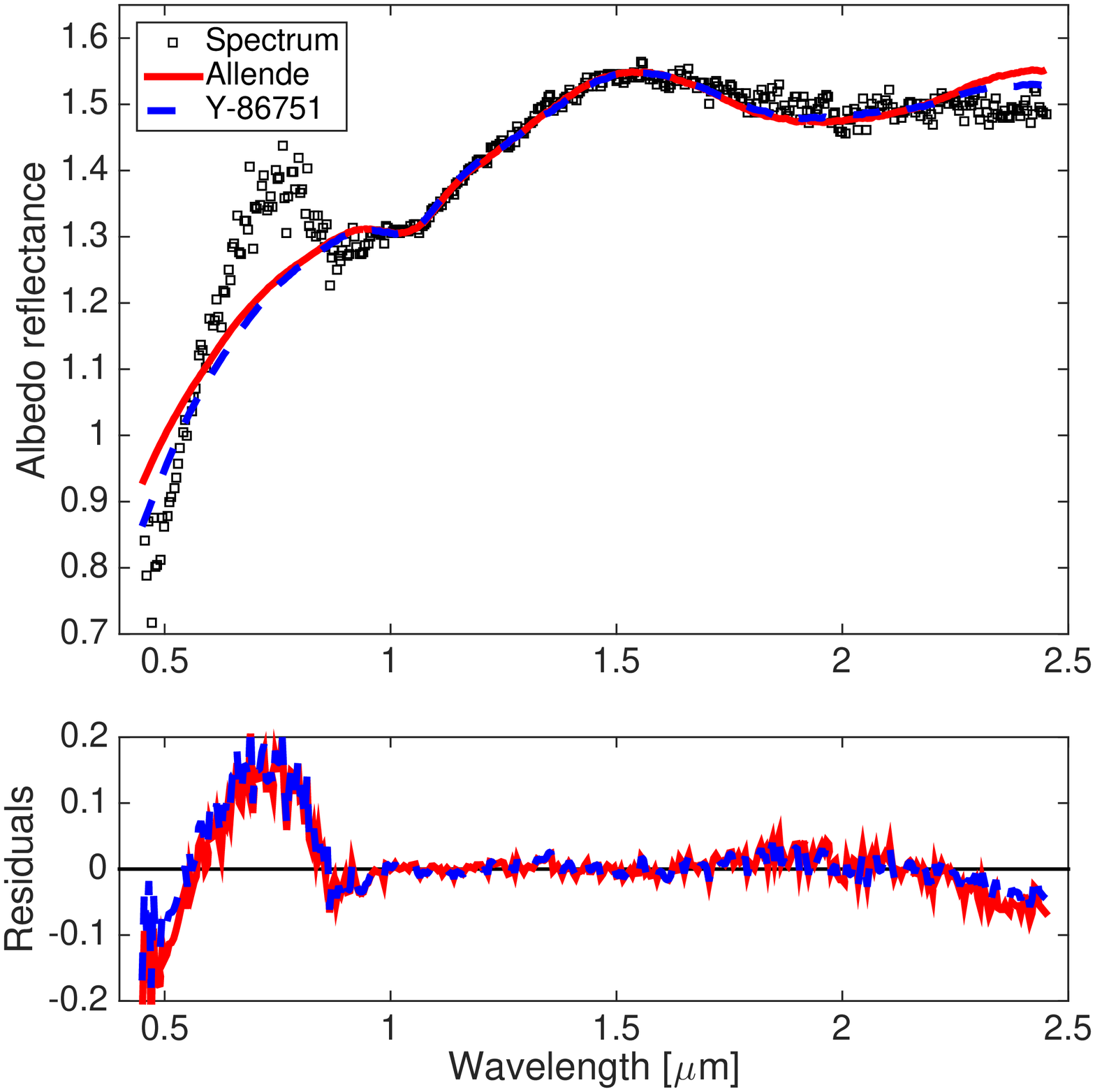}}

\\
      {(2448)~Sholokhov}

&
      {(2732)~Witt}    
     \\
     
\hline
\end{tabular}
\caption{continued}
\end{figure*}

\begin{figure*}
\addtocounter{figure}{-1}
\centering
\begin{tabular}{|c|c|}

\hline
{\includegraphics[width=7cm]{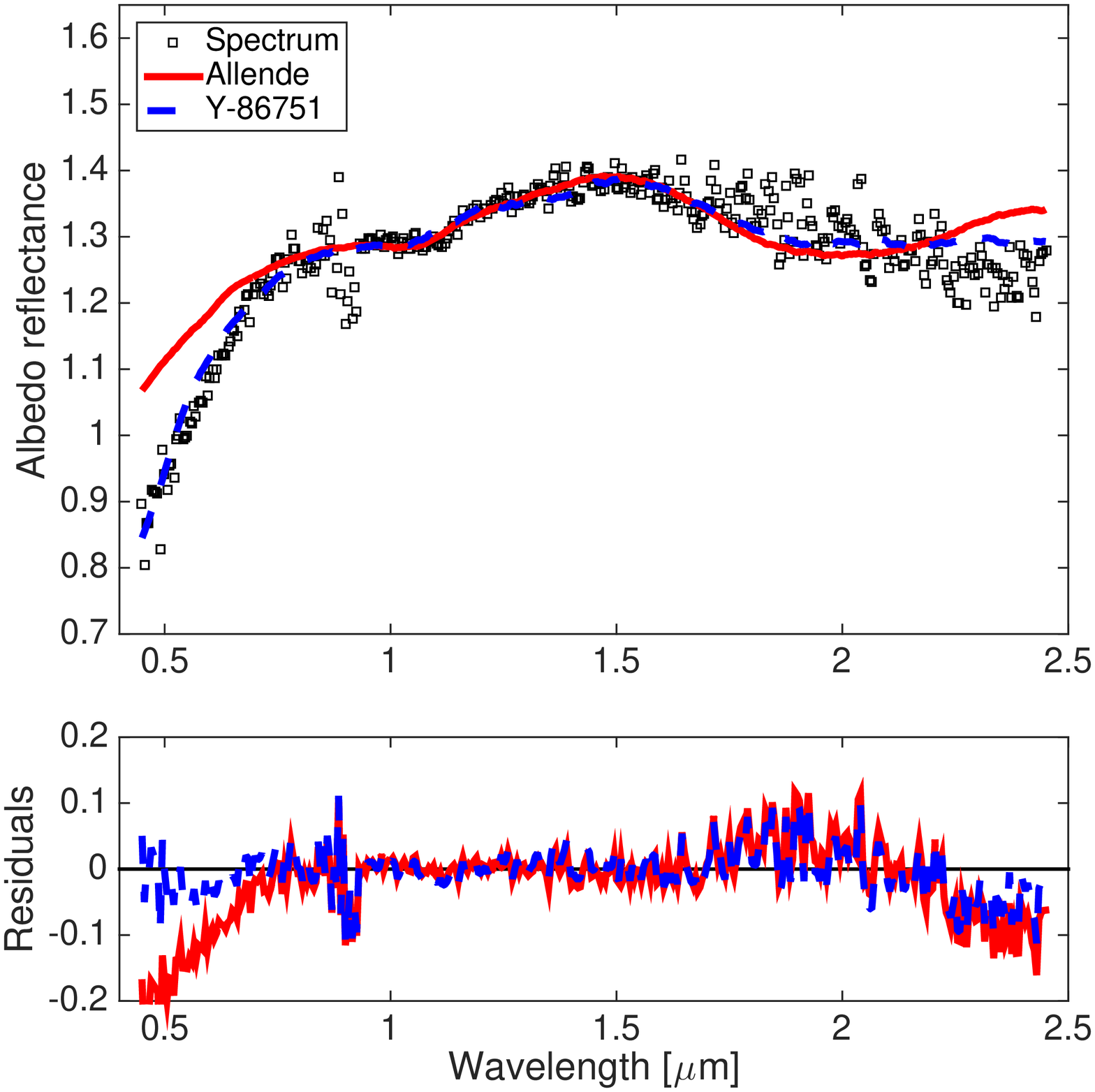}}

&
{\includegraphics[width=7cm]{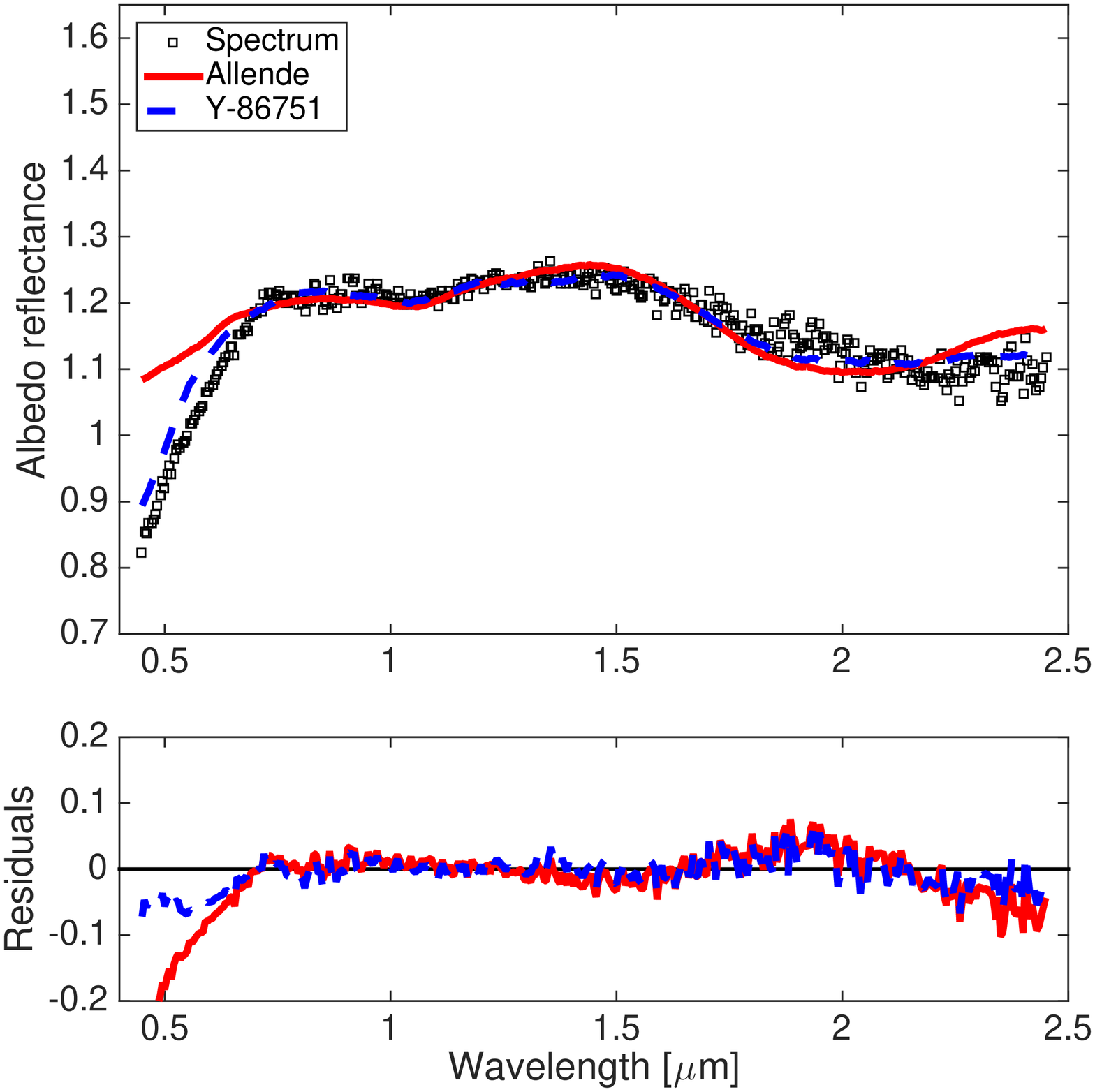}}

\\
     {(3734)~Waland}
     
& {(3844)~Lujiaxi}\\

\hline
{\includegraphics[width=7cm]{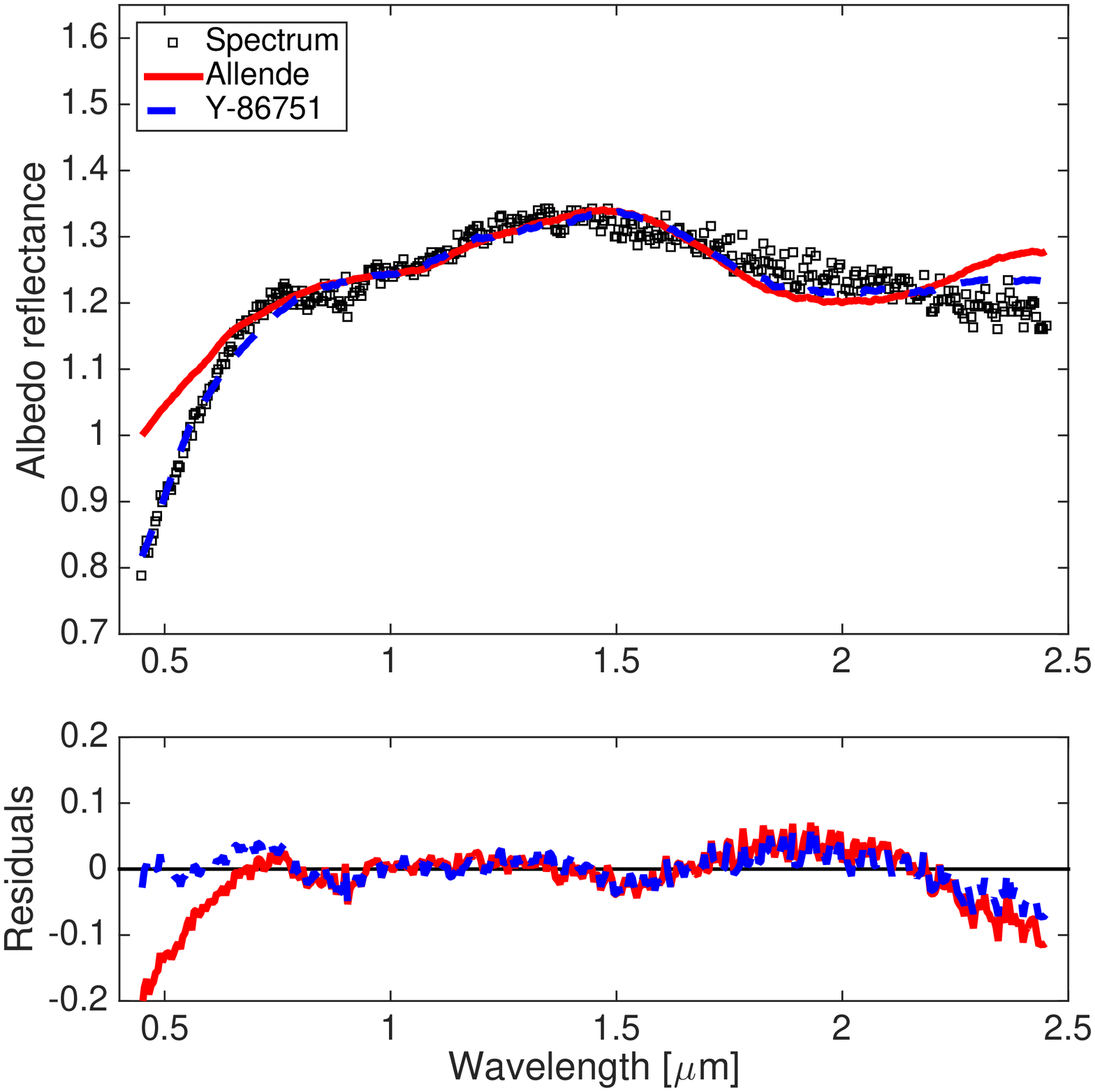}}

&
{\includegraphics[width=7cm]{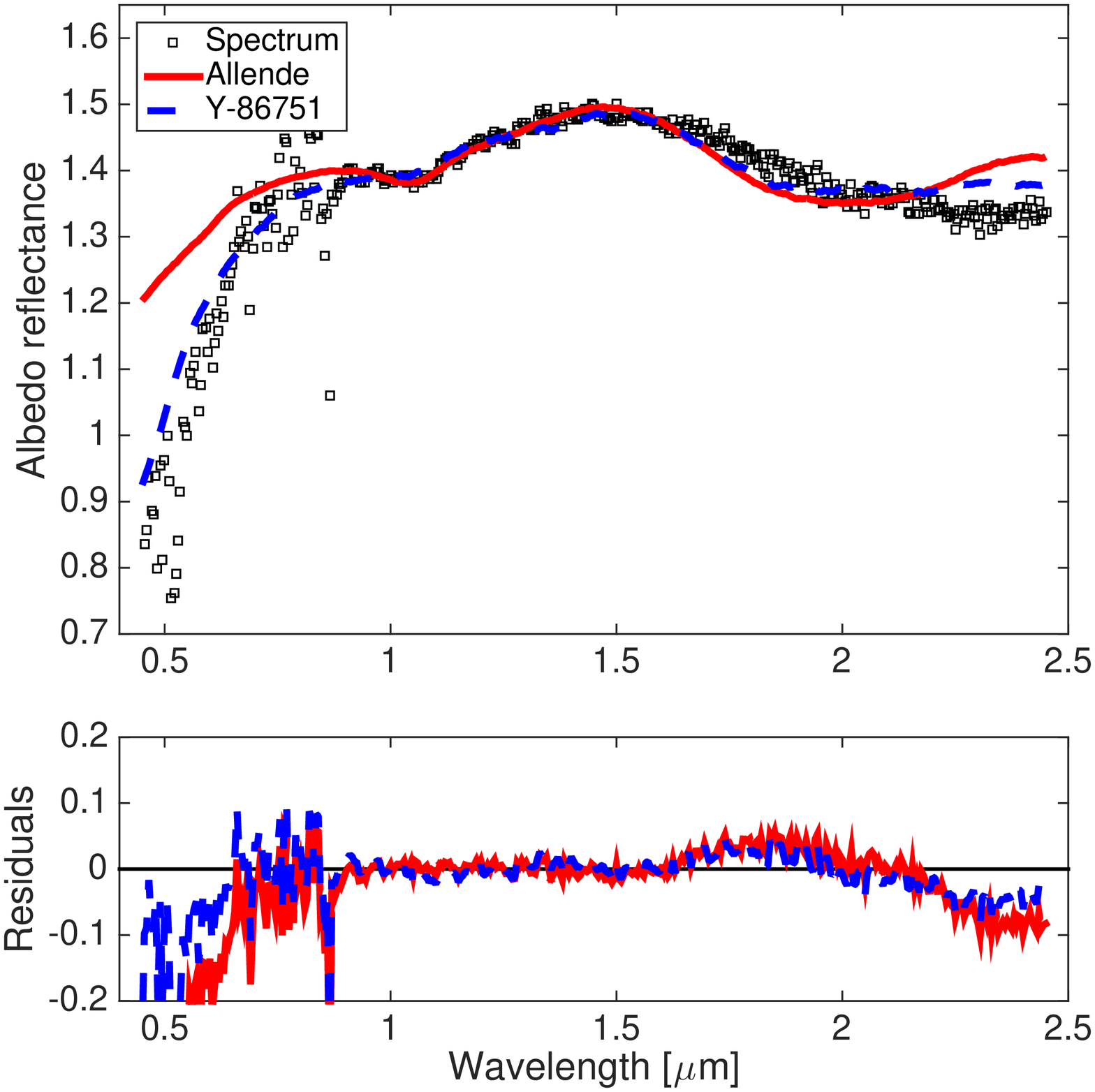}}

\\
     {(4737)~Kiladze}
     
& {(5840)~Raybrown}\\

\hline
{\includegraphics[width=7cm]{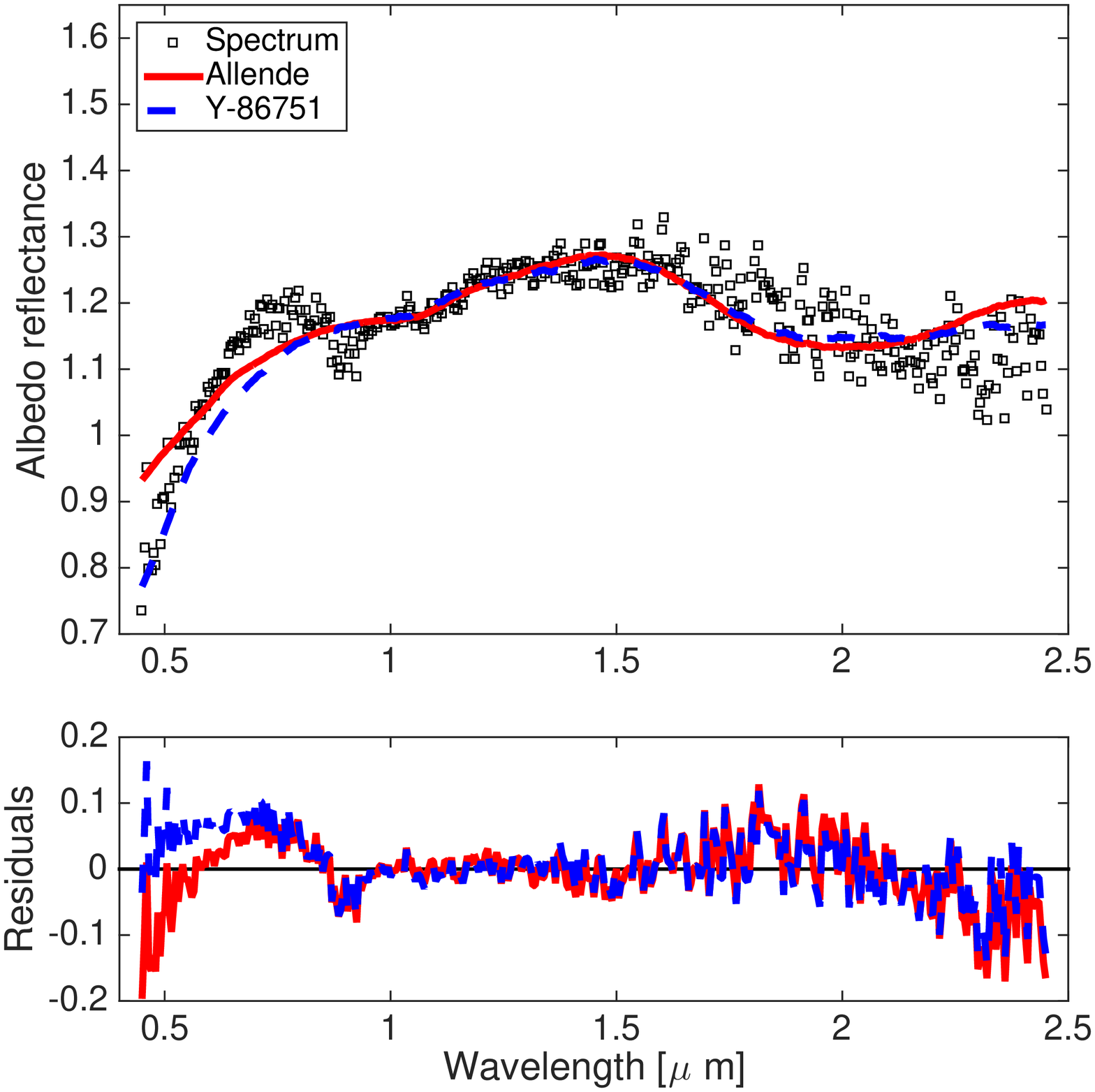}}

&
{\includegraphics[width=7cm]{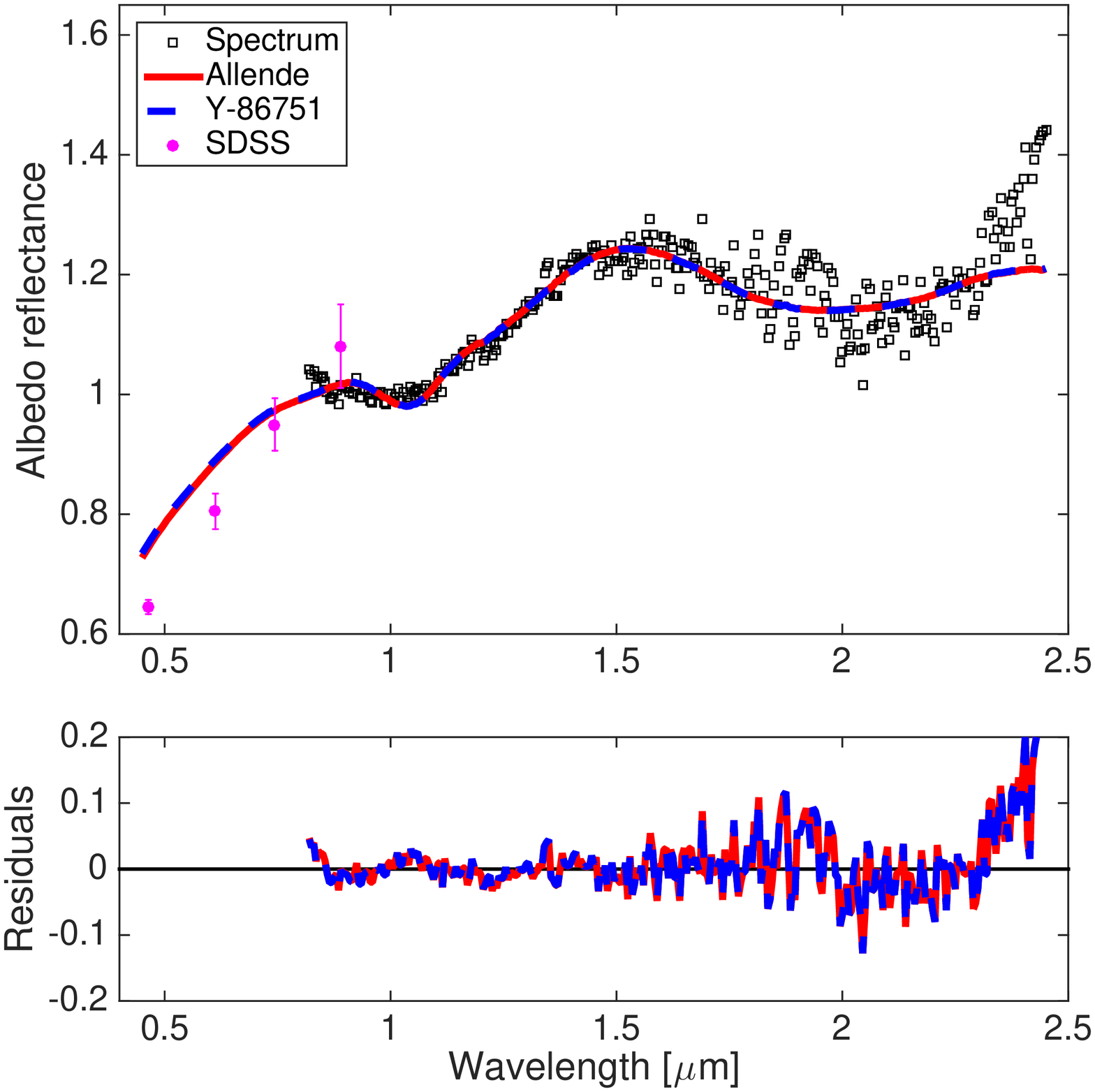}}

\\
     {(7763)~Crabeels}
     
&      {(8250)~Cornell}\\

\hline

\end{tabular}
\caption{continued}
\end{figure*}

\begin{figure*}
\centering
\addtocounter{figure}{-1}
\begin{tabular}{|c|c|}

\hline
{\includegraphics[width=7cm]{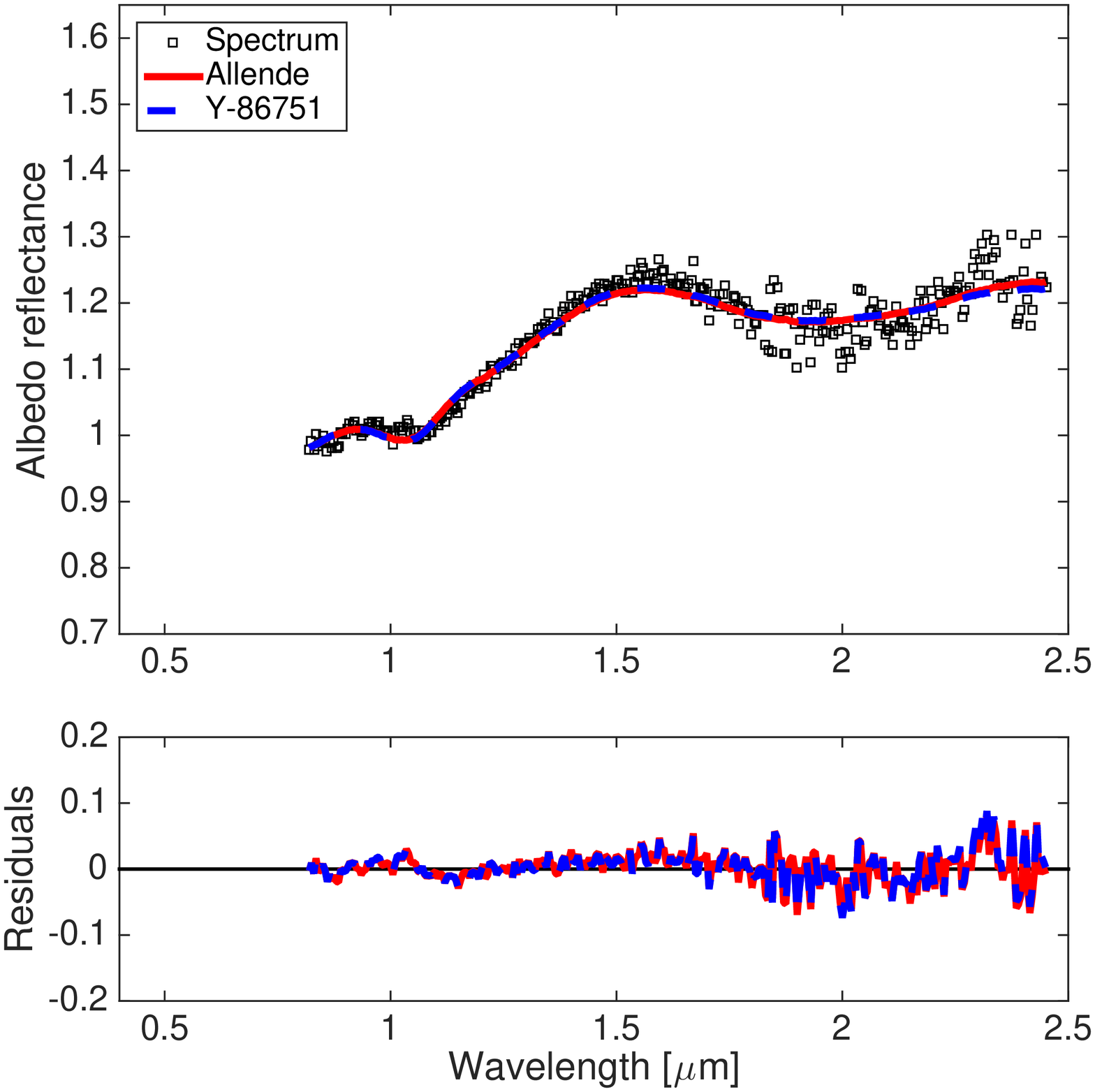}}

&
{\includegraphics[width=7cm]{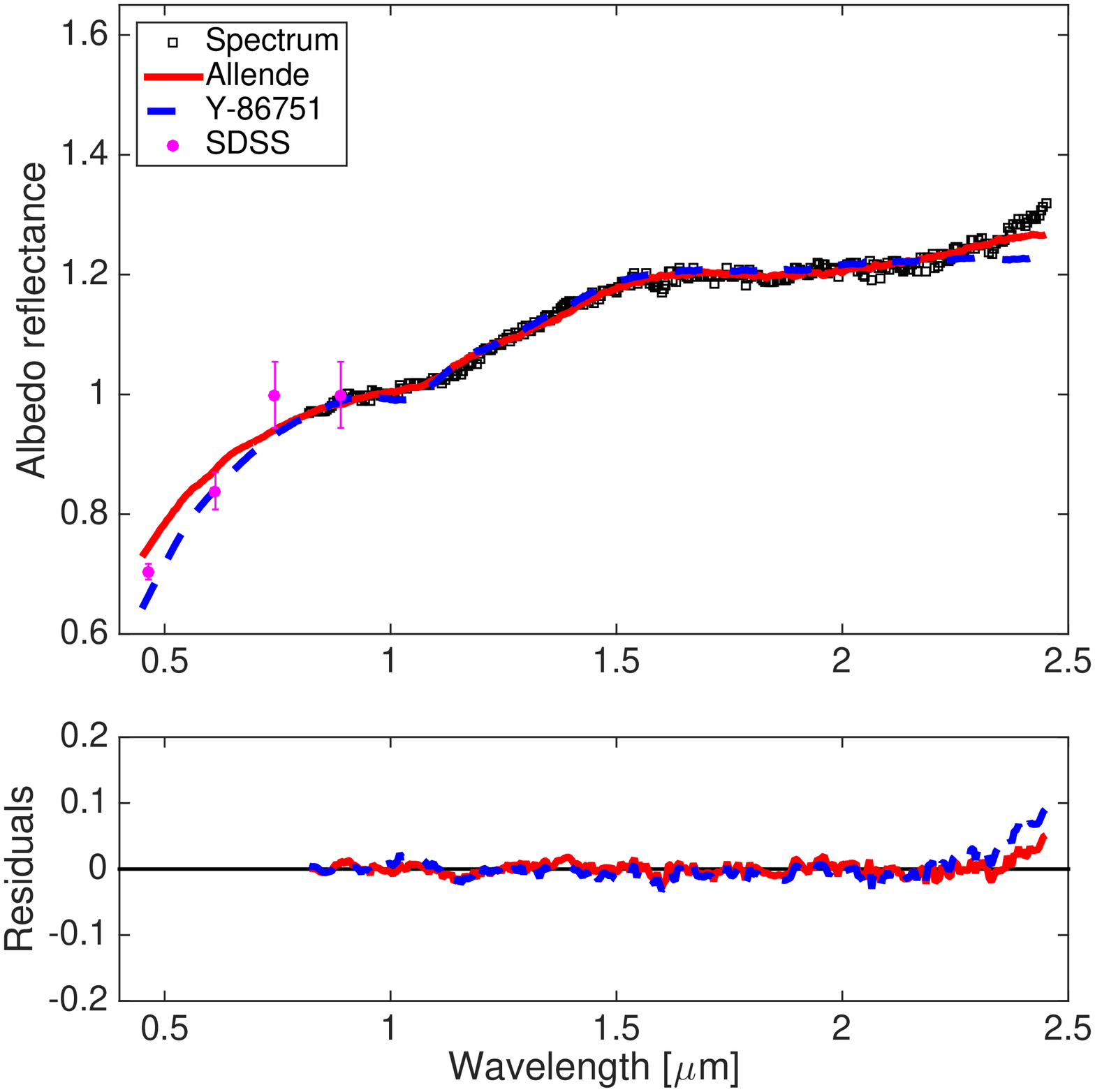}}

\\
    {(15552)~Sandashounkan}

& {(26219)~1997 YO}  \\

\hline
{\includegraphics[width=7cm]{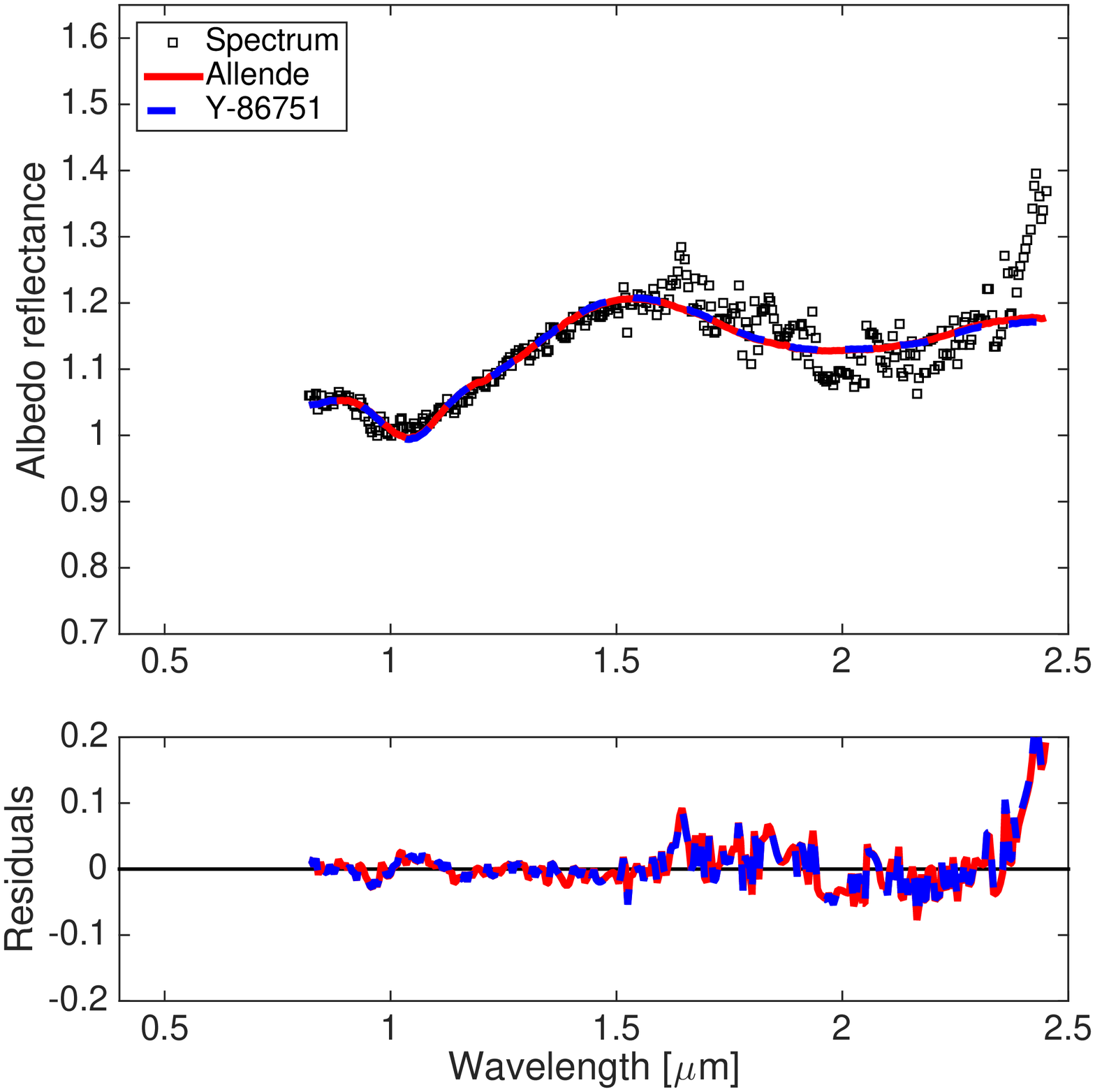}}
 &{\includegraphics[width=7cm]{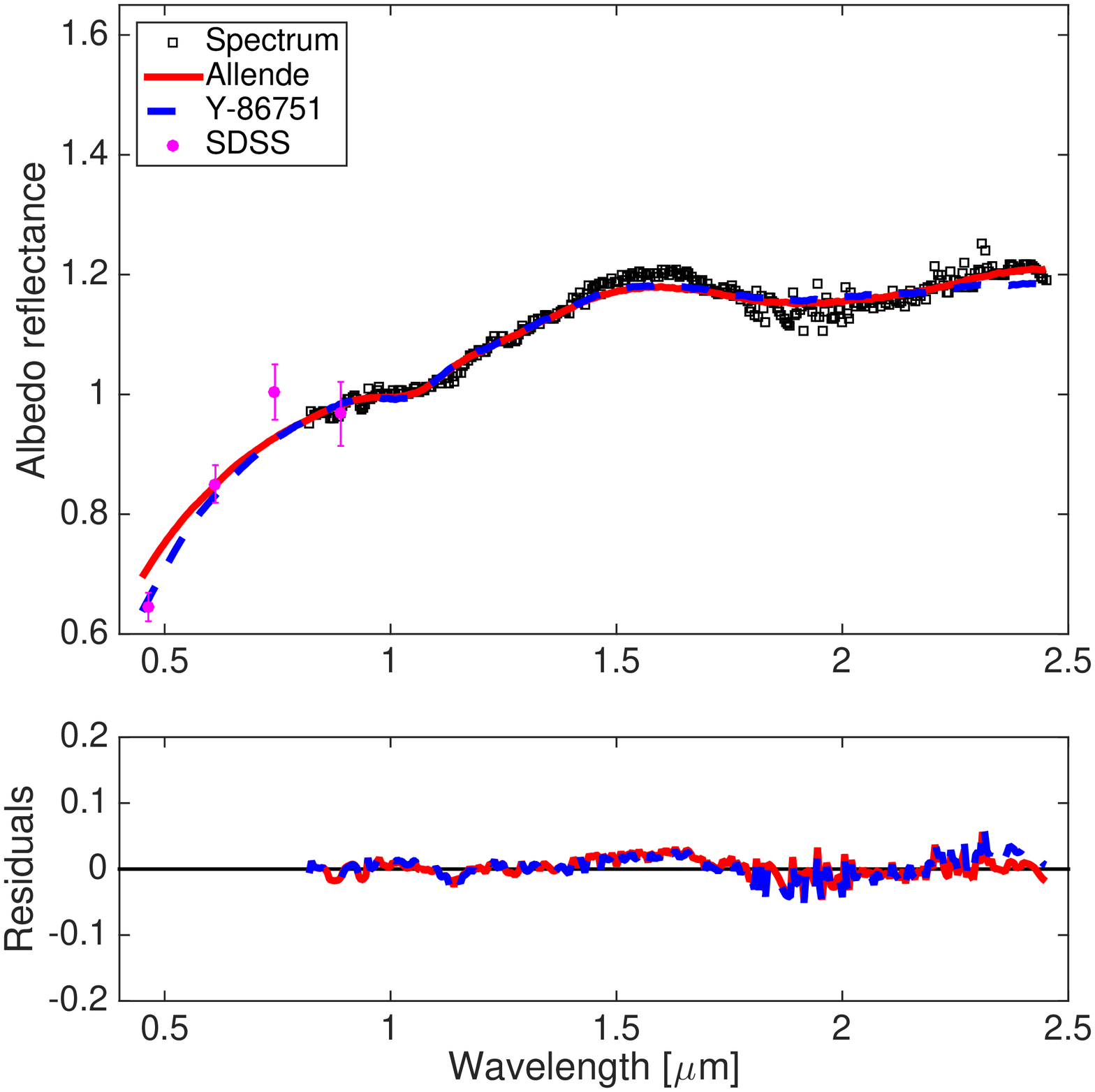}} \\

{(67255)~1997 WO21} & {(67255)~2000 ET109} 

\\

& \\

\hline

\end{tabular}
\caption{continued}
\end{figure*}

\section{Polarimetric observation presented in this work}
\label{app:pola}

\begin{table}
\centering                          
\begin{tabular}{r c c c c}         
\hline\hline                 
Object  & Date & Phase 	& $P_{\rm r}$ 	& Observatory \\ 
Number &         &  [Deg]  	& [\%] 				&         \\  
\hline                        
     12 & 04/02/2016 & 9.83  &$ -0.889 \pm 0.040$ & Calern\\
      122 &  20/05/2015 &   3.07 &$ -0.475 \pm 0.022 $ & Calern \\
   122 &  22/05/2015 &  3.45    & $-0.553 \pm 0.032 $  & Calern\\
   122 & 04/07/2016  & 7.24  & $-0.795 \pm 0.032 $ & Calern\\
   122 & 26/07/2016  & 0.61  & $-0.023 \pm 0.032 $ & Calern\\
   122 & 18/08/2016  & 7.79  &  $-0.635 \pm 0.130$ & Calern \\
      172 & 10/10/2015& 23.17& $-0.699 \pm 0.046$ & Calern   \\
   172 & 09/12/2015 &12.00 & $-1.345 \pm 0.040  $ & Calern \\
      234 &25/02/2015 &16.95 &$-1.543 \pm 0.070 $ & Calern   \\
   234 &22/05/2015 &17.38  & $-1.430 \pm 0.026  $ & Calern  \\
      234 & 19/07/2016  & 33.54 &$ 0.620 \pm   0.046$ & Calern\\
	234	&	04/12/2016	&	18.34	&	$-1.494 \pm 0.090$& Calern	\\
   236& 22/02/2015 &11.50	& $-1.314 \pm 0.040  $& Calern   \\
   236 &30/03/2015 & 0.41  & $ 0.032 \pm 0.027   $ & Calern \\
   236 &21/05/2015 &15.26 & $-1.240 \pm 0.040  $  & Calern \\
 236 & 07/06/2016 & 5.02  & $-0.873  \pm 0.064 $ & Calern \\
   236 &  05/07/2016 &  9.98  & $-1.085 \pm 0.101 $ & Calern\\
   236 & 17/07/2016 & 13.93  & $-1.302  \pm 0.069$ & Calern \\
   236 & 26/07/2016 & 16.52 & $-1.238  \pm 0.065 $& Calern \\
   236 & 28/07/2016 & 17.05  & $-1.451  \pm 0.079$& Calern \\   
   387 &10/10/2015 &16.47 & $-1.374 \pm 0.039 $ & Calern  \\
   387 &09/12/2015 &12.97  & $-1.532 \pm 0.040$ & Calern   \\
   387 & 10/04/2016 & 17.74& $-1.430 \pm 0.060$ & Calern  \\
   387 &13/04/2016 & 17.80 & $	-1.540 \pm 0.060 $& Calern  \\
   387 &01/12/2016 & 17.76 & $	-1.240 \pm 0.050 $& Calern  \\
   402 &10/10/2015 &17.35 & $-1.743 \pm 0.057  $ & Calern \\
   402 &14/10/2015 &18.05  & $-1.698 \pm 0.068 $ & Calern  \\
   402 &10/12/2015 &19.36  & $-1.502 \pm 0.106 $ & Calern  \\
   402 &01/12/2016 &6.78  & $-1.060 \pm 0.036 $ & Calern  \\
   402 &07/12/2016 &7.45  & $-1.163 \pm 0.041 $ & Calern  \\
   402 &09/12/2016 &8.17  & $-1.458 \pm 0.061 $ & Calern  \\
   402 &12/12/2016 &8.99  & $-1.340 \pm 0.082$  & Calern  \\
402 &27/12/2016 &14.07  & $-1.714 \pm0.047 $ & Calern  \\
   402 &28/12/2016 &14.42 & $ -1.804 \pm0.041 $ & Calern  \\
   402 &03/01/2017 &16.35 &  $-1.795 \pm0.032  $& Calern  \\
   402 & 07/01/2017 & 17.28  & $-1.985 \pm 0.091$& Calern  \\
 402 &16/01/2017& 19.90 &$-1.580 \pm 0.039 $ &Calern \\
   402& 17/01/2017 &20.15  &$ -1.391 \pm 0.052  $&Calern \\
   402 &24/01/2017 &21.58  & $-1.412 \pm 0.057  $& Calern \\
   458 & 10/05/2016 & 16.15 & $-1.854 \pm 0.080 $& Calern \\
    460 & 06/08/2016 & 3.68  & $-1.435 \pm 0.058 $& Calern \\
   460 & 14/08/2016 & 3.19  & $-1.203 \pm 0.074  $& Calern\\
   460 & 21/08/2016 & 5.51  & $-1.454  \pm 0.110 $& Calern \\          
     478 &18/07/2015 & 7.80 & $-0.732 \pm 0.099 $ & Calern  \\
   478 &21/07/2015 & 8.60  & $-0.683 \pm 0.039  $ & Calern\\
   478 &29/09/2015 &17.82  & $-0.100 \pm 0.100 $ & Calern  \\
   478 & 29/07/2016 & 14.68 & $-0.622 \pm 0.063$ & Calern \\  
   599 & 10/04/2016 & 10.09 & $-1.510 \pm 0.100$ & Calern \\

\hline                                   
\end{tabular}
\caption{Summary of our polarimetric measurements. The first column corresponds to the number of the observed asteroid. The second indicates the date of observation. The third one corresponds to the phase angle (angle between the sun-asteroid-observer). $P_{\rm r}$ is the polarization degree. Finally the Observatory column gives the observatory in which the data were acquired.}
\label{Pola_meas} 
\end{table}

\begin{table}
\centering                          
\addtocounter{table}{-1}
\begin{tabular}{r c c c c}          
\hline\hline                 
\multicolumn{1}{c}{Object} & Date & \multicolumn{1}{c}{Phase} & \multicolumn{1}{c}{$P_{\rm r}$} & Observatory \\ 
                            Number                      &         &  \multicolumn{1}{c}{[Deg]}   & [\%] &           \\  
\hline                        
     606 &10/12/2015 &19.80 & $-1.130 \pm 0.230$ & Calern \\
   606 &11/12/2015 &19.60  & $-1.299 \pm 0.114$ & Calern \\
      611& 20/12/2014 &20.28&  $-0.920 \pm 0.040$ & Rozhen \\
      611& 11/12/2015 &19.28& $ -0.885 \pm 0.057$ & Calern \\
   642 &10/10/2015 & 4.11  & $-0.950\pm0.042$  & Calern\\
   642 &01/12/2016 & 8.42  & $-1.725\pm0.070$  & Calern\\
   642 &07/12/2016 & 6.55  & $-1.328\pm0.090$  & Calern\\
   642 &12/12/2016 & 4.98  & $-1.183\pm0.094$  & Calern\\
   642 &03/01/2017  &7.50 &  $-1.556\pm 0.090$  &Calern\\  
    642 &06/01/2017  &8.49 & $-1.572 \pm 0.115$  &Calern\\
   679 &13/07/2015&  3.72 &  $-0.595 \pm 0.036$  & Calern\\
   679 &01/12/2016&  22.27 & $ -0.892 \pm 0.044$  & Calern\\
   679 &09/12/2016&  20.39 &  $-1.272 \pm 0.061$  & Calern\\
 	679& 21/12/2016 &17.01  & $-1.494\pm 0.061$ & Calern\\
   679& 27/12/2016 &14.94  & $-1.625 \pm 0.052$  & Calern\\
   679 &28/12/2016 &14.57  & $-1.656 \pm 0.038$  & Calern\\
679 &16/01/2017 & 7.26  & $-1.158 \pm 0.025$ &Calern \\
   679& 18/01/2017 & 6.52 &$ -1.246 \pm 0.051$  & Calern\\
  729 &11/12/2015& 18.10 & $-1.334 \pm 0.066$ & Calern   \\
     729 &01/12/2016& 13.86 &$ -1.196 \pm 0.055$ & Calern   \\
     729 &04/12/2016& 11.14 & $-1.040 \pm 0.134$ & Calern   \\
     729 &05/12/2016& 10.82 & $-1.240 \pm 0.066$ & Calern   \\
     729 &06/12/2016& 10.50 & $-1.171 \pm 0.062$ & Calern   \\
     729 &07/12/2016& 10.18 & $-1.177 \pm 0.060$ & Calern   \\
     729 &08/12/2016& 9.86 & $-1.183 \pm 0.122$ & Calern   \\
     729 &10/12/2016& 9.21 & $-0.882 \pm 0.133$ & Calern   \\
     729 &12/12/2016& 8.56 & $-0.843 \pm 0.100$ & Calern   \\
     729 &13/12/2016& 8.22 & $-1.057 \pm 0.050$ & Calern   \\ 
   729 &23/12/2016  &5.17 &$-0.891 \pm 0.047$  & Calern   \\ 
   729 &27/12/2016  &4.35  &$ -0.644 \pm 0.040$ & Calern   \\ 
   729 &28/12/2016 & 4.21  & $-0.705\pm 0.046$ & Calern   \\ 
   729 &03/01/2017 & 4.37 & $-0.662 \pm 0.031$  & Calern   \\ 
   729  & 06/01/2017&  4.96 &$ -0.796  \pm 0.103$   & Calern   \\
  729 & 16/01/2017  &7.99  & $-1.108 \pm 0.050$  & Calern \\
   729 & 19/01/2017 &  8.72 &  $-1.104 \pm 0.064$  & Calern \\
   753 &11/12/2015 &23.96 & $-0.123 \pm 0.100$   & Calern \\
     824 &13/07/2015 & 3.35 &$-0.795 \pm 0.051$  & Calern  \\
   824 &15/07/2015 & 3.74  & $-0.924 \pm 0.051$   & Calern \\
   	824	&	04/12/2016	&	11.41	&	$-2.154 \pm 0.119$	& Calern\\	
   	824	&	13/12/2016	&	13.76	&$	-2.589 \pm 0.212$	& Calern\\	
	824  & 09/01/2017    & 18.11    & $-1.714 \pm 0.176$  & Calern \\
   908 &08/12/2015& 22.01 &$-0.381\pm 0.033 $   & Calern \\
   908 &04/12/2016& 25.36 &$ 0.640 \pm 0.166 $  & Calern  \\
   980 &04/07/2016  &3.02  & $-0.607 \pm 0.033$& Calern \\
   980 &22/07/2016 & 6.01  &$ -0.771 \pm 0.053$ & Calern\\  
   980 &26/07/2016 & 7.98 & $-1.032 \pm 0.076$  & Calern\\
   980 &05/08/2016 &12.48  & $-1.320\pm 0.064$ & Calern\\
   980 &10/08/2016 &14.54  & $-1.175 \pm 0.107$ & Calern \\
   980 & 02/12/2016 & 22.30 & $-0.775 \pm  0.041$& Calern \\
 1040 &06/08/2016 & 4.14 & $-0.121 \pm  0.088$ & Calern \\
  1040 &12/08/2016 & 4.71 & $-0.388 \pm  0.106$ & Calern \\
  1040 &22/08/2016& 6.94 &$ -1.042 \pm  0.119$  & Calern\\
\hline                                   
\end{tabular}
\caption{continued}
\end{table} 

\begin{table}
\centering                          
\addtocounter{table}{-1}
\begin{tabular}{r c c c c}          
\hline\hline                 
\multicolumn{1}{c}{Object} & Date & \multicolumn{1}{c}{Phase} & \multicolumn{1}{c}{$P_{\rm r}$} & Observatory \\ 
                            Number                      &         &  \multicolumn{1}{c}{[Deg]}   & [\%] &           \\  
\hline                        
  1284 &10/10/2015 & 9.23  &$ -1.926 \pm 0.054$   & Calern \\
  1284 &09/12/2015 &25.20 & $-0.190 \pm 0.040$   & Calern  \\
  1284 &09/12/2016 &22.95 & $-0.531 \pm 0.082$   & Calern  \\
  1284 &10/12/2016 &22.89 & $-0.548 \pm 0.258$  & Calern  \\
  1332 & 20/12/2014 & 16.64 &$ -0.630 \pm 0.170$ & Rozhen  \\
  1332 & 10/04/2016 & 8.90 & $-2.100 \pm 0.300$ & Calern  \\
  1372 &26/02/2015 & 19.96 & $-0.911 \pm 0.124$ & Calern   \\
  1406&	04/12/2016&	11.06	&$	-0.441 \pm 0.101$	& Calern\\
  1702 &22/05/2015 & 7.39 & $-1.076 \pm 0.650$  & Calern  \\
  1702 &02/08/2016 & 12.08& $-1.007 \pm 0.108$ & Calern\\
  1702 &22/08/2016 & 5.75 & $-0.718 \pm 0.068$ & Calern	\\ 
  2085 &19/12/2014 &15.82 &$ -1.920 \pm 0.090$  & Rozhen  \\
  2354 &11/12/2015 & 3.54 & $-0.653 \pm 0.069$  & Calern  \\
  2448 &04/12/2016 & 20.27 &$ 0.055 \pm 0.129$	& Calern\\
  2448 &21/12/2016 &17.35	& $-0.643 \pm 0.119$ & Calern\\
  2732 &06/12/2016 & 8.86 &$ -0.772 \pm 0.330$	& Calern\\
  3269 &05/12/2016 & 12.48 &$ -0.919 \pm 0.275$ 	& Calern\\
\hline                                   
\end{tabular}
\caption{continued}
\end{table}

\end{appendix}

\end{document}